\documentclass[12pt]{article}

\usepackage{epsfig,rotating}   

\newcommand{\hso}{H$_2$SO$_4$}
\newcommand{\ho}{H$_2$O}

\newcommand{\degc}{$^\circ$C}
\newcommand{\wpm}{Wm$^{-2}$}
\newcommand{\cft}{$^{14}$C}
\newcommand{\beten}{$^{10}$Be}
\newcommand{\cotwo}{CO$_{2}$}

\newcommand{\lappeq}{\mathrel{\rlap{\raise.5ex\hbox{$<$}}{\lower.5ex\hbox{$\sim$}}}}
\newcommand{\gappeq}{\mathrel{\rlap{\raise.5ex\hbox{$>$}}{\lower.5ex\hbox{$\sim$}}}}

\begin{document}           

\pagestyle{empty}

\begin{titlepage}

\begin{center}

EUROPEAN ORGANIZATION FOR NUCLEAR RESEARCH

{\small
\begin{tabbing}
 \=  \hspace{117mm}  \=  \kill 
 \>   \>CERN/SPSC 2000-021 \\ 
 \> \>SPSC/P317 \\
 \> \>April 24, 2000
\end{tabbing}  }
\vspace{-1mm}

\vspace{-23mm}
\begin{figure}[htbp]
  \begin{center}
      \makebox{\epsfig{file=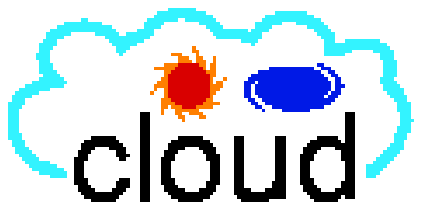,height=18mm} \hspace{0.2mm}}
  \end{center}  
\end{figure}
\vspace{-8mm}

\textbf{PROPOSAL \\[1ex] A STUDY OF THE LINK BETWEEN   
 COSMIC RAYS AND CLOUDS  \\ 
 WITH A CLOUD CHAMBER AT THE CERN PS}   \\[1ex] 
 {\footnotesize
 B.\,Fastrup, E.\,Pedersen \\ 
 \emph{University of Aarhus, Institute of Physics and Astronomy, 
 Aarhus, Denmark}  \\[1ex] 
 E.\,Lillestol, E.\,Thorn  \\
 \emph{University of Bergen, Institute of Physics, Bergen,  Norway}  
 \\[1ex]
 M.\,Bosteels, A.\,Gonidec, G.\,Harigel, J.\,Kirkby*, 
 S.\,Mele, P.\,Minginette, B.\,Nicquevert, 
 D.\,Schinzel, W.\,Seidl \\ 
 \emph{CERN, Geneva, Switzerland} \\[1ex] 
 P.\,Grunds\mbox{\o}e, N.\,Marsh, J.\,Polny, H.\,Svensmark  \\
 \emph{Danish Space Research Institute, Copenhagen, Denmark} \\[1ex]  
  Y.\,Viisanen  \\
 \emph{Finnish Meteorological Institute, Helsinki,
 Finland} \\[1ex]
 K.\,Kurvinen, R.\,Orava  \\
 \emph{University of Helsinki, Institute of Physics,  
 Helsinki, Finland}  \\[1ex] 
 K.\,H\"{a}meri, M.\,Kulmala, L.\,Laakso, J.M.\,M\"{a}kel\"{a}, 
 C.D.\,O'Dowd \\ 
 \emph{University of Helsinki, Lab.\,of Aerosol and
 Environmental Physics, Helsinki, Finland} \\[1ex]
 V.\,Afrosimov, A.\,Basalaev, M.\,Panov  \\ 
 \emph{Ioffe Physical Technical Institute, Dept.\,of Fusion
 Technology, St.\,Petersburg, Russia} \\[1ex]
 A.\,Laaksonen, J.\,Joutsensaari  \\
 \emph{University of Kuopio, Department of Applied Physics, Kuopio,
 Finland}  \\[1ex] 
 V.\,Ermakov, V.\,Makhmutov, O.\,Maksumov, P.\,Pokrevsky,
 Y.\,Stozhkov, N.\,Svirzhevsky  \\
 \emph{Lebedev Physical Institute, Solar and Cosmic Ray Research
 Laboratory, Moscow, Russia} \\[1ex] 
 K.\,Carslaw, Y.\,Yin  \\
 \emph{University of Leeds, School of the Environment, Leeds, United
 Kingdom} \\[1ex]
 T.\,Trautmann  \\
 \emph{University of Mainz, Institute for Atmospheric Physics,
  Mainz, Germany} \\[1ex]
 F.\,Arnold, K.-H.\,Wohlfrom  \\
 \emph{Max-Planck Institute for Nuclear Physics (MPIK), Atmospheric
Physics Division,
 Heidelberg, Germany} \\[1ex]
 D.\,Hagen, J.\,Schmitt, P.\,Whitefield  \\
 \emph{University of Missouri-Rolla, Cloud and Aerosol Sciences
 Laboratory,  Rolla, USA} \\[1ex] 
 K.\,Aplin, R.G.\,Harrison  \\
 \emph{University of Reading, Department of Meteorology,
 Reading, United Kingdom} \\[1ex]
 R.\,Bingham, F.\,Close, C.\,Gibbins, A.\,Irving, B.\,Kellett,
M.\,Lockwood \\
 \emph{Rutherford Appleton Laboratory, 
 Space Science \& Particle Physics Depts., Chilton, United
 Kingdom} \\[1ex] 
 D.\,Petersen, W.W.\,Szymanski, P.E.\,Wagner, A.\,Vrtala  \\ 
 \emph{University of Vienna, Institute for Experimental Physics,
 Vienna, Austria}  \\[1ex]   }                        
 {\normalsize CLOUD$^\dagger$  Collaboration} \\
\end{center}
\vspace{-5mm}

\vfill

\noindent \rule{60mm}{0.1mm} \\ {\footnotesize
$*$) spokesperson  \\
$\dagger$) Cosmics Leaving OUtdoor Droplets }
\date{}


\end{titlepage}

\pagestyle{plain}     

\pagenumbering{roman}  
\setcounter{page}{2}  
\newpage \tableofcontents 

 
\newpage
\pagestyle{plain}     
\pagenumbering{arabic}  
\setcounter{page}{1}  

%
%
\section{Summary}  \label{sec_summary}

In 1997 Svensmark and Friis-Christensen \cite{svensmark97} announced a
surprising discovery that global cloud cover correlates closely with
the galactic cosmic ray intensity, which varies with the sunspot
cycle.   Although clouds retain some of the Earth's warmth, for most
types of cloud this is more than compensated by an increased reflective
loss of the Sun's  radiation back into space.  So more clouds in
general mean a cooler climate---and fewer clouds mean global warming.
The Earth is partly shielded from cosmic rays by the magnetic
disturbances carried by the  solar wind.  When the solar wind is
strong, at the peak of the 11-year sunspot cycle, fewer cosmic rays
reach the Earth. The observed variation of cloud cover was only a few
per cent over the course of a sunspot cycle.  Although this may appear
to be quite small, the possible long-term consequences on the global
radiation energy budget are not.

Beyond its semi-periodic 11-year cycle, the Sun displays unexplained
behaviour on longer timescales.  In particular, the strength of the
solar wind and the magnetic flux it carries have more than doubled
during the last century \cite{lockwood}. The extra shielding has reduced
the intensity of cosmic rays reaching the Earth's atmosphere by about
15\%, globally averaged. This reduction of cosmic rays over the last
century is independently indicated by the light radioisotope record in
the Greenland ice cores.  If the link between cosmic rays and clouds is
confirmed it implies  global cloud cover has decreased during the last
century. Simple estimates indicate that the consequent global warming
could be comparable to that presently attributed to greenhouse gases
from the burning of fossil fuels.

These observations suggest that solar variability may be  linked to
climate variability by a chain that involves the solar wind, cosmic rays
and clouds. The weak link is the connection between cosmic rays and
clouds.  This has not been unambiguously established and, moreover, the
microphysical mechanism is not understood.  Cosmic rays are the dominant
source of ions in the free troposphere and stratosphere and they also
create free radicals.  It has been proposed
\cite{arnold82}--\cite{yu00} that ions may grow via clustering to form
aerosol particles which may ultimately become cloud condensation nuclei
(CCN) and thereby seed clouds.  Recently a search for massive ions in
the upper troposphere  and lower stratosphere was started by
MPIK-Heidelberg \cite{eichkorn99} using aircraft-based ion mass
spectrometers. Preliminary results indeed indicate the presence of
massive positive and negative ions. In addition to their effect on
aerosol formation and growth,  cosmic rays may also possibly enhance 
the formation of ice particles in clouds \cite{tinsley}.

We therefore propose to test experimentally the link between cosmic rays
and clouds and, if confirmed, to uncover the microphysical mechanism. We
propose to make the measurements under controlled laboratory conditions
in a beam at the CERN Proton Synchrotron (PS), which provides an
adjustable source of ``cosmic rays". The experiment, which is named
CLOUD (Cosmics Leaving OUtdoor Droplets), is based on a cloud chamber
that is designed to  duplicate the conditions prevailing in the
atmosphere.   To our knowledge, cloud chamber data under these
conditions have never been previously obtained.

This document is organised as follows. First we provide an overview of
the scientific motivation for the experiment, including a summary of the
most recent satellite observations of clouds.   We present the
scientific and experimental goals of CLOUD in Section~\ref{sec_goals}. 
These are followed in Sections
\ref{sec_detector}--\ref{sec_accelerator_and_beam} by descriptions of
the detector and its performance, of the data interpretation and cloud
modelling, and of the accelerator requirements.  The planning for the
experiment is summarised in Section  \ref{sec_planning}.    Since this
proposal concerns several different scientific
disciplines---atmospheric, solar-terrestrial and particle physics---we
provide fairly extensive background information, mostly in footnotes and
in Appendices~\ref{sec_cloud_physics}--\ref{sec_cloud_models} which
cover, respectively, cloud physics, aerosol-cloud-climate interactions,
classical operation of a Wilson cloud chamber, cosmic rays in the
atmosphere, and cloud models. 

\section{Scientific motivation; the origins of climate change}
\label{sec_motivation}

\subsection{Global warming} \label{sec_global_warming}

Global warming is a major concern of the world, with its potentially
devastating effects on coastal settlements and world agriculture.   The
steep rise in greenhouse gas emissions since the Industrial Revolution
has increased the \cotwo\ concentration in the atmosphere by about 30\%.
This is widely believed to be the dominant cause of the observed rise of
about 0.6\degc\  in the  global mean surface temperature during this
period \cite{ipcc}. 

\begin{figure*}[t]
  \begin{center}
      \makebox{\epsfig{file=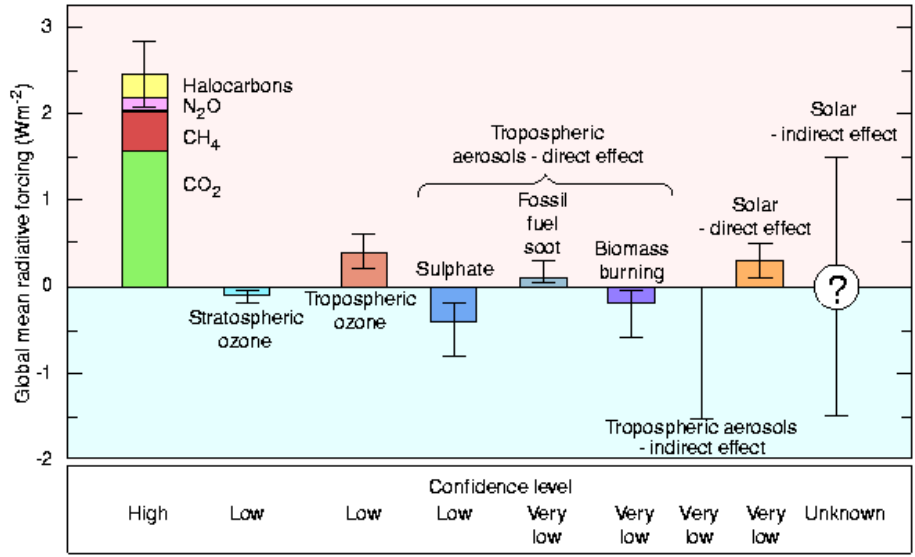,width=120mm}}
  \end{center}
  \vspace{-5mm}
  \caption{IPCC estimates  of the global annual averaged radiative
forcings due to changes in  anthropogenic greenhouse gases and aerosols
from 1850--1992 (first seven columns of the figure) \cite{ipcc}.
Positive forcings lead to a warming and  negative forcings cause a
cooling. Natural changes due to the Sun are indicated by the final two
columns; the first is the IPCC estimate of changes in solar output over
the same period and the second concerns the present CLOUD proposal to
study of the influence of galactic  cosmic rays on cloud formation.
Since the galactic cosmic ray flux is modulated by the solar wind, this
would provide a mechanism for indirect solar radiative forcing.}
  \label{fig_ipcc_forcing}    
\end{figure*}

A small systematic rise or fall in the global temperature is caused by a
net imbalance (``forcing'') in the Earth's energy radiation budget.  The
radiative forcing caused by the increase in the \cotwo\ fraction since
1750 is estimated to be 1.5 \wpm\ (Fig.~\ref{fig_ipcc_forcing}) 
\cite{ipcc}, compared with the global average incoming solar 
radiation\footnote{By convention, the incoming solar radiation of about
1366 \wpm\ is averaged over the total surface area of the Earth, i.e.
divided by a factor four, to give a global average of 342 \wpm.}  of 342
\wpm, i.e. an imbalance of only 0.4\%.   After including the effects of
all greenhouse gases (+2.45 \wpm),  aerosols\footnote{Aerosols are
0.001--1 $\mu$m diameter particles of liquid or solid suspended in the
air \cite{acc}.  Atmospheric aerosols include dust, sea salt, soot
(elemental carbon), organic compounds from biomass burning, sulphates
(especially H$_2$SO$_4$ and (NH$_4$)$_2$SO$_4$) from SO$_2$, and
nitrates  (especially HNO$_3$) from NO and NO$_2$.  Aerosol
concentrations vary typically from $\sim$100 cm$^{-3}$ in  maritime air
to $\sim1000 $ cm$^{-3}$ in unpolluted air over land masses, but there
are large variations from these values.}  \mbox{(-0.5~\wpm)} and their
indirect influence on clouds (-0.75 \wpm, but poorly known), the present
net radiative forcing from mankind is estimated to be about 1.2 \wpm. 

The climate models \cite{ipcc} upon which the predictions of greenhouse
warming depend have gradually improved as new effects and better data
have been incorporated.  They now provide a reasonable match to  the
observed variation in global temperatures over the last century. 
However they remain subject to  significant uncertainties, especially
from feedback mechanisms and from the effects of anthropogenic
aerosols.  The latter contribute directly by  scattering and absorption
of radiation and also indirectly by influencing cloud formation. 
Nonetheless it is in this climate of scientific uncertainty that major
political decisions on greenhouse gas emissions are presently being
made (Earth Summit in Rio de Janeiro, 1992, and UN Climate Convention in
Kyoto, 1997) that will have a profound effect on the economic
development of both the developed and the developing countries.  The
need for such political decisions to be based on sound scientific
grounds is self evident, and a major world-wide research effort on
climate change is underway.

\subsection{Solar variability}  \label{sec_solar_variability}

In order to determine the influence of mankind on climate change it is
first necessary to understand the natural causes of variability.   A
natural effect that has been hard to understand physically  is an
apparent link between the weather and solar activity---the
sunspot\footnote{Sunspots are areas of the Sun's  photosphere where
strong local magnetic fields (typically 2500 Gauss, to be compared with
the Earth's field of about 0.3 Gauss) emerge vertically 
\cite{foukal}.  They appear dark because their temperature is about
half of the surrounding photosphere (3,000 K compared with 5,800~K).
They are generated by the differential rotation of the Sun with respect
to latitude: one revolution takes 25~days at the equator and 28 days at
mid-latitudes.  This transforms the quiescent dipole field into a
toroidal field and eventually creates ``knots'' of strong localised
fields.  These knots may  penetrate the photosphere to form sunspots,
which appear cooler due to modification of the normal convective motions
of the plasma by the strong magnetic fields. The sunspots first appear
at high latitudes and then gradually migrate towards the equator.  They
eventually disappear by magnetic recombination, leaving a quiescent
dipole field once more (but of opposite polarity). The cycle from dipole
to toroidal and back to dipole field is known as the solar (sunspot)
cycle and takes about 11 years on average.}  cycle.

\begin{figure*}[htbp]
  \begin{center}
      \makebox{\epsfig{file=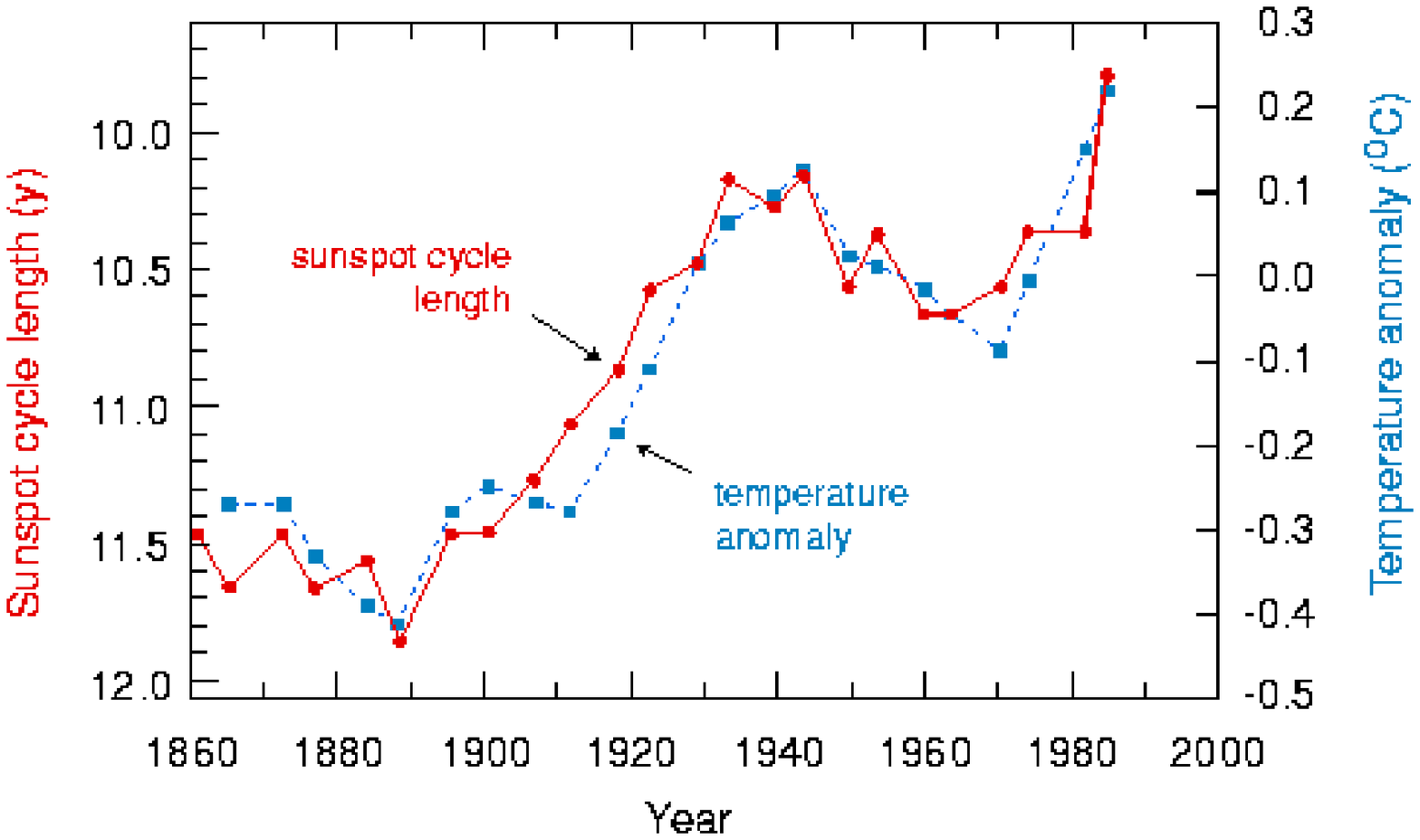,height=75mm}}
  \end{center}
  \caption{Variation during the period 1861--1989 of the sunspot cycle
length (solid curve) and the temperature anomaly of the Northern
Hemisphere (dashed curve) \cite{friis}.  The temperature data are from
the IPCC
\cite{ipcc}.}
  \label{fig_sunspots_vs_temp}  
  \begin{center}
      \makebox{\epsfig{file=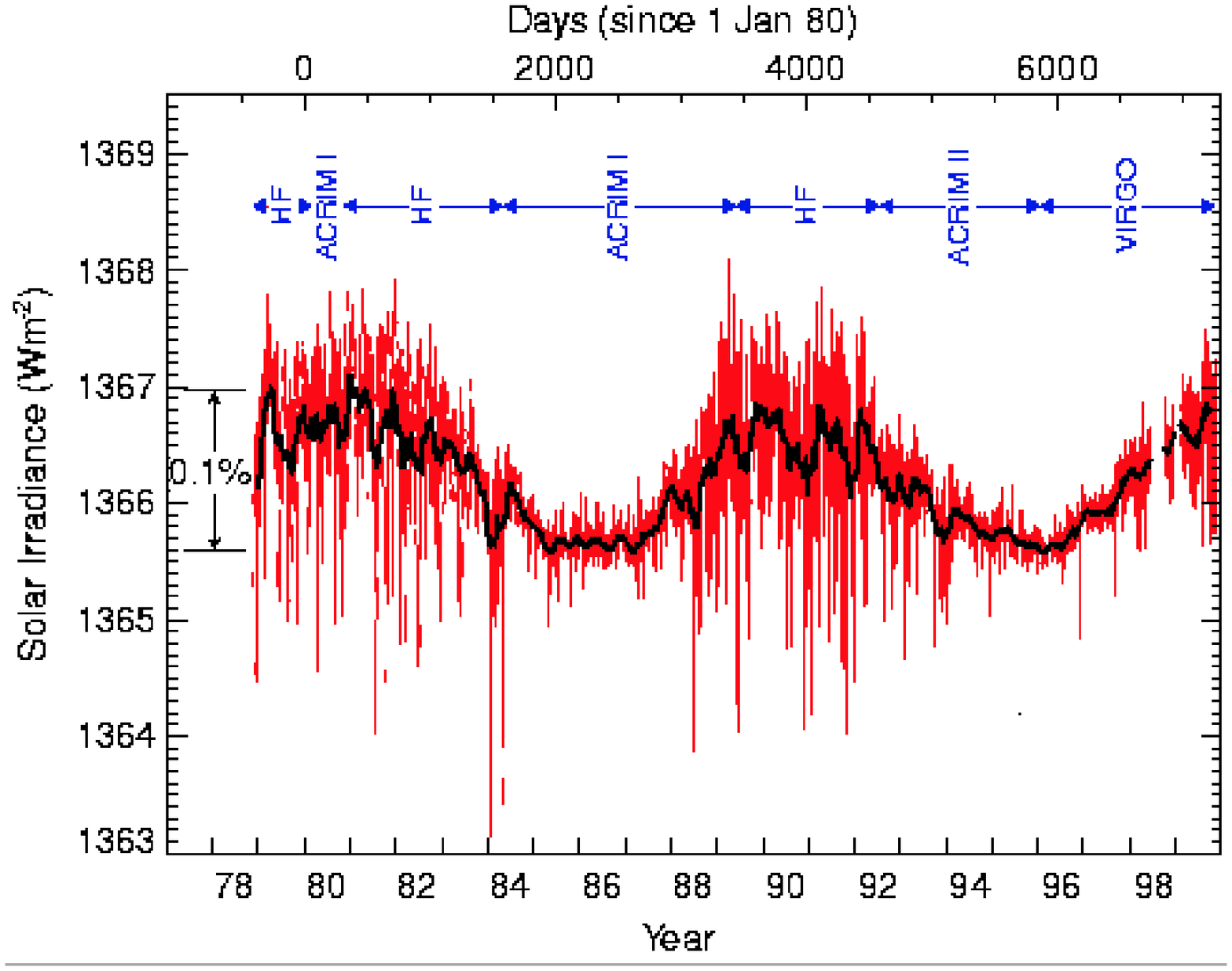,width=110mm}}
  \end{center}
    \vspace{-3mm}
  \caption{Satellite measurements of the variation of the Sun's 
irradiance over two solar cycles \cite{frohlich97}.  Sunspot maxima
occurred around the beginning of 1981 and mid 1990, and sunspot minima
occurred around the beginning of 1986 and 1996. The rapid fluctuations
are due to sunspots rotating into the field of view.  The solid line
represents the smoothed data.   The measurements are from active cavity
irradiance monitors (ACRIM I and II) on the Solar Maximum Mission
Satellites and from HF-type radiometers on the Nimbus 7 Earth Radiation
Budget (ERB) and Earth Radiation Budget Satellite (ERBS) experiments.}
  \label{fig_solar_irradiance}    
\end{figure*}

The observation that warm weather seems to coincide with high sunspot
counts and cool weather with low sunspot counts was made as long ago as
two hundred years by the astronomer William Herschel \cite{herschel} who
noticed that the price of wheat in England was lower when there were
many sunspots, and higher when there were few.   The best known example
of this effect is known as the Maunder Minimum \cite{eddy}, the Little
Ice Age between 1645 and 1715---which ironically almost exactly
coincides with the reign of Louis XIV, \emph{le Roi Soleil},
1643--1715---during which time there was an  almost complete absence of
sunspots.  During this period the River Thames in London regularly froze
across and fairs complete with swings, sideshows and food stalls were a
standard winter feature.  

Since that time there have been numerous observations and
non-observations of an apparent link between climate and the sunspot
cycle  \cite{dickinson}--\cite{hoyt}.  One example of a positive
observation  was presented by Friis-Christensen and Lassen in 1991
(Fig.~\ref{fig_sunspots_vs_temp}) \cite{friis}.  They used the sunspot
cycle length as a measure of the Sun's  activity.  The cycle length
averages 11~years but has varied from 7 to 17 years, with shorter cycle
lengths corresponding to a more magnetically-active Sun.  A  correlation
was found between the sunspot cycle length and the change in land
temperature of the Northern Hemisphere in the period between 1861 and
1989.  The land temperature of the northern hemisphere was used in order
to avoid the lag by several years of air temperatures over the oceans,
due to their large heat capacity.  The data shown in
Fig.~\ref{fig_sunspots_vs_temp} cover the period during which
greenhouse gas emissions are believed to be the major cause of the
global warming of 0.6\degc.  Of particular note is the dip between 1945
and 1970, which cannot be explained by the steadily rising greenhouse
gas emissions but seems to coincide with a decrease in the Sun's 
activity.

In the absence of sufficiently sensitive measurements, it was  suspected
that the Sun's  irradiance may be fluctuating over the solar cycle.  
However,  the steadiness of the Sun's  irradiance over almost two 
sunspot cycles has recently been established by satellite measurements
(Fig.~\ref{fig_solar_irradiance}) 
\cite{sun_in_time}-\cite{frohlich97}.  The solar irradiance is slightly
higher at sunspot maximum; although sunspots are cooler and have reduced
emission, this is more than compensated by an associated increase in 
nearby bright areas known as \emph{plages} and \emph{faculae}.   The
mean irradiance changes by about 0.1\% from sunspot maximum to minimum
which, if representative over a longer time interval, is too small  (0.3
\wpm, globally-averaged) to account for the observed changes in the
Earth's temperature.  However it is not completely negligible, and
current estimates---using the data shown in
Fig.~\ref{fig_solar_irradiance} together with the sunspot
record---attribute a net direct radiative forcing  over this century of
about +0.3~\wpm\ due to changes in  solar irradiance (indicated by the 
``solar - direct effect'' in  Fig.~\ref{fig_ipcc_forcing}). 

\subsection{Cosmic rays and cloud variability}
\label{sec_cloud_variability}

\begin{figure}[htbp]
  \begin{center}
      \makebox{\epsfig{file=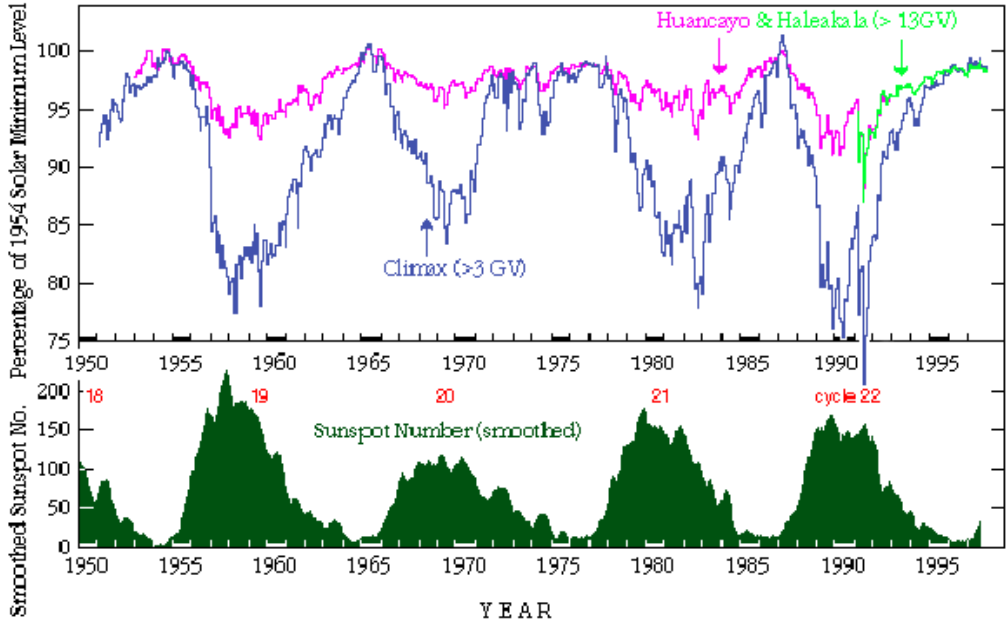,height=82mm}}
  \end{center}
  \vspace{-9mm}
  \caption{Variation with time of sunspot number and cosmic ray flux, as
measured by ground-based neutron counters.  The neutron data are from
 the University of Chicago neutron monitor stations at Climax, 
  Colorado (3400 m elevation; 3 GeV/$c$ primary charged particle
cutoff), Huancayo, Peru (3400 m; 13 GeV/$c$ cutoff) and Haleakala,
Hawaii (3030 m; 13~GeV/$c$ cutoff).  The stronger modulation of the
cosmic ray flux at higher latitudes (Climax) is due to the lower primary
cutoff energy. The neutrons are mostly produced by primary hadronic
interactions in the first 1--2 $\lambda_{int}$ of the atmosphere and
therefore measure the changes in cosmic ray intensity at altitudes above
about 13 km.   The primary cosmic radiation is about 80\% protons, 15\%
He nuclei and 5\% heavier nuclei.   At sea level the most numerous
\emph{charged} particles are muons and their fluctuation is less
pronounced---about 3\%  over a solar cycle \cite{ney}---since they are
produced from the high-energy component of cosmic radiation, which is
less affected by the solar wind.}
  \label{fig_cosmics_vs_sunspots}     
\end{figure}

It is well known that the cosmic ray intensity on Earth is strongly
influenced by the solar wind\footnote{The solar wind is a continuous
outward flow of charged particles  (mainly protons and electrons, with
5\% helium nuclei) from the plasma of the Sun's  corona.  Sources
include streams from a honeycomb of magnetic fields in the solar
atmosphere, and large and small mass ejections.   The solar wind
creates the huge heliosphere of the Sun that extends out 50--100 AU,
well beyond the orbit of Neptune.  At the Earth's orbit the solar wind
has a velocity of 350--800~km~s$^{-1}$ ($\beta = $ 0.001--0.003) and an
intensity of
\mbox{(0.5--5)$ \times 10^8$ particles~cm$^{-2}$~s$^{-1}$}, carrying
with it a magnetic field of about  $5 \times 10^{-5}$ Gauss.}
 \cite{ney},   whose strength varies with  the sunspot cycle
(Fig.~\ref{fig_cosmics_vs_sunspots} and Appendix
\ref{sec_atmospheric_cr_meaurements}). The solar wind contains
frozen-in irregular magnetic fields.  Galactic cosmic rays that enter
the solar system suffer many scatters from these irregularities and
undergo a random walk.  This has been theoretically shown \cite{gleeson}
to be equivalent to a heliocentric retarding electric potential,  which
varies over the course of the solar cycle. At times of low sunspot 
activity, the solar wind is weaker and the retarding potential at the
Earth's orbit is about 400~MV.   At times of high sunspot activity, the
retarding potential is  about 1200~MV, thereby reducing the low-energy
component of galactic cosmic radiation reaching Earth. 

\newpage

The cosmic rays are also deflected by the Earth's geomagnetic field,
which they must penetrate to reach the troposphere.\footnote{The
troposphere is the lowest level of the atmosphere and the region where
there is enough water vapour and vertical mixing for clouds to form
under suitable conditions. The troposphere has a depth of about 18 km
over the tropics, decreasing to about 8 km over the poles; it contains
about 80\% of the mass of the atmosphere.   The troposphere is divided
into the \emph{planetary boundary layer}, extending from the Earth's
surface up to about 1~km, and the \emph{free troposphere}, extending
from 1~km to the boundary with the stratosphere (the tropopause).  There
is an overall adiabatic lapse rate of temperature in the troposphere of
between 6\degc\ (moist air) and 9.7\degc\ (dry air)  per km altitude,
reaching a minimum of about -56\degc\ at the tropopause.  The
stratosphere extends up to about 50~km and has a temperature that slowly
rises with altitude due to absorption of solar UV radiation. This leads
to very little turbulence and vertical mixing and, in consequence, it
contains relatively warm, dry air that is largely free of clouds.}  This
sets a  minimum vertical momentum of primary charged particles at the
geomagnetic equator of about 15 GeV/$c$, decreasing to below 0.1~GeV/$c$
at the geomagnetic poles. (The actual cutoff at high latitudes is
determined by the atmospheric material; for example a 1 GeV/$c$ proton
can penetrate only as far as 15~km altitude.) As a result the modulation
of the cosmic ray intensity is more pronounced at higher geomagnetic
latitudes (Figs.~\ref{fig_cosmics_vs_sunspots} and
\ref{fig_cosmic_2}).  Averaged over the globe, the variation of the
cosmic ray flux is about 15\% between solar maximum and minimum
\cite{obrien,masarik}.    It represents one of the largest measurable
effects of sunspot activity near the Earth's surface.

\begin{figure*}[htbp]
  \begin{center}
      \makebox{\epsfig{file=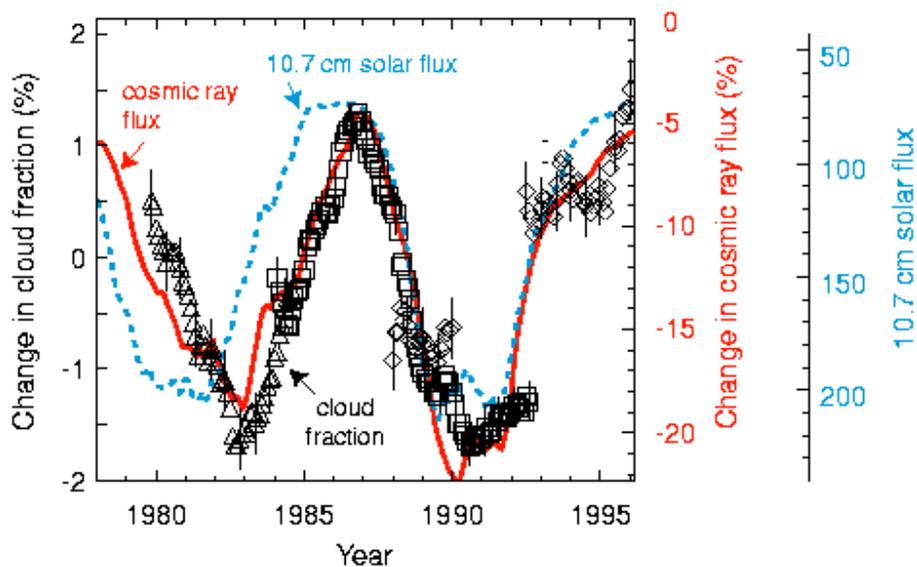,width=120mm}}
  \end{center}
  \vspace{-5mm}
  \caption{Absolute percentage variation of global cloud cover observed
by satellites (data points; left hand scale) and  relative percentage
variation of cosmic ray flux (solid curve, normalised to May 1965;
near-right hand scale) \cite{svensmark97,svensmark98}.  Also shown is
the solar 10.7 cm microwave flux (dashed curve, in units of
$10^{-22}$ Wm$^{-2}$Hz$^{-1}$; far-right hand scale).   The cloud data
are restricted to oceans; Nimbus 7 (triangles) and DMSP (diamonds) data
are for the southern hemisphere over oceans, and  ISCCP-C2 (squares)
data are for oceans with the tropics excluded. The error bars indicate
representative statistical errors. The cosmic ray data are neutron
measurements from Climax (Fig.~\ref{fig_cosmics_vs_sunspots}). All data
are smoothed using a 12-month running mean. A more recent analysis is
shown in Figs.\,\ref{fig_low_cloud_1} and \ref{fig_cloud_climax}.}
  \label{fig_svensmark_fig.1}    
\end{figure*}

But how could cosmic rays affect the Earth's weather? The energy
deposited by cosmic rays is only a few parts per billion compared with
the incident  solar energy, so a strong amplifying mechanism would be
necessary. The breakthrough was made by Svensmark and Friis-Christensen
in 1997  \cite{svensmark97} who discovered an unexpected correlation
between global cloud cover and the incident cosmic ray intensity. The
satellite data, which are shown in Fig.\,\ref{fig_svensmark_fig.1}
(taken from the later ref.\,\cite{svensmark98}), display a clear imprint
of the solar cycle on global cloud cover.\footnote{The terms cloud
cover, cloud frequency and cloud fraction are interchangeable.}  Over a
sunspot cycle, the absolute variation of global cloud cover is about
3\%, to be compared with an average total cloud cover of about 65\%,
i.e. a relative fraction of about 5\%.  

In addition to this observation of cloud variations over the timescale
of order one solar cycle, other data sets have been analysed to
investigate transient effects on clouds due to sudden changes in the
cosmic ray flux. Forbush decreases of galactic cosmic rays occur due to
solar disturbances on timescales of order days and therefore allow
study of short-term cosmic-ray induced changes. Pudovkin and Veretenenko
\cite{pudovkin} report observations of a short-term decrease of
cloudiness that correlates with Forbush decreases, using a superposed
epoch analysis on data obtained visually. Their observation is
restricted to a narrow range of latitudes (60N--64N), and they suggest
that cirrus clouds are responsible. A similar analysis of the effects of
Forbush decreases on rainfall has been carried out by the Lebedev
Physical Institute  \cite{stozhkov} using data from 50  meteorological
stations in Brazil for 47 Forbush events recorded between 1956 and 1992.
A 30\% drop in rainfall (corresponding to a 3$\sigma$ change from the
mean) is observed on the initial day of the onset of the Forbush
decrease.

A detailed mechanism modifying ice clouds has been proposed in a series
of papers (notably Tinsley and Dean \cite{tinsley}), which suggests that
significant changes in the latent heat released within supercooled
clouds can occur as a result of aerosol electrification. The proposed 
 mechanism is that charged aerosols are more effective than neutral
aerosols as ice nuclei, which are generally rare in the atmosphere.  The
aerosol electrification is due to the ionisation created by cosmic
rays. 
 Rather than a comparison with cloudiness, Tinsley's superposed epoch
analysis uses a dynamical parameter, the Vorticity Area Index (a measure
of the regional scale motion), for which he finds a correlation with
Forbush decreases. The mechanism of latent heat release via electrical
enhancement of ice nucleation can lead to substantial amplification of
the ionisation energy deposited by cosmic rays  in the atmosphere.

\subsection{Other solar-induced climate variability}
\label{sec_other_effects}

Other solar activity also follows the sunspot cycle, so in principle the
cloud variations might be attributed to (a) a direct effect of cosmic
rays on clouds, (b) a direct effect of other solar activity on clouds,
or (c) an effect of solar activity on global weather that indirectly
results in a change in cloud cover.

Although the other explanations mentioned here cannot be ruled out,
Fig.\,\ref{fig_svensmark_fig.1} contains evidence favouring (a), the
direct cosmic ray effect, which is to be investigated in the CLOUD
experiment. The solar 10.7 cm microwave flux is a good proxy both for
variations in the sunspot count and for variations in the solar output
of other electromagnetic radiation, namely visible light, X-rays and
ultraviolet rays. In Fig.\,\ref{fig_svensmark_fig.1}, the change in
cloud fraction followed the cosmic rays closely but sometimes lagged as
much as two years behind the change in the 10.7 cm flux. Delays in
cosmic ray variations occur because disturbances in the solar wind,
which scatter the cosmic rays, involve events (coronal mass
ejections---CMEs, and co-rotating interaction regions) only loosely
related to the sunspot count. Moreover the heliosphere is so large that
disturbances can take up to about a year to reach its boundary, the
heliopause. The delays show that the cloud variations correlate more
closely with the cosmic rays than with the solar electromagnetic flux.

Could the solar wind have a direct effect on clouds, while
coincidentally modulating the cosmic rays? This is unlikely because
solar wind protons have very little energy (a few keV) and if they
penetrate the Earth's magnetosphere in the auroral zone, the outer
atmosphere stops them at altitudes above 100 km. Shock waves associated
with CME disturbances colliding with the slow solar wind can accelerate
protons to 100 MeV or more, generating solar cosmic rays, but only
rarely do they have sufficient energy to reach the tropopause at 8--18
km altitude.

The \emph{sign} of the effect is a consideration in assessing possible
solar mechanisms influencing the cloud fraction. While the cloud
fraction correlates positively with the cosmic ray intensity, it
correlates negatively with the solar electromagnetic radiation and
with the strength of the solar wind. If enhancements in electromagnetic
radiation or the solar wind were responsible for a direct effect on
clouds, they would be in the counter-intuitive sense of reducing the
cloud fraction.

As to whether changes in cloud fraction might be an indirect result of
changes in global weather due to solar effects that coincide with the
cosmic ray modulation, none of the other mechanisms on offer for a solar
role in climate change suggests an effect on cloud fraction. Mechanisms
currently discussed include:
\begin{itemize}

\item Increases in visible light, causing warming at the Earth's
surface.\\[-4ex]

\item Increases in ultraviolet (UV), causing warming in the 
stratosphere \cite{haigh}.\footnote{Only wavelengths above 300 nm
penetrate to the troposphere and the Earth's surface.  These show a tiny
peak-to-peak variation of less than 0.1\% over the solar cycle.  However
the variation is more pronounced in the UV.  Wavelengths below 300 nm,
which account for only about 2\% of the total solar irradiance, vary by
about 5\% (200--300 nm) to 50\% (100--150~nm), or even more at shorter
wavelengths \cite{hunten}.  The UV radiation is absorbed by ozone in the
upper stratosphere, which warms as a result.   This has the potential to
influence large-scale dynamics of the troposphere
\cite{haigh}-\cite{shindell}, although vertical mixing between the
stratosphere and troposphere is weak.} \\[-4ex]

\item Solar-induced turbulence in the Earth's outer atmosphere that
scatters gravity waves back to the lower stratosphere, again with
warming effects \cite{geller}. 

\end{itemize}

All of these mechanisms seem likely to affect the geographical
distribution of clouds, in particular by a poleward shift of the
mid-latitude jet streams and depressions, during high solar activity.
There is no obvious short-term link with cloud fraction.

Solar activity has many manifestations, and all of the means by which it
may influence the Earth's climate deserve further investigation. It is
no part of our case to suggest that the link between cosmic rays and
clouds is the only important solar-climatic mechanism. For the reasons
given in this section we nevertheless consider that the preferred
interpretation of Figure~\ref{fig_svensmark_fig.1} is the simplest,
namely that cloud behaviour is directly influenced by cosmic rays. The
opinion is reinforced by recent studies summarised in the next section.

\subsection{Improved satellite measurements of clouds}
\label{sec_improved_measurements}

Currently the best continuous satellite observations of cloud
properties are from the International Satellite Cloud Climate Project
(ISCCP) D2 data, which covers the period from July 1983 to September
1994 \cite{rossow91,rossow96}.  The D2 data are constructed by combining
uniform analysis results from several satellites---up to five
geostationary satellites and two polar orbiting satellites---to obtain
complete global coverage every 3 hours.  Cloud measurements are made at
visible ($\lambda \sim$ 0.6 $\mu$m), near infra-red (3.7 $\mu$m) and
infra-red (IR) wavelengths (10--12 $\mu$m).  The IR measurements have
the advantage that they provide  continuous detection through day and
night.   As well as cloud frequency, the cloud-top temperatures and
pressures are also determined. The temperatures and pressures are
obtained by assuming an opaque cloud, i.e. an emissivity
$\epsilon = 1$,  and adjusting the cloud's pressure level (effectively
the cloud-top altitude) in the model until the reconstructed outgoing IR
flux matches that observed.  The clouds are classified into 3 altitude
ranges according to the pressure at their top surface:  low,
$>$680~hPa (approximately $<$3.2 km); middle,  680--440~hPa
(3.2--6.5~km); and high,
$<$440 hPa ($>$6.5~km).

\begin{figure*}[htbp]
  \begin{center}
      \makebox{\epsfig{file=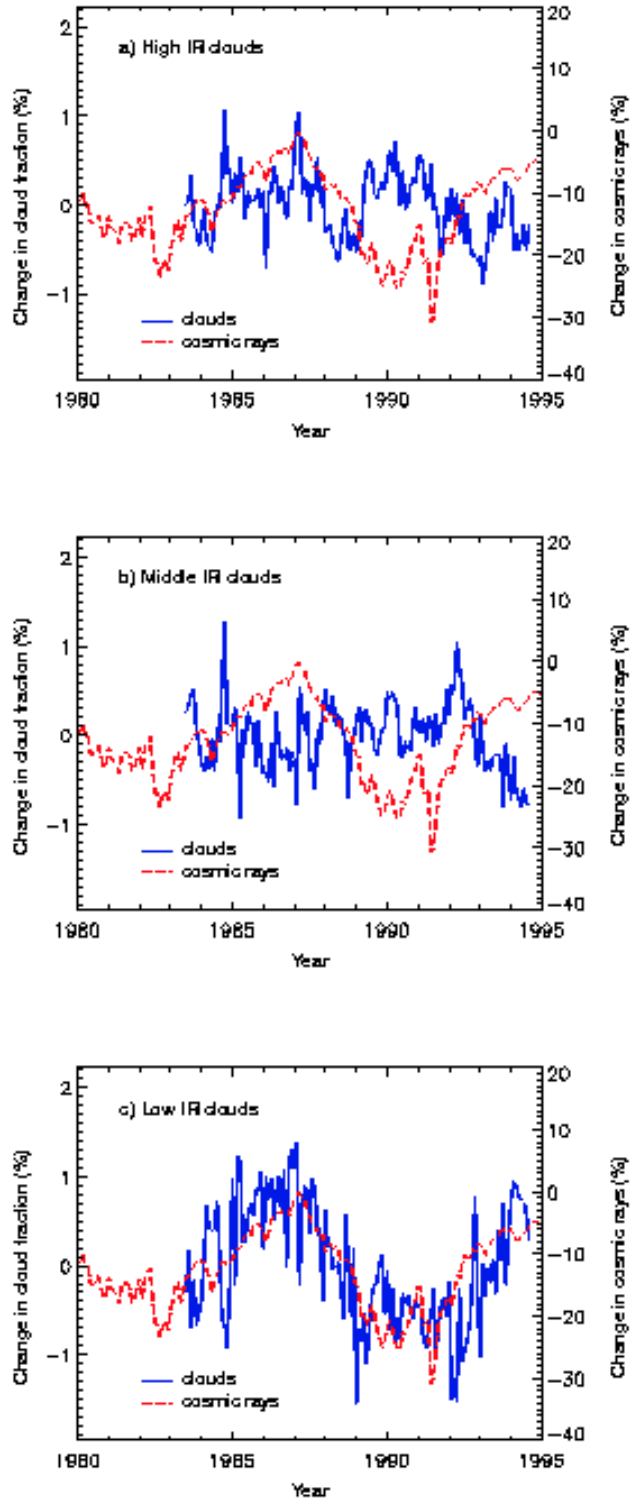,height=200mm}}
  \end{center}
 \vspace{-5mm}
  \caption{Monthly mean values for the global absolute variations of IR
cloud coverage for a) high
    ($<$440~hPa), b) middle (440--680 hPa), and c) low ($>$680 hPa)
clouds (solid lines) \cite{marsh}. Cosmic rays, measured by the Climax
neutron monitor, are indicated by the dashed lines, normalised to May
1965. The mean global cloud fraction over this period for high, middle
and low IR clouds is 13.5\%, 19.9\%, and 28.0\% respectively.  The cloud
measurements are obtained from the ISCCP-D2 IR dataset
\cite{rossow91,rossow96}.}
  \label{fig_low_cloud_1}    
\end{figure*}

The cloud frequency for high, middle and low IR clouds is shown in
Fig.\,\ref{fig_low_cloud_1} together with the cosmic ray variation over
the same period \cite{marsh}.   In contrast with the previous analyses
\cite{svensmark97,svensmark98}, these data are spatially and temporally
unrestricted; they include clouds over the entire globe, during both day
and night. The data indicate the presence of a significant correlation
between cosmic ray intensity and the frequency of
\emph{low}  clouds, below about 3.2 km, but none with clouds at higher
altitudes. Since the cosmic ray and ionisation intensities---and their
variations over the solar cycle---are largest above about 10 km altitude
(Appendix \ref{sec_cosmic_rays} and Fig.\,\ref{fig_cosmics}), this
would suggest mixing from the upper to lower troposphere may be
involved.  Indeed, vertical mixing is a prominent feature of
tropospheric dynamics, where large-scale vertical transport of air,
chemical species and ions can occur on timescales as short as a few
hours via strong convective updrafts and the accompanying downdrafts. 

\begin{figure*}[htbp]
  \begin{center}
      \makebox{\epsfig{file=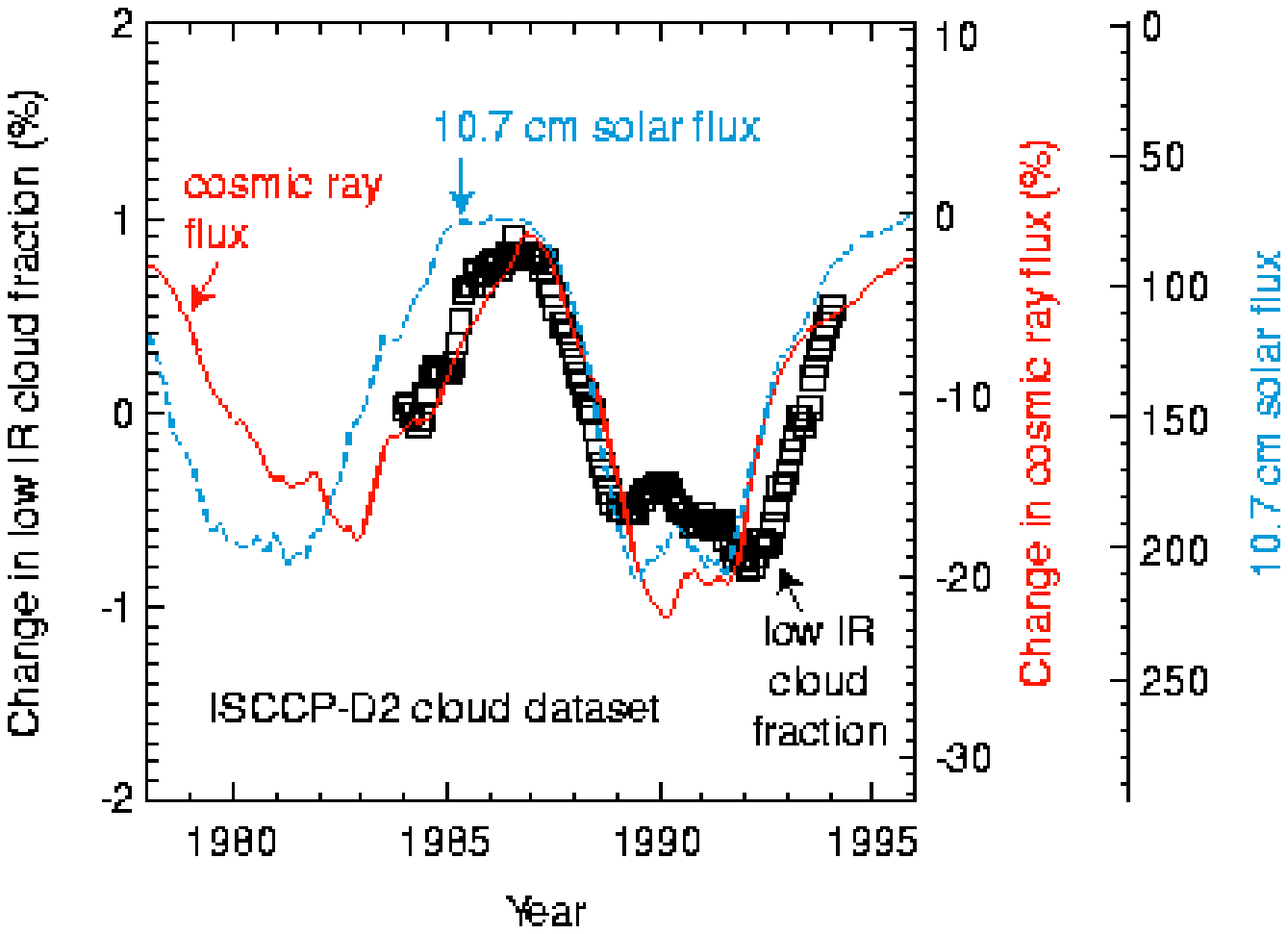,height=70mm}}
  \end{center}
 \vspace{-8mm}
  \caption{Variation over the period 1983-1994 of the low IR 
   (10--12 $\mu$m) cloud fraction in the ISCCP-D2 dataset (from
Fig.\,\ref{fig_low_cloud_1}c) in comparison with the changes of cosmic
ray flux (solid line) and 10.7 cm solar irradiance (dashed line).  The
cloud data have complete global coverage, day and night, and are
smoothed using a 12-month running mean to allow easy  comparison with
Fig.\,\ref{fig_svensmark_fig.1}.}
  \label{fig_cloud_climax}    
 \vspace{-5mm}
  \begin{center}
      \makebox{\epsfig{file=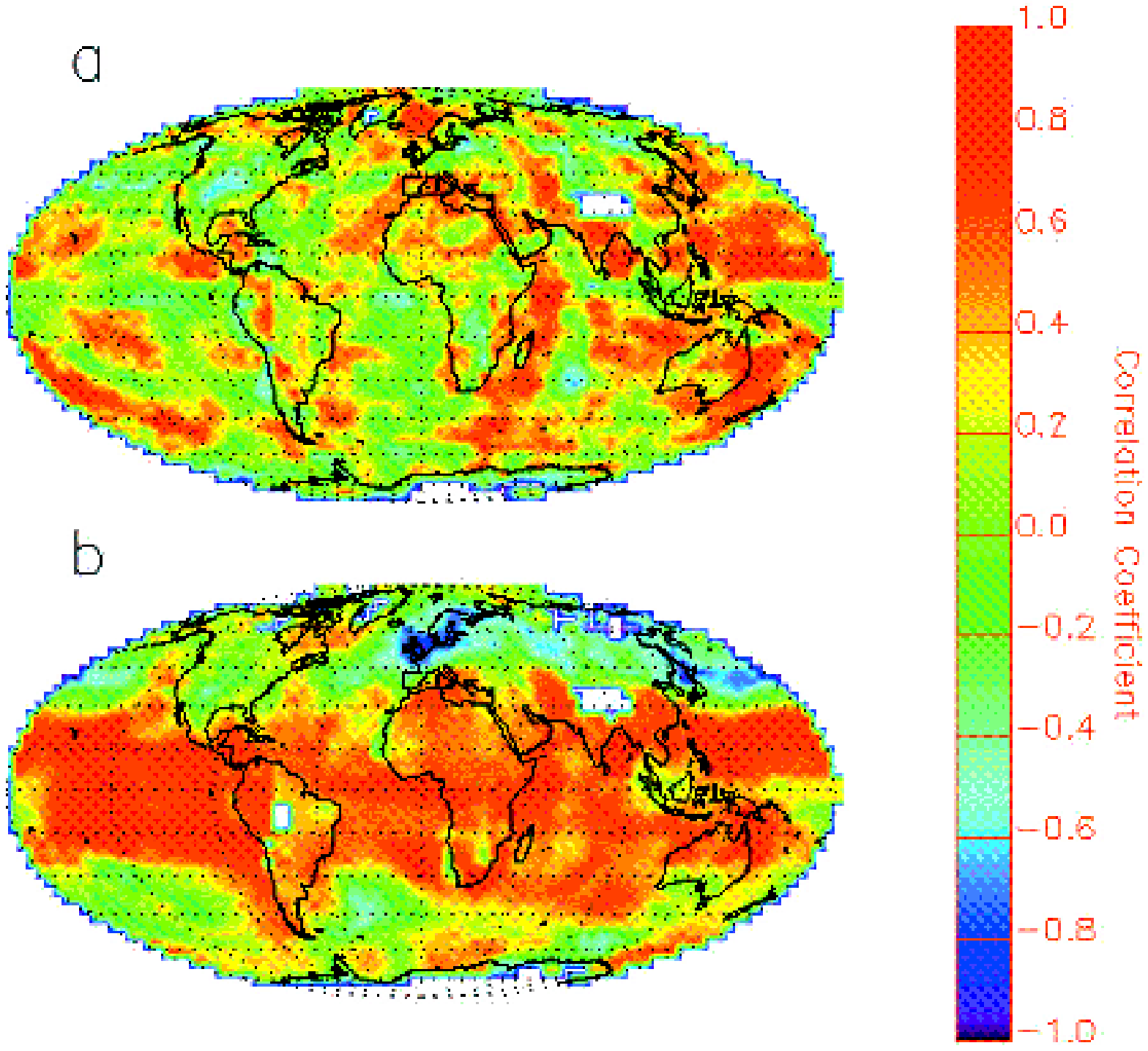,height=100mm}}
  \end{center}
\vspace{-7mm}
  \caption{Global maps of the correlation between cosmic ray intensity 
and a) low IR cloud fraction and b) low IR cloud-top temperature 
 \cite{marsh}.  The low IR cloud fractions are calculated as in 
Fig.\,\ref{fig_low_cloud_1}c), while the low cloud-top temperatures are
obtained from the ISCCP-D2 IR  model.  White pixels indicate regions
with either no data or an incomplete monthly time series. The
correlation coefficients are calculated from the 12-month running mean
at each grid point. Fractions of the Earth with a correlation
coefficient $\geq$ 0.6 are a)~14.2\%, and b) 29.6\%, respectively. The
probability  of obtaining a correlation coefficient $\geq$ 0.6 from a
random signal is $< 0.01\%$ per pixel. }
  \label{fig_low_cloud_2}    
\end{figure*}

Figure \ref{fig_cloud_climax} shows the low IR cloud fraction together 
with the variations of cosmic ray flux (solid line) and  10.7 cm solar
irradiance over this period.  The data are smoothed using a 12-month
running mean to allow easy comparison with the earlier analysis shown in
Fig.\,\ref{fig_svensmark_fig.1}.  The data confirm the presence of a
solar cycle modulation of the cloud fraction, and continue to favour the
cosmic ray interpretation.   

The global  map of the low cloud frequency correlation is shown in
Fig.\,\ref{fig_low_cloud_2}a) \cite{marsh}.  The fraction of the Earth's
surface with a correlation coefficient above 0.6 is 14.2\%. 
Figure~\ref{fig_low_cloud_2}b) shows the correlation of low IR cloud-top
temperature and cosmic ray flux.  A strong and continuous band of high
correlation ($>$0.6) extends throughout the tropics, covering 29.6\% of
the globe.   This is a surprising result  since the solar modulation of
the cosmic  ray intensity is a minimum near the geomagnetic equator,
with a peak variation about 5\% (Fig.\,\ref{fig_cosmics_vs_sunspots}).
No significant correlations are observed between cosmic rays and the
cloud-top temperatures of middle and high clouds (these data are not
mapped here).  

In summary, the ISCCP-D2 IR cloud data show a clear correlation of the
cosmic ray intensity with low clouds, below about 3 km altitude (largely
comprising stratocumulus and stratus cloud types), and no correlation
with higher clouds. (As before, these data could alternatively be
interpreted as a possible  solar-cloud link via the UV irradiance.)  The
correlation appears in two distinct parameters: a) the cloud frequency
(coverage), and b) the cloud-top temperature.  The spatial distribution
of the correlation across the globe is different in the two cases; the
regions of high correlation of cloud frequency are widely scattered
whereas the cloud-top temperature correlation is essentially continuous
over the entire tropics.  At present the reason for these different
spatial distributions is not known, although we note that these are two
separate cloud properties.  Cloud frequency measures the presence or
absence of a cloud, reflecting changes in the cloud lifetimes,  cloud
sizes, or cloud number.  On the other hand, the inferred cloud-top
temperature depends on its altitude and on the microphysical properties
of existing clouds.

The signs of the two correlations are that more cosmic rays give more
clouds and a higher cloud-top temperature.  Under the assumption of
opaque clouds, the cloud-top temperature effectively measures the
altitude at the top of the cloud; a higher cloud-top temperature implies
a lower cloud.  However the observed properties of low maritime clouds
suggest that they are not opaque \cite{heymsfield}. If the assumption of
opaque clouds is relaxed, an alternative interpretation is that the
cloud altitude does not change with increasing cosmic ray flux, but that
the emissivity of the cloud increases.  The latter could be caused by
microphysical changes such as an increase in the droplet number
concentration.

\subsection{Effect of clouds on the Earth's radiation energy budget}
\label{sec_energy_budget}  

The net radiative properties of a  cloud are mainly dependent on its
altitude and optical thickness.  Optically-thin clouds at high and
middle altitudes cause a net warming due to their relative transparency
at short wavelengths but opacity in the IR region, whereas thick clouds
produce a net cooling due to the dominance of the increased albedo of 
shortwave solar radiation.  Since the data of 
Fig.\,\ref{fig_low_cloud_1} indicate that the solar modulation appears
in the low clouds, the \emph{sign} of the cosmic-climate effect is now
known: increased cosmic rays are associated with increased low-clouds
and therefore with a cooler climate.

Estimates from the Earth Radiation Budget Experiment (ERBE) indicate,
overall, that clouds reflect more energy than they trap, leading to a
net cooling of about 28~\wpm\ from the mean global cloud cover of 63\%
(Table \ref{tab_erbe})  \cite{hartmann}. The observed absolute variation
in low cloud cover of about 2\%  over a solar cycle
(Fig.\,\ref{fig_low_cloud_1}c) corresponds to about 7\% relative
variation. From Table \ref{tab_erbe}, this would imply a solar
maximum-to-minimum change in the Earth's radiation budget of about
1.2~\wpm\  (0.3\% of the global average incoming solar radiation).  
This is a  significant effect---comparable to  the total estimated
radiative forcing  of 1.5~\wpm\  from the increase in CO$_2$
concentration during the last century.

\begin{table*}[htbp]
  \begin{center}
  \caption{Global annual mean forcing due to various types of clouds, 
from the Earth Radiation Budget Experiment (ERBE) \cite{hartmann}. The
sign is defined so that positive forcing increases the net radiation
budget of the Earth and leads to a warming;  negative forcing decreases
the net radiation and  causes a cooling.}
  \label{tab_erbe}
  \begin{tabular}{| l  l | r r | r r | r | r |}
  \hline
   Parameter& & \multicolumn{2}{|c|}{High clouds} &
 \multicolumn{2}{|c|}{Middle clouds} &   Low clouds & Total \\
 \cline{3-7}  
   & & Thin & Thick & Thin & Thick & All &  \\
  \hline
  \hline
  &  &  & & & & &  \\[-3ex]
 Global fraction & (\%) & 10.1 & 8.6 & 10.7 & 7.3 & 26.6 & 63.3 \\
\hline
  \multicolumn{2}{|l|}{Forcing (relative to clear sky):}   &  & & & & & 
\\
 Albedo (SW radiation) & (\wpm) & -4.1 & -15.6 & -3.7 & -9.9
 & -20.2 & -53.5  \\
 Outgoing LW radiation & (\wpm) & 6.5 & 8.6 & 4.8 & 2.4 & 3.5  &
 25.8 \\
\hline Net forcing & (\wpm) & 2.4 & -7.0 & 1.1 & -7.5 & -16.7 & -27.7
\\[0.5ex]
  \hline
  \end{tabular}
  \end{center}
\vspace{-10mm}
\end{table*}

\subsection{History of cosmic rays and climate change}
\label{sec_history}

If the periodic 11-year cycles of the sunspots and the associated cosmic
ray flux  were the end of the story, then it would be of limited concern
since there would be no resultant long-term change in the Earth's
weather but simply another cyclic ``seasonal'' change (albeit with a
period of 11 years and therefore heavily damped by the thermal mass of
the oceans).   However there is clear evidence of longer-term and
unexplained changes both in the Sun's  and in the Earth's magnetic
behaviours  and these, in turn, seem to have had long-term effects on
the Earth's climate.

\subsubsection{Global warming during the past century}
\label{sec_century}

\paragraph{The light radio-isotope record:}

Direct measurement of cosmic rays with particle detectors has been
systematically carried out only during the last 50 years.  However there
exists another reliable record of cosmic ray fluxes on Earth, which
stretches back for at least 200 millennia: the light radio-isotope
record  \cite{obrien,beer,masarik}.  Light radio-isotopes are
continually created in the atmosphere by spallation of N and O atoms. 
The spallation interactions are induced mainly by low energy neutrons
created in the secondary reactions of cosmic rays: protons, alphas and
heavier particles.  The two long-lived radioisotopes with the highest 
production rates are \cft\ (half life =  ($5730\pm40$) years, and global
mean production rate  $\sim$2.5 atoms cm$^{-2}$s$^{-1}$) and  \beten\
(half life = 1.5M years, global mean production rate
$\sim$3.5$\times 10^{-2}$ atoms cm$^{-2}$s$^{-1}$).

In the case of \cft\ atoms, they are rapidly oxidised to form 
\cft O$_2$.   The turnover time of \cotwo\ in the atmosphere is quite
short---about 4 years---mostly by absorption in the oceans and
assimilation in living plants.  However, recirculation from the oceans
has the result that changes in the \cft\ fraction on timescales less
than a few decades are smoothed out.  Plant material originally contains
the prevailing atmospheric fraction of \cft\ and, subsequently, since
the material is generally not recycled into the atmosphere, the fraction
decreases with the characteristic half life of
\cft.

\begin{figure*}[htbp]
  \begin{center}
      \makebox{\epsfig{file=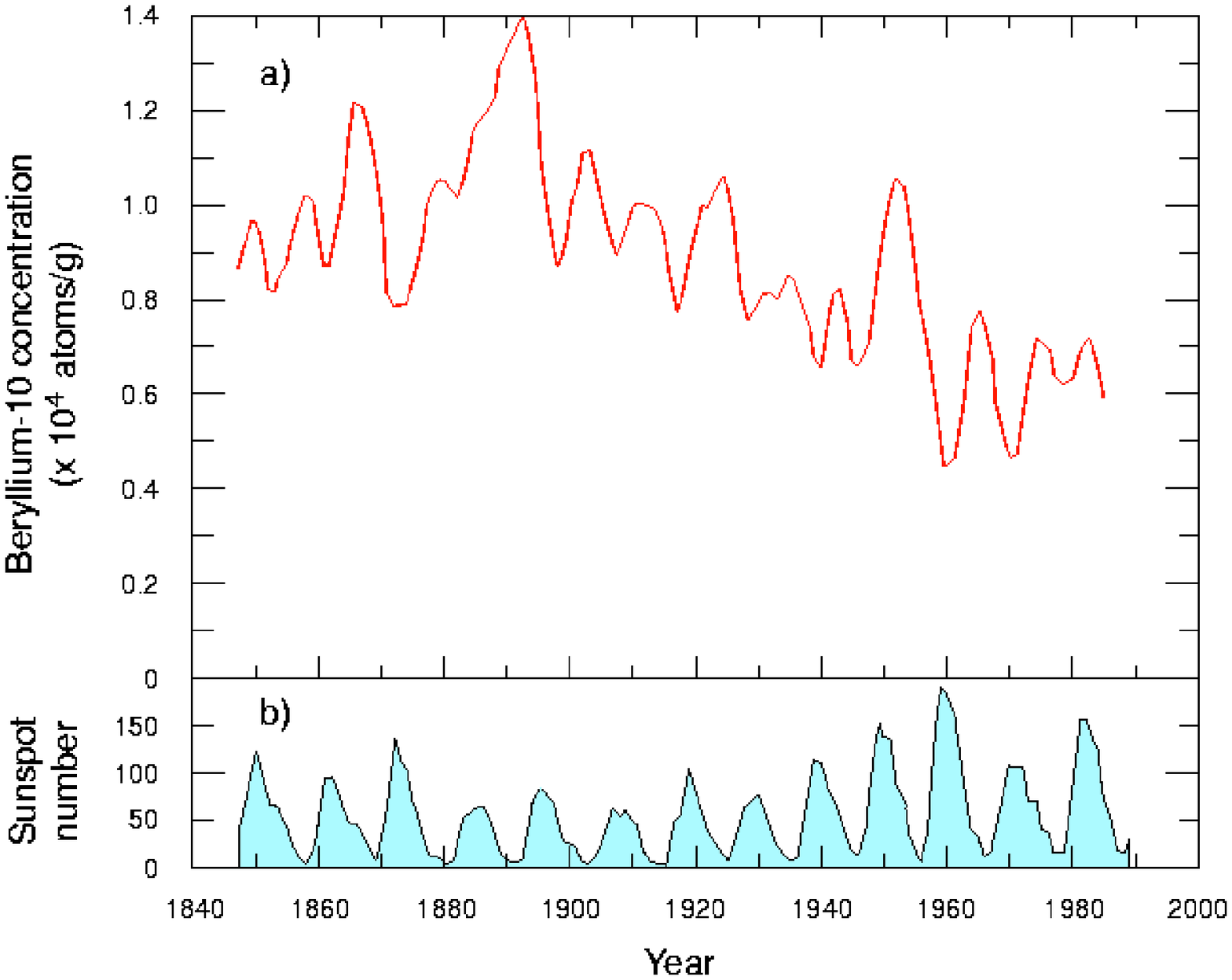,width=100mm}}
  \end{center}
  \vspace{-5mm}
\caption{a) Concentration of \beten\ in a 300 m ice core from Greenland
spanning the last 150 years \cite{beer}.  The data are  smoothed by
 an approximately 10 year running mean and have been shifted earlier  by
2~years to account for settling time.  b)~The sunspot cycle over the
same period, which shows a negative correlation with the short-term
($\sim$ 11 year) modulation of the  \beten\  concentration. }
  \label{fig_be10}    
  \begin{center}
      \makebox{\epsfig{file=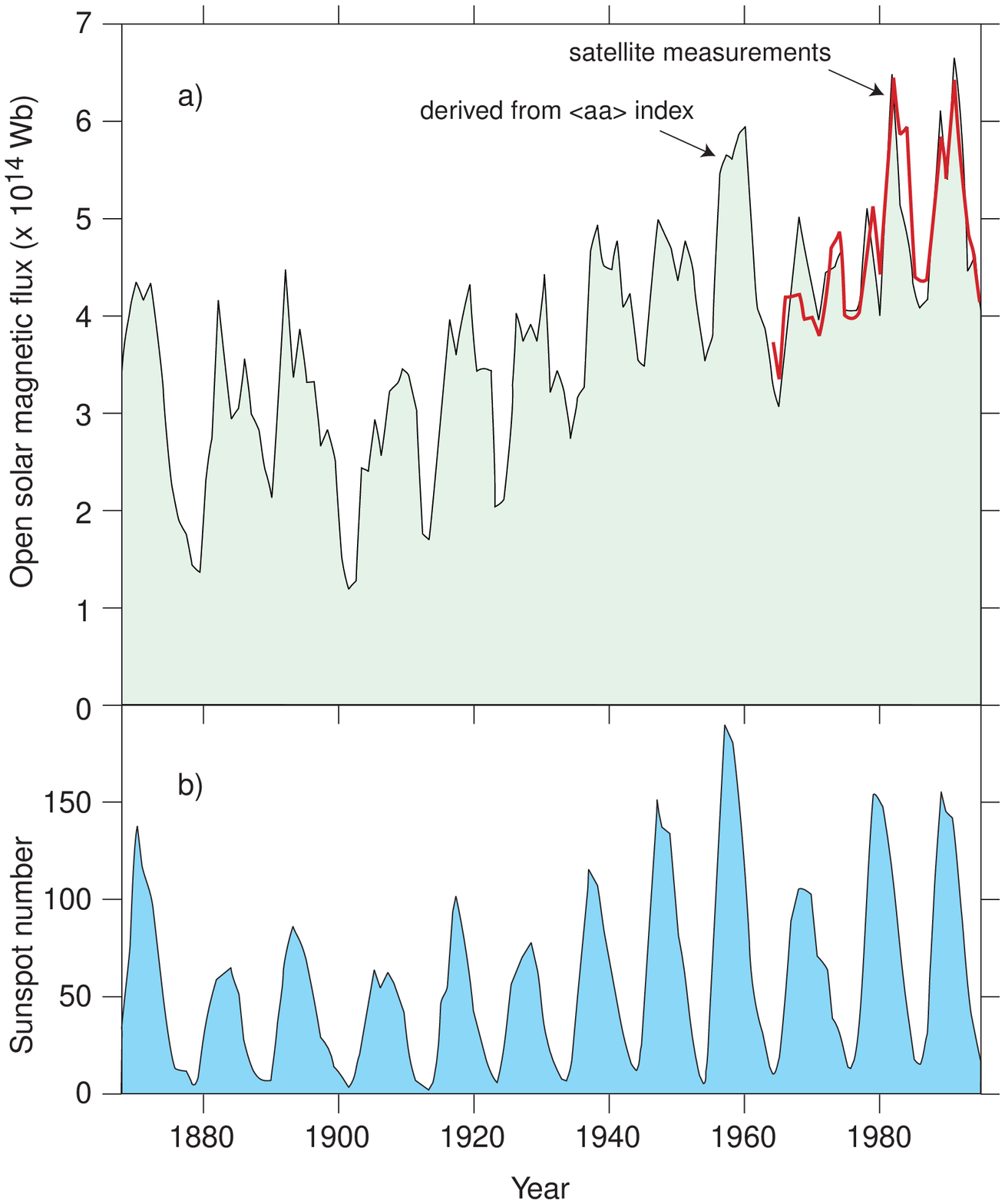,width=80mm}}
  \end{center}
    \vspace{-5mm}
  \caption{a) The total solar open magnetic flux (coronal source flux) 
   derived from interplanetary observations for 1964--1996 (thick solid
line) and derived from the $<$aa$>$ geomagnetic index for 1868--1996
(shaded curve)
\cite{lockwood}. b) The variation of the annual mean sunspot number.}
  \label{fig_solar_magnetic_flux}    
\end{figure*}

In the case of \beten, after production it rapidly attaches  to aerosols
(solid or liquid) and follows the motion of the surrounding air masses. 
Since the production of \beten\ follows the intensity profile of the
hadronic cosmic ray showers, about 2/3  is produced in the stratosphere
and 1/3 in the troposphere, globally averaged.  Due to the tropopause
barrier, aerosols in the stratosphere take about 1--2 years to  settle
to the  Earth's surface, whereas the mean residence time in the
troposphere is only days or weeks.  The removal mechanism is rain and
snow, and so seasonal variations of precipitation  may distort any
measurements on timescales less than a year or so.  In summary, despite
a factor 100  lower production rate than
\cft,  the advantages of \beten\ are that it settles out relatively
rapidly ($\lappeq$~2~years) and it is not  re-circulated into the
atmosphere.  It is therefore sensitive to changes in the cosmic ray flux
on short time scales of only a few years. On the other hand, \cft\ has
the advantages that it is not polar-centric and is independent of
precipitation variability. 

\paragraph{The change of cosmic ray intensity during the past century:}

Analysis of the \beten\ concentration in a Greenland ice core
(Fig.~\ref{fig_be10}) \cite{beer} reveals that the cosmic ray flux has
been steadily decreasing over the course of the last century; it is
weaker today at its \emph{maximum} during the sunspot cycle than it was
at its \emph{minimum} around 1900. From refs.\,\cite{obrien} and
\cite{masarik}, we estimate the global average reduction of cosmic ray
intensity to be about 15\% over the last century (with a  range of
\mbox{10--25\%}).  This estimate is supported by direct measurements of
cosmic rays over the last 40 years made by the Lebedev Physical
Institute using balloon-borne detectors  (Table
\ref{tab_gcr_flux} and Appendix \ref{sec_atmospheric_cr_meaurements})
\cite{stozhkov00}.

\begin{table}[htbp]
  \begin{center}
  \caption{Decreases of the galactic cosmic ray intensity measured over
the period 1957--1999 from balloon observations by the Lebedev Physical
Institute \cite{stozhkov00}.  An $\sim$11-year smoothing of the data was
applied to filter out the periodic solar cycle modulation.}
  \label{tab_gcr_flux}
  \vspace{2mm}
  \begin{tabular}{| l l c c |}
  \hline
  \textbf{Station} & \textbf{Location} & \textbf{Geomagnetic cutoff} & 
\textbf{Cosmic ray} \\
       &  & \textbf{rigidity, R$_c$} & \textbf{decrease} \\
       &  & \textbf{[GeV/$c$]} & \textbf{/10 years} \\
  \hline
    Murmansk & 68.57N, 33.03E  &  0.6 & 2.3\%  \\
    Moscow   & 55.56N, 37.11E  &  2.4  & 1.8\% \\[0.5ex]
  \hline
  \end{tabular}
  \end{center}
\vspace{-5mm}
\end{table}

The cause of this systematic decrease in the cosmic ray flux during the
last  century has been a marked strengthening of the solar wind and the
interplanetary magnetic field it carries into the heliosphere.  This is 
revealed by  the $<\!aa\!>$ geomagnetic index,\footnote{The $<\!aa\!>$
geomagnetic index is a sensitive measurement  by two antipodal stations
of short-term (3-hour interval) variations of the geomagnetic field at
the Earth's surface \cite{mayaud}, which is affected by the interactions
of the solar wind with the Earth's magnetosphere.} for which there is a
continuous  record extending back to 1868, covering 12 sunspot cycles.  
Lockwood \emph{et al.} \cite{lockwood} have estimated the source
magnetic flux, $F_s$, that leaves the corona and enters the heliosphere,
from the level of geomagnetic activity seen at Earth in the $<\!aa\!>$ 
index. This method to derive the coronal source flux has been
successfully tested against near-Earth interplanetary space
measurements made since 1963, during which time the coronal source  flux
of the Sun has been observed to rise a factor 1.4. In the period since
1901, Lockwood \emph{et al.} \cite{lockwood} calculate the increase  to
have been a factor 2.3  (Fig.~\ref{fig_solar_magnetic_flux}).  The
reason for this dramatic increase in the Sun's  magnetic activity is a
mystery.

\begin{figure}[htbp]
  \begin{center}
      \makebox{\epsfig{file=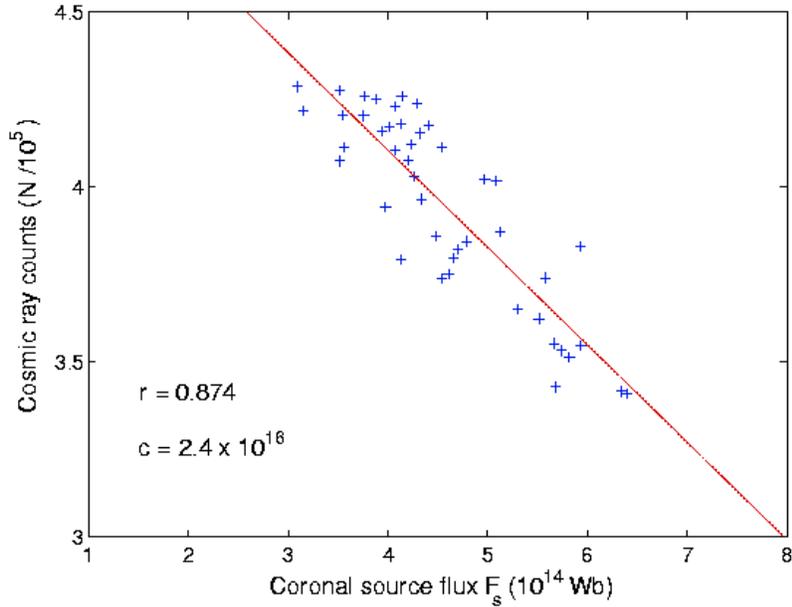,height=80mm}}
  \end{center}
  \vspace{-5mm}
  \caption{The negative correlation of annual means cosmic ray counts,
$N$, observed at Climax (cut-off 3 GeV) and the coronal source flux,
$F_s$ (as estimated from geomagnetic observations). The peak
correlation coefficient is $r = -0.874$, with the cosmic ray data
lagged by one month. The probability of arriving at this result by
chance is 
$c = 2.4 \cdot 10^{-16}$. The solid line is the best linear regression
fit.}
  \label{fig_gcr_vs_csf}    
\end{figure}

\begin{figure}[htbp]
  \begin{center}
      \makebox{\epsfig{file=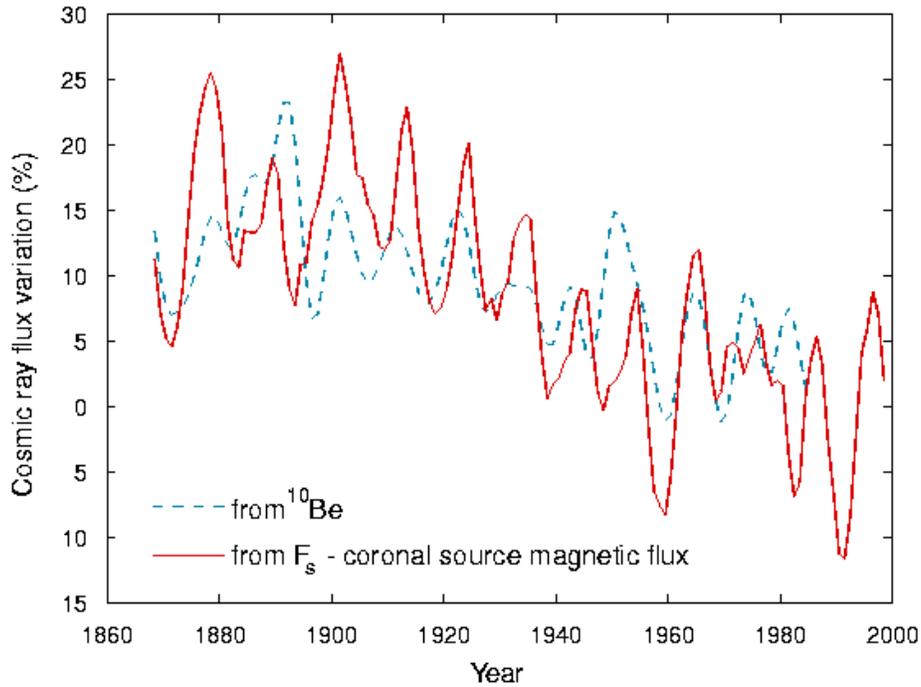,height=90mm}}
  \end{center}
    \vspace{-5mm}
  \caption{The estimated changes in the cosmic ray flux over the last
140 years. The solid line is the per cent variation (relative to the
mean value for solar cycle 21) derived from the linear regression of
cosmic ray fluxes with the coronal source flux. The dashed line is the
variation derived from observations of the \beten\ isotope concentration
found in a Greenland ice core.}
  \label{fig_gcr_1870_2000}    
\end{figure}

We have estimated the change in cosmic ray intensity over the last 140
years using these coronal source flux data.  Scattering, gradient and
curvature drifts caused by the heliospheric magnetic field are the major
contributor to the shielding of cosmic rays from the Earth and a very
strong anti-correlation between cosmic ray fluxes with the heliospheric
field near Earth was reported for the recent sunspot  cycles 21 and 22
(Fig.\,\ref{fig_cosmics_vs_sunspots}).  Thus an anti-correlation of the
source flux $F_s$ with cosmic ray fluxes is expected.  This is seen in
Fig.\,\ref{fig_gcr_vs_csf} which shows monthly mean counts, N, detected
by the Climax neutron monitor in the interval 1953-1998 as a function of
the annual estimates of $F_s$ made by Lockwood \emph{et al.}. The peak
correlation (r = -0.874) is, essentially, 100\% significant and is
obtained with the cosmic ray fluxes lagged by one month. The line shows
a linear regression fit to these data which can be used, along with the
$F_s$ data sequence, to extrapolate the cosmic ray flux variation back
to 1868. The percent variation of the cosmic ray flux (relative to the
average value for solar cycle 21) derived this way is shown by the solid
line in Fig.\,\ref{fig_gcr_1870_2000}. This extrapolation predicts a
fall in the average cosmic ray fluxes of about 20\% since 1900 for
Climax (3~GeV cutoff), which  implies about 15\%, globally averaged, in
agreement with the estimates given above. This change can be compared
with the \beten\ isotope record in ice sheets. The inferred cosmic ray
variation is plotted as a dashed line in Fig.\,\ref{fig_gcr_1870_2000} 
and the long-term changes agrees well with the variation inferred from
$F_s$.

If the cosmic-cloud link is real then this reduction of cosmic ray
intensity implies a net positive radiative forcing equivalent to about
one solar cycle, i.e. +1.2 \wpm\ (Section~\ref{sec_energy_budget}) over
the past century.  In short, a systematic  decrease in the cosmic ray
flux of the magnitude indicated by the \beten\ and coronal magnetic flux
measurements  could have  caused  a reduction in cloud cover and
consequent warming of the Earth comparable to the observed rise of
0.6\degc\ in global temperatures last century, which is presently
attributed predominantly to anthropogenic greenhouse gases.

\subsubsection{Climate change over the last millennium}
\label{sec_millenium}

The observations of a correlation between the Sun's  activity, cosmic
rays and the Earth's climate can be extended to earlier times with
either
\beten\ or  \cft\ data.  The latter agree well with \beten\ data after
accounting for a lag of $\sim$50 years in the \cft\ data due to the
effects of re-circulation from the oceans \cite{beer}.

\begin{figure*}[tbp]
  \begin{center}
      \makebox{\epsfig{file=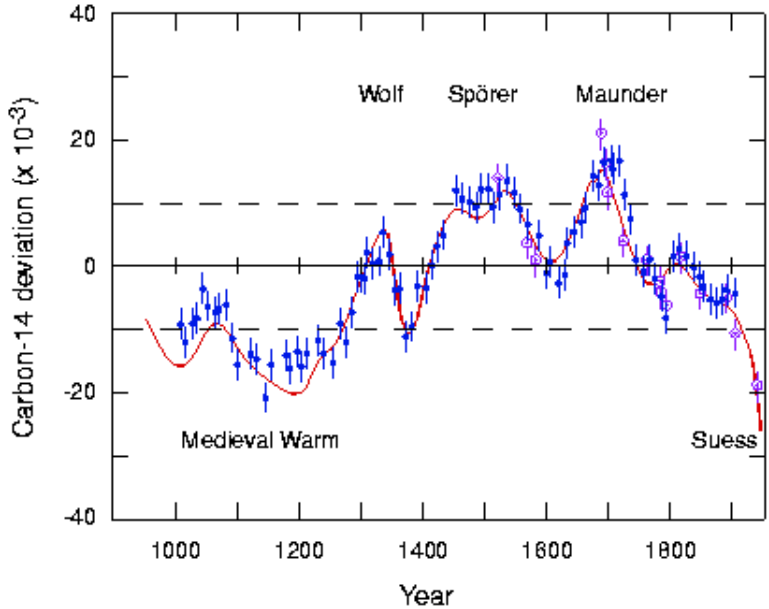,height=80mm}}
  \end{center}
  \vspace{-8mm}
  \caption{History of deviations in the relative atmospheric \cft\
concentration from tree-ring analyses for the last millennium 
\cite{damon}.  The data points (dots and open circles) are two
independent high-precision measurements.  The  solid curve represents a
combined fit to a large number of other measurements of medium
precision.  The dashed lines indicate \cft\ deviations of 10 parts per
mil.  The first four labelled periods correspond to recorded climatic
anomalies.   The sharp negative
\cft\ deviation during the present century is the Suess effect, due to
the burning of
\cft-depleted fossil fuels. }
  \label{fig_c14_1k_damon}    
\end{figure*}

  By analysing the \cft\ content in the rings of long-lived trees such
the Californian bristlecone pine, a year-by-year record has been
assembled of the cosmic ray flux on Earth over the past several thousand
years.  The data for the last 1000 years are shown in
Fig.~\ref{fig_c14_1k_damon} \cite{damon}.  The periods where the
\cft\ deviation approaches or exceeds 10 parts per mil correspond to
recorded climatic anomalies: a)~1000--1270, the so-called Medieval Warm
period,  b) 1280--1350, the Wolf minimum, c)~1420--1540, the Sp\"{o}rer
minimum, and d) 1645--1715, the Maunder minimum.  The warm period that
lasted until about 1300 enabled the Vikings to colonise Greenland and
wine making to flourish in England.  It was followed by a period of
about 500 years during which---save for a few short interruptions---the
glaciers advanced and a cooler, harsher climate predominated.

The Maunder Minimum, when there was an almost complete absence of
sunspots, corresponded to a high cosmic ray flux on Earth and therefore,
under the present hypothesis, to an increased cloudiness.  This provides
a consistent explanation for the exceptionally cold weather during this
period. Indeed, in every case the lack (a) or excess (b--d) of \cft\ is
consistent with the hypothesis of a higher cosmic ray flux leading to
more clouds and cooler temperatures, and vice versa.

\begin{figure}[tbp]
  \begin{center}
      \makebox{\epsfig{file=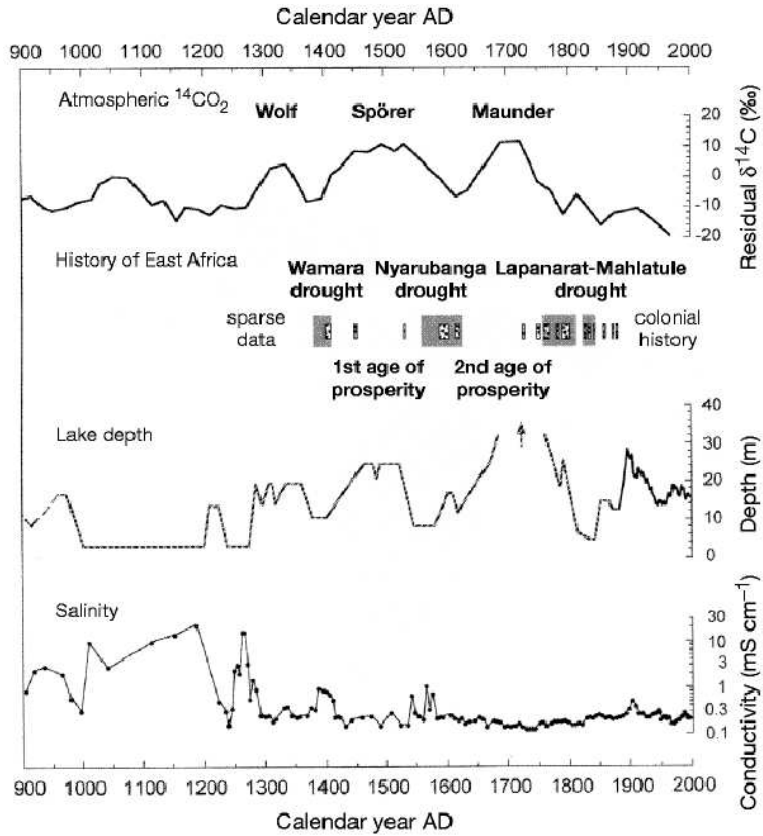,height=110mm}}
  \end{center}
  \vspace{-5mm}
  \caption{History of rainfall and drought in equatorial east Africa
during the last 1100 years \cite{verschuren}. The central and lower
figures show the  reconstructed depth and salinity, respectively, of
Crescent Island Crater lake (Kenya). The radiocarbon dating error for
the lake data is $\pm 50$ y.  The upper figure shows the atmospheric
\cft\ deviation over the same period.  Grey bars indicate evidence of
drought-related political upheaval recorded in oral tradition,
genealogically dated using a 27-yr dynastic generation. Dotted bars
compile the evidence of severe drought periods from various archival
records.}
  \label{fig_africa}       
\end{figure}

Evidence has recently been presented that this climate pattern extended
into the equatorial regions, and is therefore likely to have been a
global phenomenon. Figure \ref{fig_africa} shows the correlation of the
\cft\ record with the  depth and salinity of a lake in equatorial East
Africa over the last 1100 years \cite{verschuren}. The reconstruction is
based on three independent palaeolimnological proxies: sediment
stratigraphy and species compositions of fossil diatoms and midges. 
These data not only confirm the presence of the major climatic anomalies
associated with the  Medieval Warm period and the Wolf, Sp\"{o}rer and
Maunder minima but also identify three extended drought periods between
the minima: AD 1390--1420, 1560--1625 and 1760--1840.  The agreement
with the \cft\ record is striking. The cultural history of the region,
recorded in documents and oral tradition, coincides with the
experimental data (see Fig.\,\ref{fig_africa}).  In the present
hypothesis, the periods of high cosmic ray flux would have corresponded
to increased cloudiness.  Under the assumption that increased
cloudiness implies increased cloud lifetime, the rainfall would have
decreased but at the same time the evaporative losses would have
decreased.  Which of these opposing  effects would dominate depends on
latitude and other regional effects.

These and earlier historical examples of climate anomalies  provide
strong evidence for solar variability and its coupling with the
climate.  One possible interpretation of the \cft\ data is that they
provide  a proxy for changes in the solar irradiance, and that this was
the actual cause of the climate change.   The alternative possibility is
the mechanism addressed in this proposal, namely that the coupling is
through the solar wind, cosmic rays and clouds. Regardless of the
mechanism, however, there is little doubt that the Earth has experienced
several extended warm and cold spells over the last 1000 years---and
indeed at earlier times---with climate swings comparable to the  recent
warming but which could not be due to  anthropogenic greenhouse gases. 
Whatever mechanism caused those earlier changes in the climate could
perhaps be at work today. 

In summary, there are two main conclusions to be drawn from the
historical record of cosmic rays and climate change. Firstly, the
pattern of systematic change in the global climate over recorded history
seems to follow the observed changes of cosmic ray flux; and it is
consistent with the explanation that a low cosmic ray flux corresponds
to fewer clouds and a warmer climate, and vice versa. Secondly, there
has been a systematic decrease of the cosmic ray flux by about 15\% 
 over the course of the last century, caused by a doubling of the solar
coronal source magnetic flux.  The rise of about 0.6\degc\ in global
temperatures over the last 100 years is consistent in magnitude and time
dependence with the observed changes in cosmic ray flux---and thereby
cloud cover---over the same period.   If the cosmic-cloud link is
confirmed then it  provides a new mechanism for climate change that may
significantly revise the estimated contribution to global warming from
anthropogenic greenhouse gases.    A clear and compelling case exists 
to investigate the causal link between cosmic rays and clouds.

\section{Goals of the CLOUD experiment} \label{sec_goals}

\subsection{Scientific goals} \label{sec_scientific_goals}

The primary scientific goals of the CLOUD experiment are as follows:
\\[-2ex]

\begin{enumerate}

\item To study the link between cosmic rays and the formation of large
ions, aerosol particles, cloud droplets, and ice crystals.\\[-2ex]

\item To understand the microphysical mechanisms connecting cosmic rays
to changes in aerosol and cloud particle properties. \\[-2ex]

\item To simulate the effects of cosmic rays on aerosol and cloud
properties under atmospheric conditions. \\[-2ex]

\end{enumerate}
 
Concerning the last item, particular care will be taken to assess
whether cosmic ray variations could have a significant effect on cloud
properties within the natural variability of other factors. Where
significant effects are found, we will attempt to provide the climate
modelling community with simplified parameterisations of changes in key
properties of individual clouds. This will enable the influence of
cosmic rays on clouds to be incorporated into GCMs (general circulation
models) and an evaluation made of their contribution to the global
radiative forcing over the last century (see
Fig.~\ref{fig_ipcc_forcing}).

\subsection{Paths connecting cosmic rays and clouds} 
\label{sec_cosmic_cloud_paths}

Atmospheric clouds are highly complex and can be affected by a wide
range of environmental parameters. At the simplest level, the
development of a single small cumulus cloud is influenced by small
changes in the temperature structure of the atmosphere, changes in
humidity and surface heating rate. At a more complex level, the number
of cloud droplets that form and their size distribution are affected by
the composition and sizes of the aerosol particles upon which the water
vapour condenses (Appendix~\ref{sec_cloud_physics}).  There is also an
enormous variety of cloud types, whose properties depend on the specific
environmental conditions under which they form and develop.

\begin{figure*}[tbp]
  \begin{center}
     \makebox{\epsfig{file=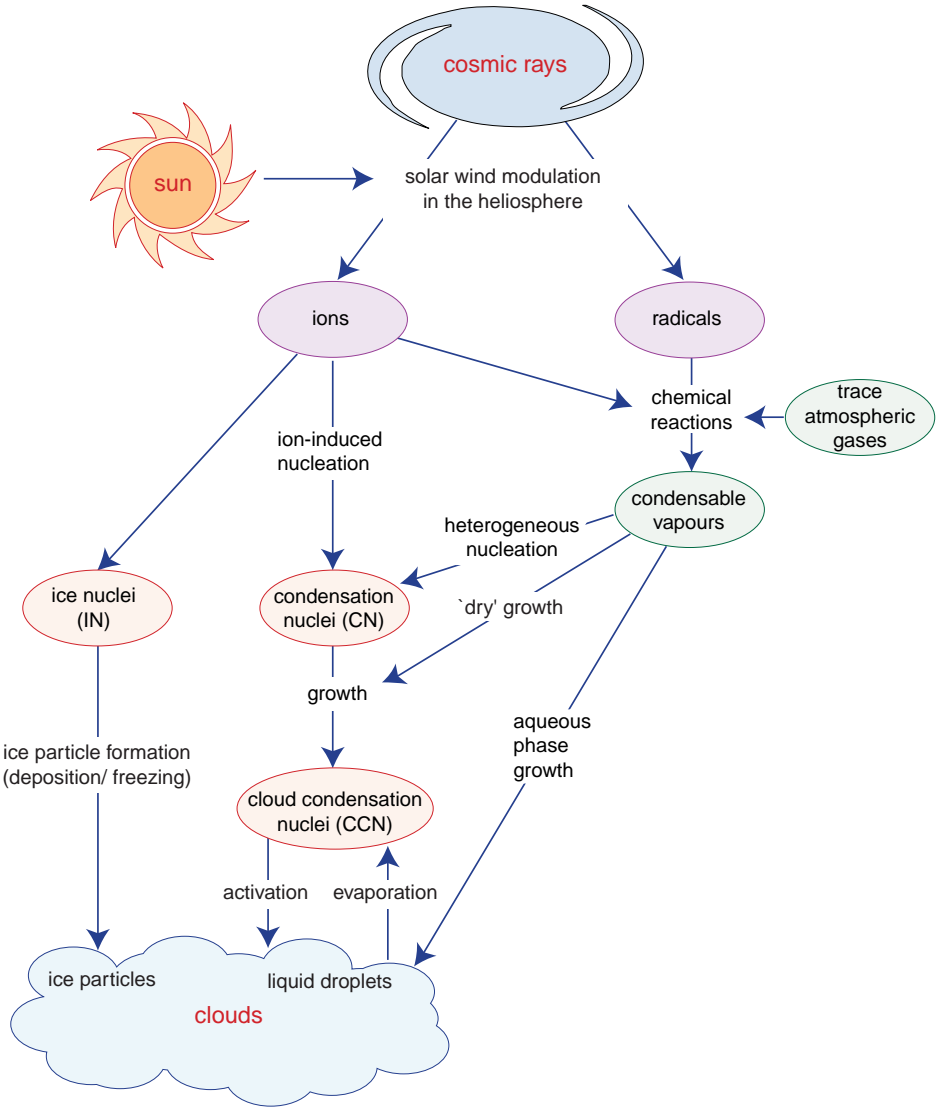,width=125mm}}
  \end{center}
  \vspace{-5mm}
  \caption{Paths that may connect cosmic rays to clouds and hence link
solar and climate variability through the solar wind modulation.  The
processes are described in Section~\ref{sec_cosmic_cloud_paths}.}
  \label{fig_cosmic_cloud_paths}    
\end{figure*}

This complexity makes it difficult to make a straightforward connection
between cosmic ray flux and cloud behaviour. Our approach in the CLOUD
experiment is to:
\begin{enumerate}

\item Identify a number of key aerosol and cloud properties that might
be affected by cosmic rays.

\item Experimentally determine the effects of cosmic rays  under
constrained laboratory conditions relevant to the atmosphere.

\item Incorporate the experimental results in computer models of aerosol
and cloud processes.

\item Simulate the behaviour of natural clouds with a variable cosmic
ray rate to determine which cloud properties are sensitive to these
variations.

\end{enumerate}

If cosmic rays can influence clouds, it is likely to be through their
effects on aerosols or on ice nucleation. Aerosols are found throughout
the atmosphere and constitute efficient cloud condensation nuclei (CCN)
for activation of cloud droplets (Appendix \ref{sec_cloud_physics}). 
The presence of a largely abundant supply of CCN ensures that the
maximum water vapour supersaturations in the atmosphere rarely exceed
values of about 1\% since higher values are arrested by the removal of
water vapour during droplet growth.

We have identified four distinct ways in which cosmic ray ionisation
could, either directly or indirectly, affect clouds in the troposphere.
The mechanisms are summarised in Fig.\,\ref {fig_cosmic_cloud_paths} and
described individually in  
Sections~\ref{sec_nucleation}--\ref{sec_ice_formation} below. It is
also possible that cosmic rays influence aerosol processes in the
stratosphere, as discussed in Section~\ref{sec_stratosphere}. 

\subsubsection{Enhanced aerosol nucleation and growth into cloud
condensation nuclei}\label{sec_nucleation}

\begin{figure}[tbp]
  \begin{center}
      \makebox{\epsfig{file=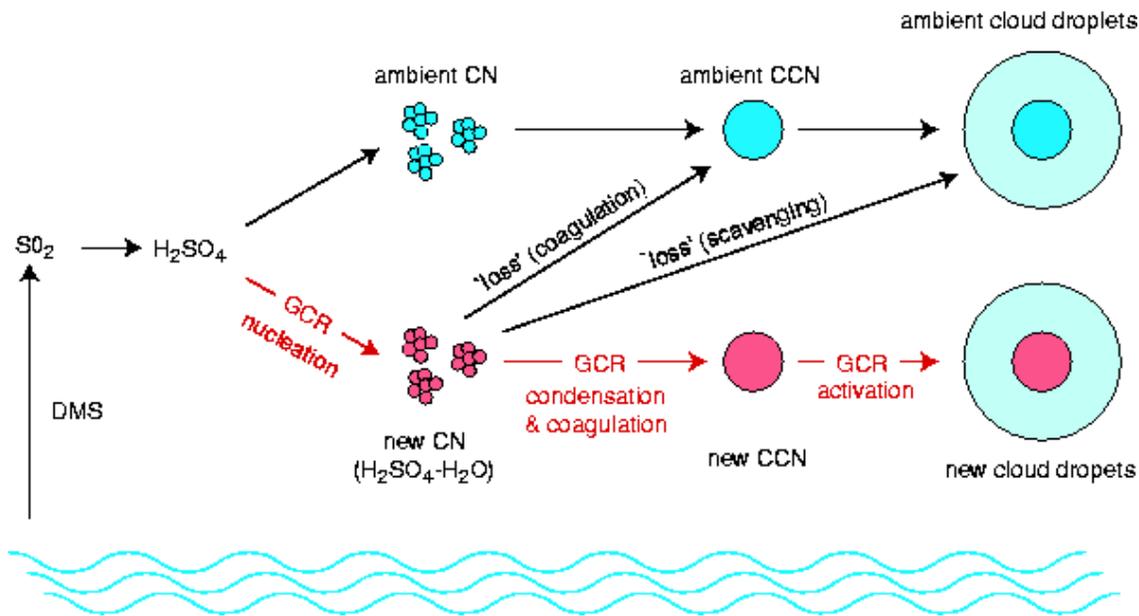,width=150mm}}
  \end{center}
  \caption{Possible influence of galactic cosmic rays (GCRs) on the
nucleation of new condensation nuclei (CN) and on the growth into cloud
condensation nuclei (CCN), ultimately causing an increase in  the
concentration of cloud droplets. The arrows labelled `GCR' are processes
whose rate may be affected by cosmic rays. Dimethyl sulphide (DMS) from
plankton is the major source of sulphur dioxide---the precursor of
sulphuric acid---in remote marine environments.}
  \label{fig_gcr_to_cn}    
\end{figure}

Cosmic rays create ions in the troposphere, which may affect aerosol
microphysical processes. Figure~\,\ref{fig_gcr_to_cn} shows a
simplified set of atmospheric pathways that might connect variations of
atmospheric ionisation with changes in the CCN abundance. The unbroken
arrows indicate processes that are known to occur in the atmosphere.
The arrows labelled `GCR' are processes whose \emph{rate} may be
affected by ionisation. The most important processes affected are
likely to be aerosol nucleation and aerosol particle growth either by
condensation or by coagulation.\footnote{\emph{Nucleation} refers to
the creation from vapours of a small ($\sim$1 nm diameter), stable
molecular cluster (aerosol) of a few tens or hundreds of molecules.
\emph{Coagulation} is the growth of an aerosol population by particles
colliding and sticking together.} Clouds that form in air containing
high CCN concentrations tend to have high droplet concentrations,
which enhances the shortwave (solar) albedo. Increase in the CCN
concentration also inhibits rainfall and therefore increases cloud
lifetimes (cloud coverage).  These effects---which are due to more,
smaller droplets at a fixed liquid water content---are particularly
significant in marine air, where the CCN concentrations are generally
quite low.

The possible cosmic ray influence on CCN abundance indicated in 
Fig.~\ref{fig_gcr_to_cn} is similar to, but not the same as, the
so-called
\emph{aerosol indirect effect} on clouds and climate. The aerosol
indirect effect concerns changes in the \emph{supply} of the aerosol
material (such as  SO$_2$ in Fig.~\ref{fig_gcr_to_cn}), which
subsequently causes a change in the CCN abundance and hence cloud
droplet concentrations. In the \emph{cosmic ray--aerosol--cloud}
indirect effect, which we propose to study, the cause of a change in CCN
abundance would be changes in the \emph{rates}  
 of certain aerosol transformation processes.

It has been long speculated that atmospheric sulphate particles are
formed by ion-induced nucleation (see, for example,
refs.\,\cite{raes85,raes86} and references quoted therein). Also, 
processes such as ion-ion recombination have been proposed which lead to
the production of stable molecular clusters in the lower stratosphere
\cite{arnold80} and in the troposphere \cite{turco98,yu00}. Since the
nucleation rate of new aerosols from these processes is proportional to
ion concentration, variations of the cosmic ray flux may translate into
variations of aerosol particle concentrations and, ultimately, CCN
concentrations. Recent experimental data suggest that ions may indeed be
involved in gas-to-particle conversion.  H\~{o}rrak \emph{et al.}
\cite{horrak} reported the spontaneous formation of bursts of
intermediate size ions in urban air, which they suggest may be due to
ion-induced nucleation. The Helsinki group has made similar observations
of aerosol bursts in unpolluted marine air \cite{odowd} and in remote
continental air \cite{kulmala00}.

Although the ion-induced nucleation phenomenon has been studied for more
than 100 years, it remains poorly understood, both experimentally and
theoretically. For example, the classical theory describes the charge
effect on nucleation by an electrostatic interaction term between the
ion and condensing molecules. However, this term does not explain the
ion sign preference exhibited by many molecules (e.g.\,water molecules
nucleate more easily on negative ions) which was experimentally found
already by Wilson in 1899
\cite{wilson99}. The sign preference has been attributed to surface
orientation of dipolar molecules \cite{loeb,talanquer}, and a partial
theoretical understanding has been proposed \cite{rusanov}.

There is new evidence that increases in CCN concentration can indeed
suppress rainfall, and hence increase cloud lifetime. A recent study
\cite{rosenfeld} used NOAA satellite data to investigate clouds that
formed downwind of industrial sites located in pristine areas.  The
otherwise uniform cloud data from these regions was streaked with bright
(highly reflective) plumes from the industrial sites.  The droplets in
these plumes were found to be more numerous than the nearby regions and
of a smaller diameter---typically less than 10$\mu$m and therefore below
the threshold size for them to coalesce efficiently and precipitate.  In
contrast the droplets outside the plumes measured more than 25 $\mu$m in
diameter.  The high reflectivity of the plumes resulted from the high
droplet number density at fixed liquid water content
(Fig.\,\ref{fig_cloud_reflectivity}).  Independent analysis of data from
the Tropical Rainfall Measuring Mission confirmed that these plumes did
indeed produce less rain and therefore had a longer lifetime than clouds
in the nearby regions. These observations suggest that, if increases in
cosmic ray flux could be translated into increases in CCN abundance, the
observed increases in cloudiness (Fig.\,\ref{fig_cloud_climax}) could
be due to decreases in precipitation efficiency.

\subsubsection{Enhanced cloud condensation nucleus activation by charge
attachment}\label{sec_activation}

Aerosol activation is the rapid growth of an existing aerosol into a
large ($\gappeq$1~$\mu$m) liquid droplet by condensation of water from a
supersaturated vapour. If aerosol charging (induced by cosmic
ionisation) could decrease the supersaturation needed for activation,
this would lead to an increase in droplet number densities in clouds,
with consequences for all cloud microphysical processes.

The possible effect of charges on aerosol activation has usually been
ignored since the conventional electrostatic interaction term is
expected to be appreciable only for very small aerosols of $\sim$1 nm
diameter for small charges (see Appendix~\ref{sec_wilson_chamber}). 
However, a recent study \cite{chen} on heterogeneous nucleation of
n-butanol vapour on charged and uncharged insoluble particles shows a
surprisingly large charge effect. The charges have a clear effect on the
nucleation process even when the seed particles are as large as 90 nm in
diameter and the supersaturation is only 0.5\%. For these experiments
the classical theory predicts that the difference in the nucleating
ability of charged and uncharged seed particles should vanish for
particle diameters above 20 nm, corresponding to supersaturations of
about 100\%.  If this result is correct, it indicates that charges may
play a role in atmospheric cloud drop activation processes, at least
when the CCN are partially insoluble (e.g.\,carbon) and carry multiple
charges.  Furthermore, the charges could influence the condensation of
low vapour pressure trace gases on aerosol particles, and thereby affect
the processes transforming CN into CCN.

\subsubsection{Formation of condensable vapours and the effect on cloud
condensation nuclei}\label{sec_vapours}

The ions and radicals, together with trace atmospheric gases, may
promote the formation of condensable vapours or enhance the condensation
of vapours already present, which can lead to the growth of existing
aerosols (Appendix \ref{sec_aerosol_cloud_climate}).  The condensation
may occur on unactivated aerosols or on cloud droplets (aqueous phase
growth), which would result in larger aerosol mass after evaporation of
the water.  Growth of aerosols can lead to a higher CCN number
concentration, and to CCN that activate into droplets at lower
supersaturations. The free radicals created by cosmic rays may also be
able to influence atmospheric chemistry under certain conditions,
especially if catalytic reactions are involved or where cosmic rays
constitute a significant source of a chemical species 
\cite{crutzen86,crutzen93}.

\begin{figure}[htbp]
  \begin{center}
      \makebox{\epsfig{file=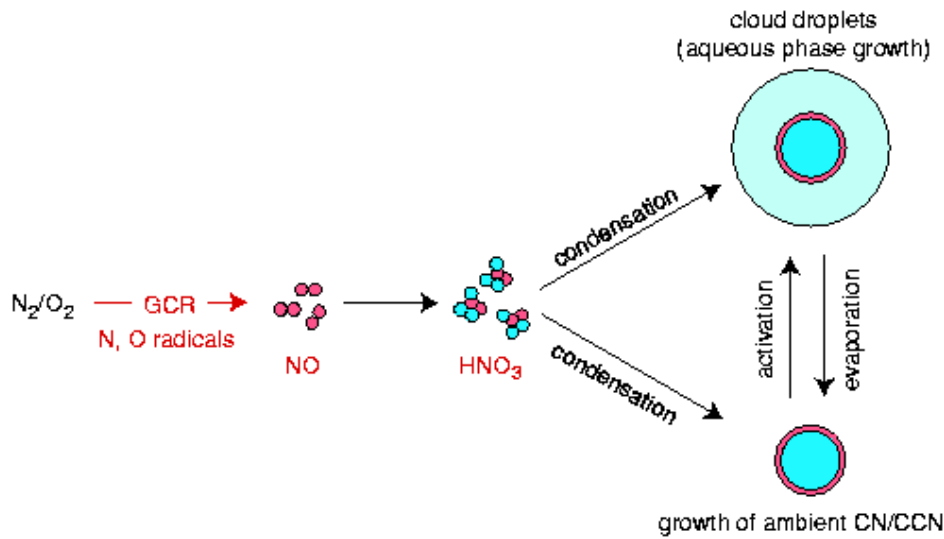,height=70mm}}
  \end{center}
  \vspace{-5mm}
  \caption{Possible influence of galactic cosmic rays on the production
of new condensable vapours and subsequent growth of existing CN and
CCN.}
  \label{fig_gcr_to_cv}    
\end{figure}

The most important radicals created by cosmic rays are N, O, and OH 
from the dissociation of the primary active constituents of air: N$_2$,
O$_2$ and H$_2$O.  The estimated  production rates are about 1--2
hydroxyl (OH)  molecules per ion-pair \cite{crutzen86} and 1.5 nitric
oxide (NO) molecules per ion-pair \cite{nicolet,jackman,crutzen75}. 
These rates imply mixing ratios of about 0.7 pptv NO are generated per
day by cosmic rays  directly in the upper troposphere.  After oxidation 
to nitric acid this may affect the growth of both CN and CCN
(Fig.\,\ref{fig_gcr_to_cv}).

Although the production of NO by cosmic rays is small on a global basis,
it may be the dominant source in the upper troposphere of remote regions
\cite{nicolet,jackman}.  Experimental evidence for the importance of
particle ionising radiation for NO production is demonstrated by a clear
nitrate signal in the Greenland ice core associated with solar proton
events and the solar cycle \cite{dreschhoff,kocharov}.  Calculations
indicate that galactic cosmic rays may be responsible for about half of
the NO in the polar stratosphere, and may be the dominant source during
the polar winters when N$_2$O oxidation is suppressed
\cite{crutzen93,jackman}. 

\subsubsection{Creation of ice nuclei}\label{sec_ice_formation}

\begin{figure}[htbp]
  \begin{center}
      \makebox{\epsfig{file=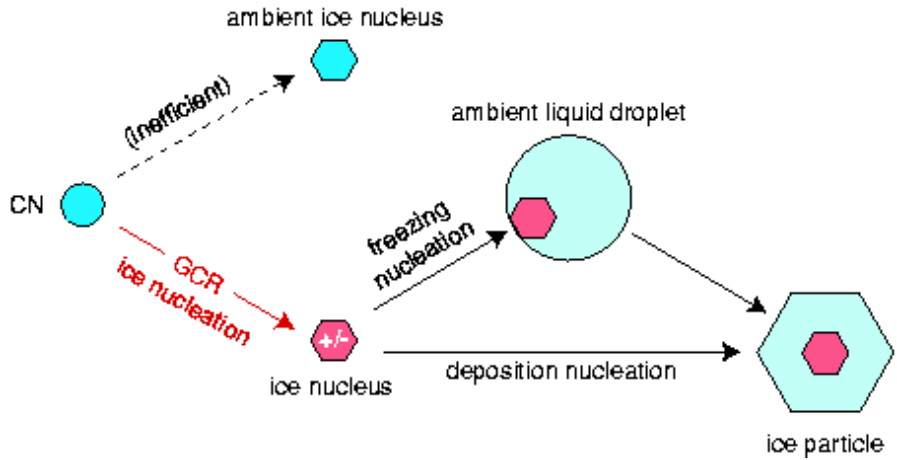,height=60mm}}
  \end{center}
  \caption{Possible influence of galactic cosmic rays on the creation of
ice nuclei (IN) and ice particles at temperatures below 273~K.}
  \label{fig_gcr_to_in}    
\end{figure}

The presence of ice in clouds has an important influence on their
radiative properties and it can also lead to precipitation.  However the
apparent lack of ice nuclei (IN) in the atmosphere is at present a
mystery: there is about 1 IN per litre at 253 K (compared with about
$10^6$ aerosol particles per litre) whereas the observed concentrations
of ice particles in clouds at these temperatures vary between 10 and 300
per litre.  Two nucleation processes can lead to ice formation: a)
direct sublimation of vapour to the solid phase (deposition nucleation)
and b) freezing of a liquid droplet (freezing nucleation). It has been
suggested that the  ionisation produced by cosmic rays may be able to
affect both deposition and freezing nucleation
(Fig.~\ref{fig_gcr_to_in}).

\begin{figure}[tbp]
  \begin{center}
      \makebox{\epsfig{file=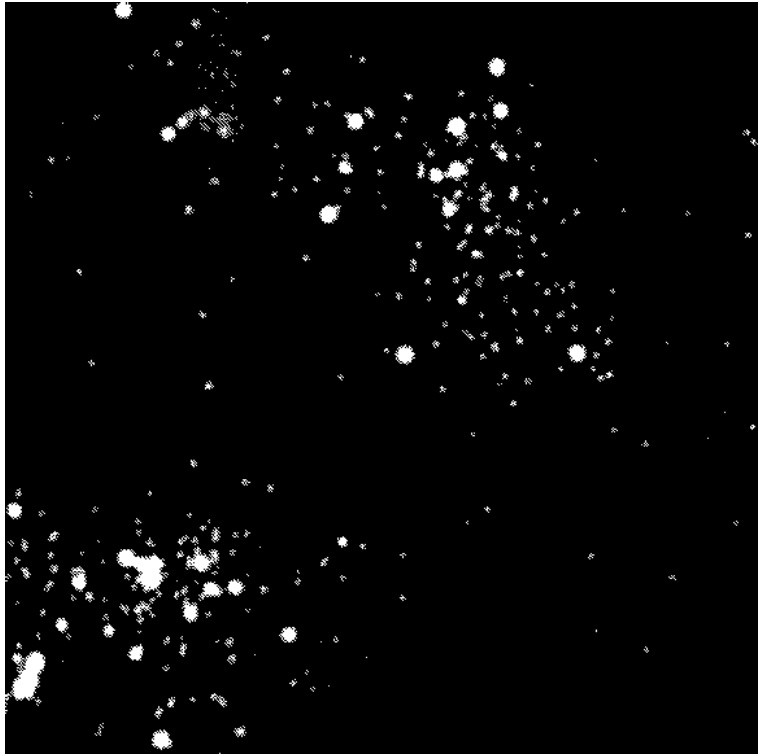,height=100mm}}
  \end{center}
 \vspace{-5mm}
 \caption{Ice and liquid water droplets observed together in a cloud
expansion chamber at 230 K. The two clusters correspond to regions of 
ionisation due to cosmic rays that crossed roughly along the direction
of view. The ice droplets visibly scatter more light than the liquid
droplets, and appear larger. In fact the large apparent size of the ice
crystals is probably the result of light scattering in the emulsion
film.  The field of view in this image is about 17 mm across and about
10~mm in depth. The data were obtained at the University of
Missouri-Rolla in 1980
\cite{anderson}.}
  \label{fig_ice_nuclei}   
\end{figure}

The possible effect of charges on ice nucleation has been studied to
some extent (see Pruppacher and Klett \cite{pruppacher} for a review),
and effects have been seen in a number of experiments. Especially, the
ice nucleating ability of various types of seed particles seems to be
influenced by ionisation.  It is interesting to note that cloud chamber
experiments at the University of Missouri-Rolla in 1980
\cite{anderson} found an enhancement of frozen droplets in the regions
where cosmic rays traversed the cloud chamber
(Fig.\,\ref{fig_ice_nuclei})---although it was not possible to determine
if the cosmic rays preceded or followed the droplet formation.  The
presence of ions was observed to raise the threshold temperature for
homogeneous ice nucleation by about 2 K.

Enhancing heterogeneous ice nucleation by electrification has been
proposed by several workers in the context of solar-terrestrial climate
connections. It is supported by very little experimental work so far
\cite{pruppacher}. If the electrofreezing effect
\cite{tinsley} is real, possible influences of cosmic rays on cirrus
formation and on the properties of middle clouds are implied.
Electrofreezing could also affect glaciation (which triggers rain) in
convective clouds. This may in some way be related to the correlation
between cosmic rays and precipitation \cite{stozhkov}.

\subsubsection{The effect of cosmic rays on stratospheric clouds and
ozone depletion} 
\label{sec_stratosphere}

The previous four processes have concerned aerosols and clouds in the
troposphere. However, cosmic ray intensities are even higher in the
stratosphere, so it is conceivable that they also have an influence on
aerosol processes in that part of the atmosphere 
\cite{krieger94}.

Chemical reactions on stratospheric aerosols are a prerequisite for
seasonal ozone depletion in both the Arctic and Antarctic stratosphere
\cite{solomon}. These reactions convert relatively inert inorganic
chlorine compounds into photochemically labile forms that can enter
into ozone-destroying gas-phase catalytic cycles. The formation of {\it
polar stratospheric clouds} in the low temperature polar stratosphere
dramatically accelerates the rate of these reactions, leading to the
formation of the ozone hole.

Significant advances have been made in our understanding of polar
stratospheric cloud formation in the last 10--15 years
\cite{peter,carslaw97}. However, there remain important uncertainties
that prevent a reliable prognosis of ozone depletion rates in a given
winter or in future years. Chief among these uncertainties is what
controls the \emph{phase} (liquid or solid) of polar stratospheric
cloud particles. At temperatures above the ice frost point the particles
may be either liquid solutions of nitric acid, sulphuric acid and water
or else solid hydrates of nitric acid. The formation of solid hydrates
is important because it allows the selective growth of a small number of
particles that subsequently become large enough to sediment out of the
stratosphere. This process leads to {\it denitrification} of the polar
stratosphere and often to strongly enhanced ozone loss (see
ref.\,\cite{waibel} and and references therein). 

Several mechanisms are recognised to be important for the formation of
solid polar stratospheric clouds particles \cite{peter,koop97a}.
However, persistent and optically very thin clouds, often seen over
several thousand kilometer regions cannot be explained by any recognised
mechanism. These tenuous hydrate clouds appear to be formed by a
mechanism that operates on a very large scale with little spatial
variability, thus excluding formation mechanisms involving localised
cooling \cite{carslaw99}. Laboratory experiments also exclude the
possibility that these solid particles form by crystallisation of the
liquid aerosols \cite{koop97b} by large scale cooling. An intriguing
possibility---so far unexplored---is that these clouds form by
deposition nucleation of nitric acid and water directly on cosmic
ray-generated ions or ion clusters. However, to date, there have been no
experiments that can confirm or dispute this possibility. An
understanding of such particles appears to be critical to a complete
understanding of denitrification and ozone loss.

\subsection{Experimental concept}  \label{sec_experimental_concept}

The essential approach of CLOUD is to test the link between cosmic rays
and cloud formation under controlled laboratory conditions in a particle
beam, which provides an adjustable and collimated source  of ``cosmic
rays''. The detector is based on an expansion cloud chamber that is
designed to duplicate atmospheric conditions.  This requires the
capability of producing the very low water vapour
supersaturations\footnote{Water vapour supersaturation, 
 $SS = S - 1 = p/p_0 - 1$, where $S$ is the saturation ratio, $p$ is the
ambient partial pressure of water vapour and $p_0$ is the saturated
vapour pressure over a plane surface of water at this temperature. 
Supersaturation is frequently expressed as a percentage.}---typically a
few tenths of a percent---found in clouds, so that the precise
cloud-forming properties of the air parcel under study can be
measured.  

The advantage of this approach is that all the experimental conditions
can be precisely controlled and measured.  In particular, by comparing
measurements at several beam intensities from zero up to high values, we
can determine if there is a causal relationship between ionising
radiation and cloud formation.  Such measurements are difficult to
perform with cosmic rays in the atmosphere since the natural intensity
variations are modest and they follow the slow 11-year solar cycle.
Furthermore, the CLOUD experiment provides full control of the initial
gas mixture and aerosol content and, in addition, complete physical and
chemical analysis of the final products after beam exposure. This will
greatly facilitate an understanding of the microphysics and chemistry of
any observed effects.  The challenges of the laboratory-based approach
are to precisely duplicate the atmospheric conditions and to ensure that
the finite volume (walls) of the cloud chamber does not affect the
measurements. 

Surprisingly, cloud chamber data under atmospheric conditions in a
particle beam have never been previously obtained.  C.T.R.\,Wilson's
cloud chamber\footnote{Wilson had the inspiration for the cloud chamber
while observing meteorological phenomena on the mountain of Ben Nevis in
1894.  The phenomena were not particle tracks, however, but ``coronas''
around the Sun and glories, where the Sun glows around shadows in the
mist.  He developed the cloud chamber to try to reproduce these effects
in the laboratory.  Although his expansion cloud chamber was crucial to 
the development of nuclear  and particle physics for the next half
century and earned him the 1927 Nobel Prize in Physics, Wilson remained
fascinated by atmospheric phenomena throughout his life \cite{galison}.
Indeed, he devoted a large part of his later research life to seeking a
connection between cosmic rays and clouds.  We are therefore glad to
propose a Wilson cloud chamber for this experiment.} 
\cite{wilson,galison}  was extensively used for experimental particle
physics in the first half of the 20th century, but was mostly operated
under conditions far removed from those of the atmosphere.   In order to
grow droplets on the small ions produced by ionising radiation, Wilson
cloud chambers were operated with water vapour supersaturations of
500--600\%, to be compared with maximum
 values of about 1\% in atmospheric clouds (see Appendices
\ref{sec_cloud_physics} and \ref{sec_wilson_chamber}). However, it is
interesting to note observations made in cloud chambers in the 1960's  
 of backgrounds due to ``hypersensitive condensation nuclei''
\cite{watson}.  These activated into droplets at very low water vapour
supersaturations---of order 1\%---and  were attributed to trace amounts
of NO$_2$ vapour created by electrostatic discharges.

\begin{table*}[tbp]
  \begin{center}
  \caption{Initial CLOUD experimental programme and operating 
conditions.}
  \label{tab_experimental_conditions}
  \begin{tabular}{| l  l  l | r | r  r  r |}
  \hline
  & \textbf{Experiment} & 
  \textbf{Aerosols}$^\dagger$ \textbf{/} & 
   \textbf{Size [nm]} or &
\multicolumn{1}{|c}{\textbf{P}} & \multicolumn{1}{c}{\textbf{T}} 
 & \multicolumn{1}{c|}{\textbf{H$_2$O}}  \\ 
 & 
 & \textbf{Trace gases} & \textbf{Conc.\,[ppb]}  &
\multicolumn{1}{|c}{\textbf{[mb]}} &
\multicolumn{1}{c}{\textbf{[K]}} &  \\
  \hline
  \hline
 &  &  &  &  &  &  \\[-2ex]
\multicolumn{3}{| l |}{\textbf{I. Creation \& growth of aerosols:}}& & &
&
\\[1ex] 
 1. & Sulphuric acid & H$_2$SO$_4$ & 0.0001-1 ppb & 1000 & 270-300 &
20-80$^\ddagger$ \\ 
 2. & "  + Ammonia & " + NH$_3$ & +0.001-0.1 ppb & 1000 & 270-300 &
20-80$^\ddagger$ \\ 
 3. & Sulphur dioxide & SO$_2$ &  0.0001-1 ppb  & 1000 & 270-300 &
20-80$^\ddagger$\\
 4. & Sodium chloride & NaCl & 5-50 nm   & 500-1000  & 260-300 &
0.1-1$^\S$  \\ 
  5. & Ammonium sulphate & (NH$_4$)$_2$SO$_4$ & 5-50 nm & 500-1000 &
260-300 & 0.1-1$^\S$ \\
 6. & " + Nitric acid & " + HNO$_3$ & 0.1-100 ppb & 500-1000 & 260-300 &
0.1-1$^\S$ \\[0.5ex]
\hline
 &  &  &  &  &  &  \\[-2ex]
\multicolumn{3}{| l |}{\textbf{II. Activation of aerosols into
droplets:}}& & & &
\\[1ex] 
 7. & Sodium chloride & NaCl & 50-100 nm   & 500-1000  & 260-300 &
0.1-1$^\S$  \\ 
  8. & Ammonium sulphate & (NH$_4$)$_2$SO$_4$ & 50-100 nm & 500-1000 &
260-300 & 0.1-1$^\S$ \\
 9. & " + Nitric acid & " + HNO$_3$ & 0.1-100 ppb & 500-1000 & 260-300 &
0.1-1$^\S$ \\[0.5ex]
\hline
 &  &  &  &  &  &  \\[-2ex]
\multicolumn{3}{| l |}{\textbf{III. Formation of condensable
vapours:}}& & & &
\\[1ex] 
 10. &  NO$_x$ & N$_2$:0$_2$  & pure 80\%:20\% & 500-1000  & 260-300 &
0.1-500$^\S$  \\ 

\hline
\multicolumn{3}{| l |}{\textbf{IV. Creation of ice nuclei:}}& & &
&\\[1ex] 
 11. &  Pure water & none & 10 $\mu$m droplets  & 200-1000  & 213-273 &
1-50$^\star$ \\ 
  12. & Ammonium sulphate& (NH$_4$)$_2$SO$_4$ & 10 $\mu$m droplets  &
200-1000 & 213-273 & 1-50$^\star$ \\ 
 13. & Sulphuric acid & H$_2$SO$_4$ & 10 $\mu$m droplets  & 200-1000 &
213-273 & 1-50$^\star$ \\
\hline
 &  &  &  &  &  &  \\[-2ex]
\multicolumn{3}{| l |}{\textbf{V. Stratospheric aerosols:}}& & & &
\\[1ex]
 14. & Nitric acid & HNO$_3$ & 10-20 ppb  & 10-100 & 190-200 & 5 ppm
\\[0.5ex]
\hline
  \end{tabular}
  \end{center}
$^\dagger$ Aerosol number concentration: $\sim$500 cm$^{-3}$. \\
$^\S$  Supersaturation [\%] relative to liquid water, 
 during activation. \\
$^\ddagger$ Relative humidity [\%] during beam exposure.\\
$^\star$ Supersaturation [\%] relative to ice.
\end{table*}

\subsection{Initial experimental programme}
\label{sec_experimental_programme}

The initial CLOUD experimental programme and operating conditions are
summarised in Table~\ref{tab_experimental_conditions}.  It is important
to note that our experimental search is quite broad since at present
there is no clear microphysical explanation of the possible cosmic-cloud
link.  We therefore expect to adapt our investigations according to our
experimental observations and to the current experimental and
theoretical developments at the time of taking data.  The initial
experimental programme described here should therefore be considered as
representative rather than definitive. 

Each experiment will in general  be performed with two different carrier
gases: a) pure Ar (inert carrier), b) pure artificial air (80\% N$_2$, 
20\% 0$_2$).  The cloud-forming properties and other physical and
chemical  characteristics of the aerosols will be measured under
beam/no-beam conditions using a range of equipment described in Section
\ref{sec_detector}.

In the following subsections we outline some of the experiments 
required to explore the mechanisms described in  Sections
~\ref{sec_nucleation}--\ref{sec_stratosphere}.

\subsubsection{Aerosol nucleation and growth experiments}
\label{sec_aerosol_creation}

These studies concern the formation of aerosols from the vapour phase,
via ion-induced nucleation, and their subsequent growth by vapour
condensation and coagulation.

We will investigate the clustering of trace gas molecules onto ions. The
trace gas molecules will either be formed by the ionising particle beam
in pure artificial air or be introduced directly into the chamber. Such
trace gas molecules include, in particular, H$_2$O, H$_2$SO$_4$,
HNO$_3$, NH$_3$ and certain volatile organic compounds. These trace
gases will be measured by CIMS (Chemical Ionisation Mass Spectrometry),
and the ions will be measured by ion mass spectrometers and ion mobility
analysers. During activation measurements in the cloud chamber a fairly
high supersaturation (large expansion) will be required, depending on
the size and nature of the nucleated aerosols.

A major goal of these studies is to find out what fraction of the small
ions created by ionising radiation become stable aerosol particles. 
This fraction $f$ will increase with decreasing temperature $T$ and
with  increasing abundance of the clustering trace gas molecules. For
example, in the case of H$_2$SO$_4$/H$_2$O ion clusters, $f$ is
expected to increase with increasing relative humidity and relative
acidity.  However, for low H$_2$SO$_4$ concentrations a kinetic
limitation becomes important which is related to the limited ion-ion
recombination lifetime
$t_{IR}$. It is therefore important to work under well-controlled
conditions with respect to the total ion concentrations $n_i$ and
thereby $t_{IR}$ on the one hand and the sulphuric acid concentration
[H$_2$SO$_4$] on the other. Both $n_i$ and [H$_2$SO$_4$], along with
[H$_2$O] and $T$, must be precisely known and adjustable.

The link proposed in section 3.2.1 between aerosols and clouds includes
also a charge-enhanced growth of aerosols from small size (Aitken mode,
20--100~nm diameter) into the cloud-condensation nuclei (CCN) mode
($\gappeq$100 nm) where they can efficiently activate to form cloud
droplets. We will start with a study of aerosols that are common in the
atmosphere and known to be important in cloud formation such as NaCl,
(NH$_4$)$_2$SO$_4$ and H$_2$SO$_4$.  Both ``dry'' and aqueous-phase
growth will be studied.  We will begin with well-known, simple systems
to first confirm that the apparatus is well understood. This involves
measurements with monodisperse NaCl or (NH$_4$)$_2$SO$_4$ aerosols of
diameter 10~nm, followed by 20 nm and then 40 nm. Then a second or third
vapour component will be added and more complex systems studied.

Comparisons will be made with and without beam.  Since diffusion of
small ions is significant for beam exposures longer than about a minute
(Section~\ref{sec_beam_diffusion}), the with/without beam measurements
will be made in separate runs.   Because of their relatively small
mobility, the loss of aerosol particles to the walls of the cloud
chamber is rather slow (Section~\ref{sec_particle_losses}) and,
depending on the specific conditions, growth processes lasting 1--10
hours or more can be studied.  

We will also investigate aqueous phase growth by taking measurements
with several activation $\rightleftharpoons$ evaporation (expansion
$\rightleftharpoons$  compression) cycles. The possibility that negative
ions cause faster growth and condensation rates than positive ions will
be investigated by selecting the ion charges with the field-cage of the
cloud chamber (Section \ref{sec_electric_field}). 

\subsubsection{Cloud condensation nuclei activation experiments} These
studies concern the growth of aerosols into cloud droplets at relative
humidities greater than 100\%. Such aerosols, known as cloud
condensation nuclei, have typical sizes in the atmosphere of 50-100
$\mu$m. Aerosols of a well defined size and with typical atmospheric
composition (NaCl, (NH$_4$)$_2$SO$_4$, H$_2$SO$_4$) can be produced with
standard aerosol generation techniques
(Section~\ref{sec_aerosol_system}). Experiments will be performed to
examine the activation of these aerosols into cloud droplets. The
cooling of the cloud chamber by expansion will be sufficiently precise
to be able to induce supersaturations with respect to water that are
typical of the full range of values observed in natural clouds
(Section~\ref{sec_design_considerations}). 

Experiments with and without beam will enable the effect of aerosol
charging on activation to be investigated. The charge distribution on
the aerosols will be measured using a uniform electric field created by
a field cage. In particular, we wish to know whether aerosol charging
can reduce the critical supersaturation required to activate an aerosol
particle into a water droplet. This can  be investigated by counting the
number of activated droplets as the supersaturation is gradually
increased in separate experiments.

\subsubsection{Condensable vapour formation experiments}

These experiments will first quantify the poorly-known production rates
of a) nitric oxide (NO) and b) hydroxyl radicals (OH) by cosmic
radiation.  The former will involve pure artificial air and the latter
will involve argon and water vapour.  Subsequently we will investigate
the effects of these vapours on the nucleation and growth of aerosols. 

\subsubsection{Ice nuclei formation experiments}\label{sec_ice_nuclei}

These studies concern the formation of ice nuclei in supercooled vapours
at low temperatures. The expansion chamber will be used to create a
supercooled cloud by expansion and growth of drops at temperatures below
260~K.  The temperature of the drops can be controlled by the initial
temperature of the chamber before expansion.  The beam will be pulsed
through the supercooled cloud and data recorded.  The presence of an ice
crystal in a cloud of drops is easily identified since an ice crystal
scatters much more light than a water drop (Fig.\,\ref{fig_ice_nuclei})
\cite{anderson}.  The charge distribution on the ice nuclei will be
measured using a uniform electric field created by the field cage. In
addition to experiments with supercooled liquid droplets already present
(freezing nucleation), we will also investigate ice nucleation without
pre-existing droplets (deposition nucleation). 

\subsubsection{Stratospheric cloud formation experiments} 

These experiments concern the deposition nucleation of nitric acid and
water vapours onto ion clusters to form nitric acid hydrates. Particles
composed of such hydrates are thought to be the principal component of
the  polar stratospheric clouds that initiate the destruction of ozone.

The temperature of the cloud chamber will be reduced to typical polar
stratospheric values of between 190 and 200~K. Nitric acid and water
vapour will be introduced into the chamber at partial pressures
representative of the stratosphere (10$^{-4}$ Pa for nitric acid vapour
and $5\times 10^{-2}$ Pa for water vapour). At these pressures and
temperatures the nitric acid hydrates become supersaturated and can
condense as crystals provided a suitable nucleus is present. We seek to
establish whether ion clusters can serve as these nuclei just as they
can for the formation of sulphuric acid droplets in the stratosphere
\cite{arnold82}. A background air mixture composed of water, nitric acid
and sulphuric acid vapours will be used to represent the species most
likely to contribute to initial ion cluster formation.

\section{CLOUD detector} \label{sec_detector}

\subsection{Design considerations} \label{sec_design_considerations}

\paragraph{Choice of expansion cloud chamber:} 

The expansion cloud chamber has several important advantages over other
nucleation devices\footnote{Besides the expansion cloud chamber, other
experimental devices for studying nucleation include the thermal
diffusion cloud chamber, cooled-wall expansion chamber, shock tubes and
turbulent mixing chambers.}  for the studies proposed here.   In
particular, previous measurements with expansion chambers by our
collaboration (Helsinki, Missouri-Rolla, and Vienna) have verified that
the thermodynamic conditions after an expansion are precisely known and
reproducible, provided that the initial conditions are well-known and
the expansion ratio (pressure change) is well-measured. Moreover the
expansion chamber can provide a large volume with uniform thermodynamic
conditions where, for example, relatively slow processes can be
measured.  Expansion chambers may in fact be the only devices that can
achieve the necessary thermodynamic stability and precision to study
aerosol activation in the laboratory at  the very small water vapour
supersaturations (few
$\times$ 0.1\%) found in clouds and, in addition, cover the full range
of supersaturations up to those required to activate small ions and
nanometre-sized aerosols.

\paragraph{Size of cloud chamber:} 

A large cloud chamber ($\sim$50 cm diameter) achieves the longest time
of known thermodynamic conditions in the fiducial volume\footnote{The
fiducial volume is the central region of the chamber where the
thermodynamic and other conditions are well known,  and where the 
measurements are made.} of the chamber.   The sensitive
time\footnote{The sensitive time following an expansion refers to the
period during which no significant changes of thermodynamic conditions
occur in the central part of the chamber caused by the heating
influences of the walls.} ranges from about a  second for expansions
that produce high enough supersaturations in water vapour to activate 
ions (a large difference in temperature between the walls and the gas) 
to several tens or even hundreds of seconds for the activation and
growth of large aerosols (a small difference in temperature)
(Section~\ref{sec_sensitive_time}).  The sensitive time of an expansion
chamber increases steeply with its size since it is proportional to the
square of the ratio of the  volume to wall area, which re-heats the gas
following an adiabatic expansion.  In addition a chamber size of about
50 cm ensures that diffusion losses of the aerosols to the walls are not
significant for measurements lasting up to several hours with a single
fill (see Section 
\ref{sec_particle_losses}).

\paragraph{Water vapour supersaturation:} 

The cloud chamber is required to operate at water vapour
supersaturations ($SS$) from below zero (unsaturated) up to about
700\%.  An  important requirement is to provide precise simulation of
the conditions found in clouds, for which  $0.1\% \lappeq SS  \lappeq
1\%$ (Appendix  \ref{sec_cloud_physics}).  This supersaturation range
corresponds to a broad activation range of aerosols, namely radii from
about 1 $\mu$m (at 0.1\% $SS$) down to about 50~nm (at 1\% $SS$). In
order to probe the aerosol size distribution with sufficient resolution,
the chamber needs to achieve a $SS$ precision after expansion of about
0.1\% in the range  \mbox{$0 <  SS  < 1\%$}.  Furthermore, the cloud
chamber also needs  to  measure condensation nuclei down to small ion
dimensions (0.2 nm), which requires large expansions producing
supersaturations of up to 500\%.  The precision of the large $SS$ values
is, however, less demanding than for the small values.    

\paragraph{Expansion time:}  The time duration of the expansion pulse
needs to be short compared with the droplet growth time so that the
start time is the same for all activated aerosols.  This is especially
important for experiments involving a broad size distribution of
aerosols (to represent atmospheric aerosols), which activate over a
relatively wide range of supersaturations.  A rapid expansion ensures
that the larger aerosols do not deplete the water vapour and prevent
activation of the smaller aerosols.  In practise the fastest required
expansion time is about 200 ms.  For comparison, the expansion time for
the 25~cm Vienna chamber is 20-30 ms and, for the 38 cm Missouri-Rolla
chamber, it is 200~ms.

\paragraph{Temperature and pressure operating range:} 

The cloud chamber is required to operate over the full range of
temperatures and pressures encountered by clouds in the troposphere and
stratosphere, namely  173 K $ <  T  <$ 293 K and $0 <  P  <$ 101 kPa. The
maximum pressure of the chamber is actually 150 kPa, in order to allow
measurements to be made at 1 atm following a large expansion. It is
important to note that the full range of pressure change can be
addressed with our chamber design (Section
\ref{sec_expansion_techniques}).  In a typical expansion chamber the
pressure of the gas in the sensitive volume is used to drive the
piston.  For low pressures this becomes ineffective.  In the cloud
chamber proposed here,  the piston  is moved by a hydraulic system and
thus a fast and precise expansion and re-compression can be achieved
regardless of the pressure change in the chamber.

\begin{figure*}
  \begin{center}
      \makebox{\epsfig{file=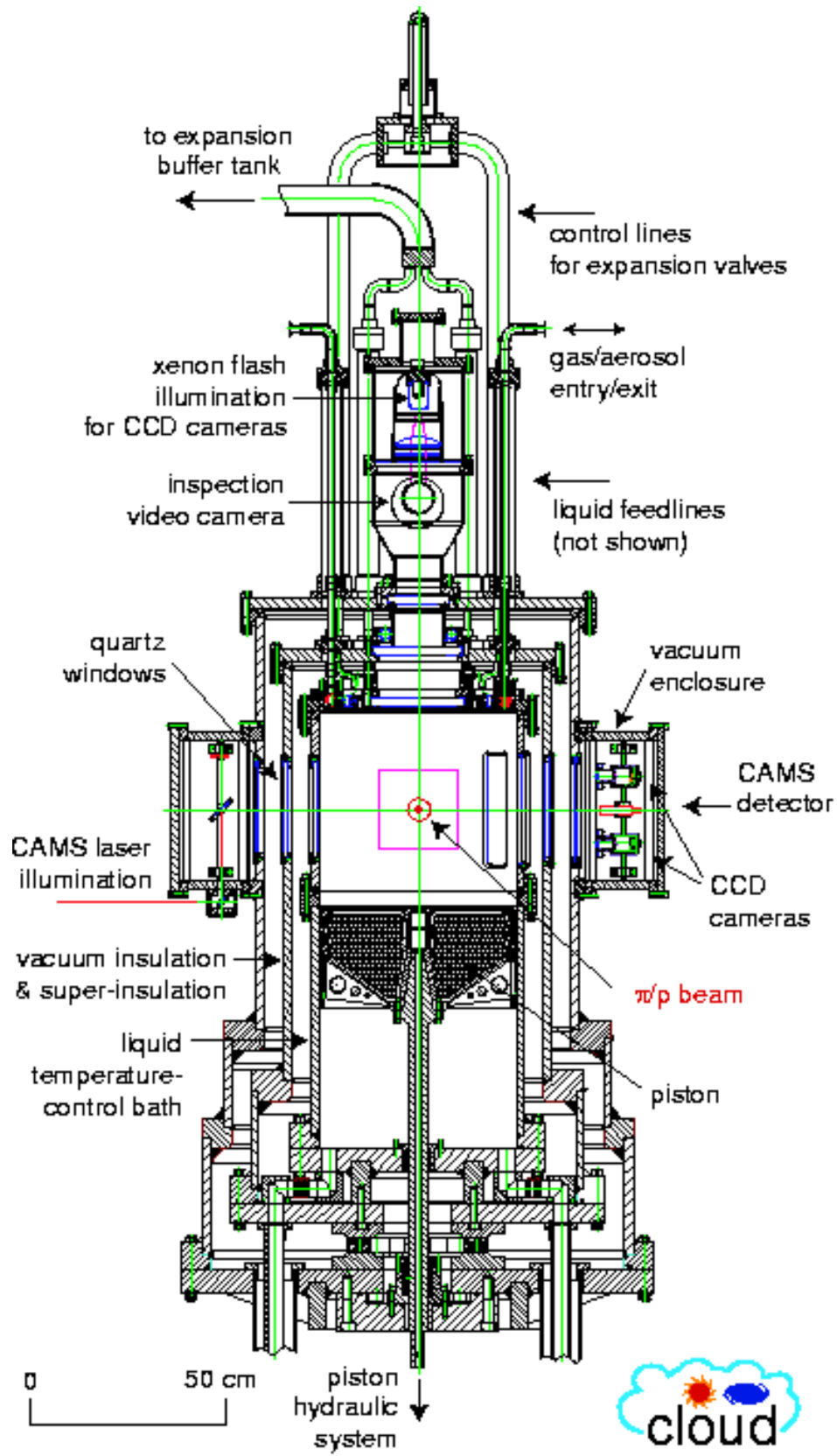,height=220mm}}
  \end{center}
  \caption{Vertical section through the CLOUD detector.}
  \label{fig_cloud_detector_v}    
\end{figure*}

\begin{sidewaysfigure}
  \begin{center}
      \makebox{\epsfig{file=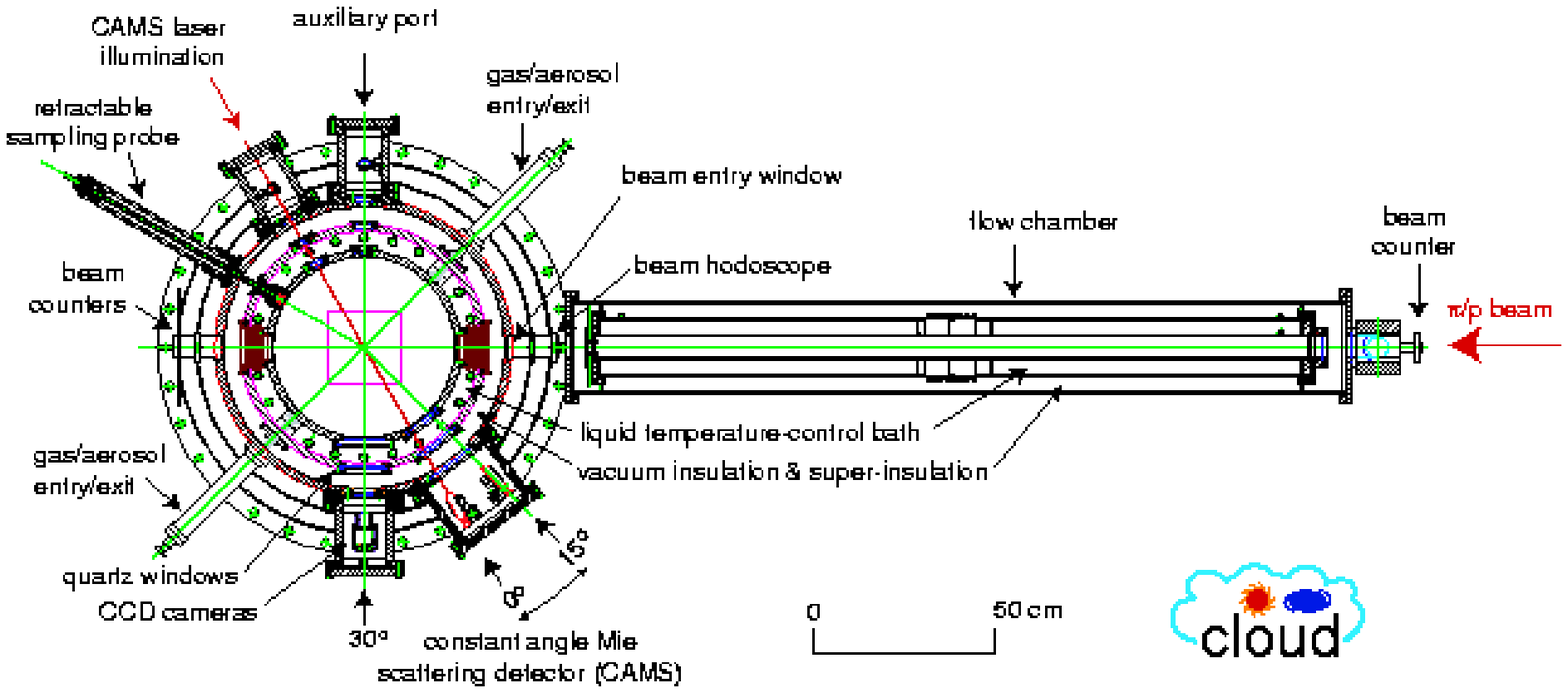,width=235mm}}
  \end{center}
  \caption{Horizontal section through the CLOUD detector.  The CAMS
detectors all lie in the same plane whereas the CCD cameras are
displaced above and below the plane of the figure.}
  \label{fig_cloud_detector_h}    
\end{sidewaysfigure}

\paragraph{Temperature and pressure stability:}

The requirement of a precision of 0.1\% on the supersaturation places
demanding requirements on the temperature control of the cloud chamber.
Taking a design value for the supersaturation error of one half this
value, i.e. 0.05\%, implies the need for  a temperature stability 
$\Delta T = $ 0.01 K and a pressure stability $\Delta P/P  = 1.3
\cdot 10^{-4}$ or, equivalently, a volume stability  $ \Delta V/V  = 0.9
\cdot 10^{-4}$ (see Section~\ref{sec_droplet_growth_time}, 
 Eq.\,\ref{eq_dp/p}). Note that these are
\emph{stability} requirements; the absolute precisions of the
temperature and pressure are less demanding.

\subsection{Cloud chamber} \label{sec_cloud_chamber}

\subsubsection{Overview} \label{sec_detector_overview}

The CLOUD detector is  shown in  Figs.\,\ref{fig_cloud_detector_v} and
\ref{fig_cloud_detector_h}.   The active volume is a cylinder of
dimensions 50~cm (height) $\times$ 50~cm (diameter) and the fiducial
volume is the central region of linear dimensions about 10--20 cm.  The
chamber is completely  surrounded by a liquid bath which maintains
precise temperature control. An outer vacuum enclosure, together with
super insulation, maintains the thermal insulation of the inner
detector. The cloud chamber volume is equipped with an electrode
structure (field cage) to provide a  clearing field or the possibility
to drift and measure charged aerosols. It also allows a method to select
positive or negative charges for separate study. 

Expansions are made by two techniques: a piston and a buffer expansion
tank.  The former is operated by a hydraulic system similar in design to
the Big European Bubble Chamber (BEBC).  The buffer expansion tank
involves an external tank with a volume ten times larger than that of
the cloud chamber. Expansions are effected by first reducing the
pressure in the external tank to the desired value and then opening
fast-acting valves connecting the external tank to the cloud chamber.

The optical readout of the cloud chamber comprises two systems: a) a
constant angle Mie scattering (CAMS) detector and b) a stereo pair of
CCD cameras.  The two systems are complementary but, nevertheless, have
a broad region of overlap where they can provide mutual cross-checks.  
The CAMS system can measure very high droplet number densities
($\sim$10--$10^7$ cm$^{-3}$) whereas the CCD cameras operate best in a
lower range \mbox{($\sim$0.1--$10^5$ cm$^{-3}$).}  The CAMS system
provides a high-resolution measurement of mean droplet radii vs.\,time,
whereas the CCD cameras provide a measurement of droplet size in coarser
time intervals, using pulse height information.  Finally, the CAMS
detector integrates over all illuminated droplets whereas the CCD
cameras reconstruct the 3-dimensional spatial positions of individual
droplets, and tracks their movements. This is important for identifying
ice nuclei (Section~\ref{sec_ice_nuclei}) and for measuring droplet
drift velocities (large droplet sizes). 

The illumination system comprises: a) a laser for illumination of a
narrow region for the CAMS detector (and, in parallel, for the CCD
cameras) and b) a xenon flash tube for the CCD cameras, mounted at the
top window (an auxiliary xenon flash system is also mounted close to the
CCD cameras). A video camera is also mounted at the top window in order
to provide a visual inspection of the chamber volume and piston surface.
The video camera and xenon illumination share the same window by means
of a partially reflecting mirror (Fig.\,\ref{fig_cloud_detector_v}).  
The side windows are designed to allow simultaneous viewing of both the
beam- and no-beam regions of the chamber.  All windows are made of
optical quality quartz.  The use of quartz allows for the possibility of
including UV irradiation to investigate reactions involving
photochemical processes. 
 
Cleaning of the chamber is very important to ensure reliable results. 
The upper window is designed to be removable to allow cleaning access to
the inside of the chamber and to minimise the transition time between
different chamber fills. The chamber is also cleanable by vacuum baking.
Vacuum evacuation is also an effective and rapid technique for removing
a gas/aerosol mixture before refilling. Finally, large chamber
expansions with a pure carrier gas can be used to confirm the
cleanliness of the chamber before filling with a new mixture.  Indeed,
droplet activation and sedimentation is a proven  technique to clean the
cloud chamber to ultra low levels of contamination.

The chamber is equipped with two sampling probes.  These provide the
possibility of extracting gas and aerosols from the fiducial region of
the cloud chamber.  They also provide the possibility for special
measurements such as, for example, injection of a special gas into a
limited region at the centre of the chamber.  The probe tubes are
straight and have a large diameter inner bore. All valves and pipes
leading to and from the chamber are designed to provide efficient
transmission of aerosols (large diameters and gentle curves). 
 
A  flow chamber (2m-length $\times$ 6cm-diameter) is integrated with the
cloud chamber assembly and exposed to the same beam as the cloud
chamber.  It provides the source for external measurements of the
physical and chemical properties of the trace gases, aerosols and ions
with mass spectrometers, ion detectors, condensation particle counters
(CPC) and differential mobility particle sizers (DMPS). 

\begin{figure}[tbp]
  \begin{center}
      \makebox{\epsfig{file=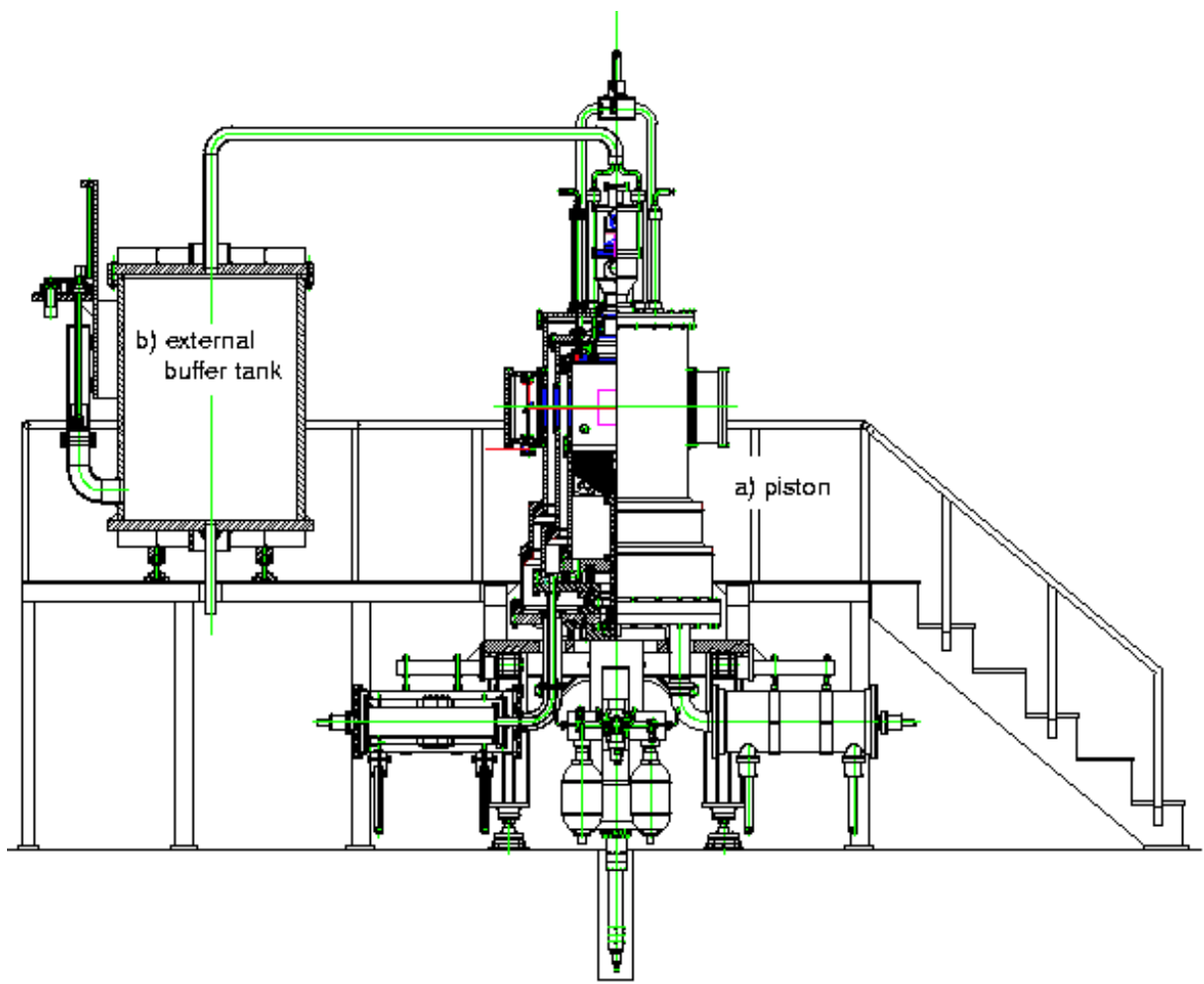,width=160mm}}
  \end{center}
\vspace{-5mm}
  \caption{Cloud chamber expansion systems: a) piston and b) external
buffer tank.}
  \label{fig_expansion_systems}    
\end{figure}

\subsubsection{Expansion techniques} \label{sec_expansion_techniques}

The cloud chamber has a flexible choice of expansion techniques and
expansion/ re-compression cycles. There are two expansion techniques
(Fig.\,\ref {fig_expansion_systems}):

\paragraph{a) Piston:} 

Piston movement involves active control of the piston connecting rod by
means of a hydraulic system similar in design to that used for the Big
European Bubble Chamber (BEBC).  An important advantage of piston
expansions is that they minimise the gas turbulence, an effect that
becomes more significant as the expansion ratio increases. The piston
also provides a flexible choice of precise expansion and re-compression
cycles, with piston movements reproducible to very high
precision---better than 5~$\mu$m.  The range of piston movement is
5~$\mu$m to 200 mm i.e. volume expansion ratios of between
$10^{-5}$ and 0.4. The very small expansions will provide
supersaturations characteristic of those found in atmospheric clouds,
while the largest expansions will activate aerosols of 
molecular/small-ion dimensions. The largest expansions also provide an
effective method to clean trace condensation impurities from the cloud
chamber by activation and then sedimentation.

\paragraph{b) External buffer tank:}  
  
This is a second method to produce small expansions. It involves a tank
with a volume
$\sim$10 times larger than the active volume of the  cloud chamber, i.e.
$\sim$1 m$^3$. The tank is not cooled but has good thermal insulation to
ensure that any temperature fluctuations are slow over the period of a
particular measurement.  It is directly connected to the active gas
volume via synchronised  fast-acting valves  at the top of the chamber.
Expansions are effected by first achieving a pressure equilibrium  with
the valves open, then closing the valves, reducing the pressure in the
buffer tank by the required (small) amount and finally re-opening the
valves.  The valves are left open during the sensitive time of the 
cloud chamber.  The piston remains in a fixed location throughout. An
important advantage of this method is that it extends the sensitive time
since it provides passive compensation for the pressure rise following
an adiabatic expansion.  The pressure rise occurs due to heating of the
layer of gas near the walls and it causes a temperature rise throughout
the volume by an adiabatic re-compression of the gas.    

\subsubsection{Operational experience with cloud chambers}
\label{sec_operational experience}

Some operational experience of our collaboration with cloud expansion
chambers is summarised below:

\begin{enumerate}

\item \emph{Reproducibility:} The expansion ratio is determined from the
initial and final pressures, since these quantities can be measured
precisely---better than $10^{-4}$ absolute.  Other quantities such as
temperature change and supersaturation are calculated from the pressure
change and the known initial conditions.  In the 38 cm Missouri-Rolla
chamber the pressure change is highly reproducible from pulse to pulse
(a spread of about $2 \cdot 10^{-4}$ for a large pressure change of 90
kPa).

\item \emph{Cycle time:}  The 25 cm Vienna chamber requires about 
 1 hr between expansions to allow time for thermodynamic equilibrium to
be re-established.  The 38 cm Rolla chamber requires about 5 min for
equilibrium to be re-established following a deep expansion producing a
temperature change $\Delta T$ = -40~K, during continual operation. The
cycle time is largely determined by how quickly the walls of the chamber
can be brought back to the operational temperature following an
expansion cycle.

\item \emph{Conditioning time:} The 25 cm Vienna chamber requires 1--2
days conditioning prior to taking data.

\item \emph{Chamber cleaning:} Based on 20 years experience with the
operation of expansion cloud chambers at Missouri-Rolla, the following
technique will clean a cloud chamber sufficiently for the demanding
requirements of homogeneous nucleation measurements:

\begin{itemize}

\item  The chamber is first  fogged with liberal quantities of pure high
quality water to dilute and remove water soluble components,.  A fog
nozzle sprays water drops over the chamber so that all surfaces are
wetted and water runs down the walls. (In the CLOUD chamber the water
will be injected through the sampling probe tubes.) After the
application of about 4 litres, the water at the bottom of the chamber is
removed with a tube suction device.  After about 3 applications, a
vacuum is applied to remove residual water.

\item Acetone is then applied and is very effective at removing the last
traces of water (particularly any water that has a surfactant preventing
evaporation) as well as traces of any hydrocarbons etc., for which it is
an excellent solvent.  The acetone is removed from the bottom of the
chamber and the remaining (high vapour pressure) acetone is removed by
vacuum evaporation.

\item Finally, the chamber is warmed under vacuum to remove the
residual  volatile contamination. 

\end{itemize}

\end{enumerate}

\subsubsection{Piston and hydraulic system}
\label{sec_hydraulic_system}

The piston expansion system comprises two main parts: the main piston
and connecting rod, and the hydraulic  system
(Fig.\,\ref{fig_expansion_systems}a).  The latter is based on the same
design as was used to control the 2~m-diameter piston of 
BEBC~\cite{herve1}.

The main piston head is made of a stainless steel envelope containing an
inner stiffening structure. The piston has several grooves for holding
spring-loaded PTFE seals which provide the gas seal with the walls of
the cylinder. The upper surface of the piston is   light-absorbing to
reduce reflections from the top-window illumination
(Section~\ref{sec_ccd}).  The outer region of the upper surface of the
piston is shaped to provide a film of water for establishing 100\%
relative humidity in the active chamber volume.  The inner structure of
the piston is a sandwich assembly made of honeycomb and resin polymer
plates reinforced with carbon fibre. This provides a good structural
rigidity while maintaining a low mass.  A low mass is important since
the required acceleration rates are large (up to 30 g) to achieve a fast
expansion time for piston strokes up to the maximum of 200 mm. The main
piston rod is made of stainless steel.  The upper end of the rod is
linked to the piston through a kernel which is dismountable for assembly
and maintenance purposes. The lower end of the rod is fitted with a
hydraulic piston which is driven by the servo-mechanism.

\begin{figure}[tbp]
  \begin{center}
      \makebox{\epsfig{file=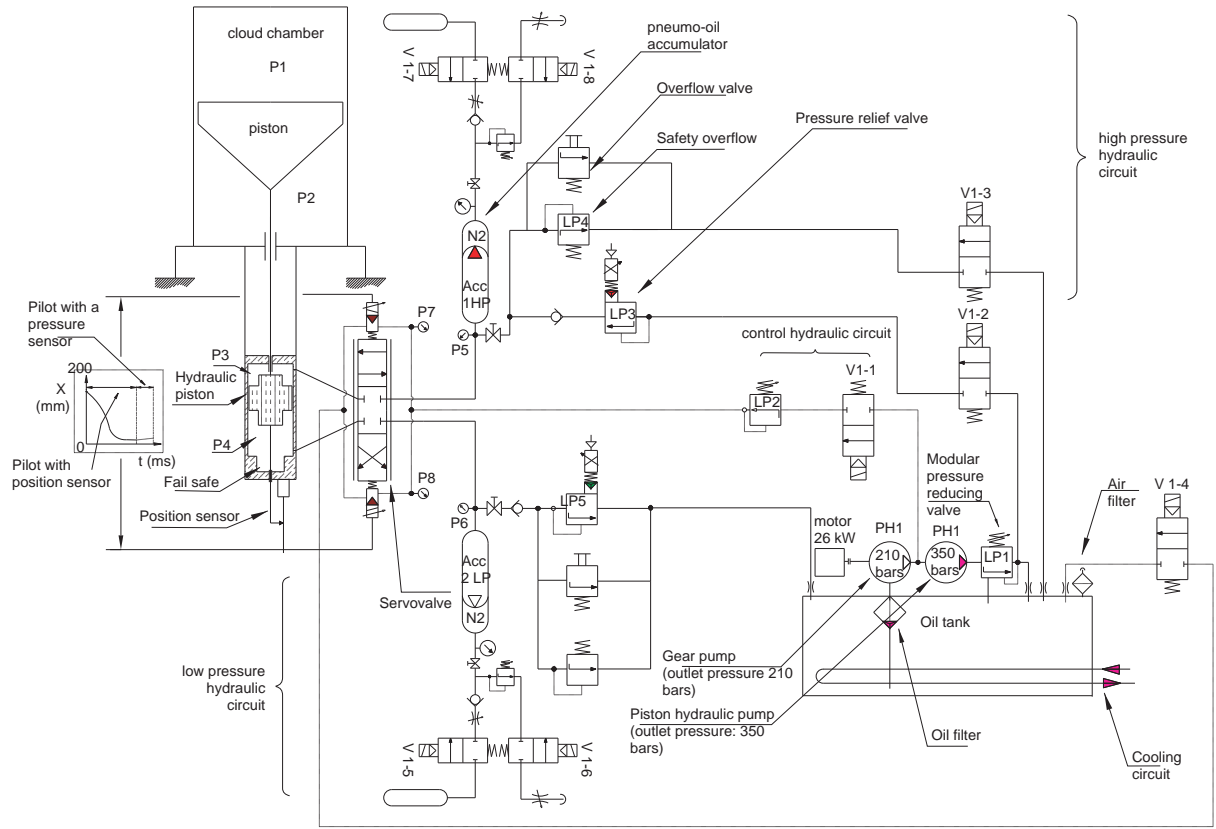,width=160mm}}
  \end{center}
  \vspace{-5mm}
  \caption{Schematic diagram of the hydraulic system for piston
expansions and re-compressions.}
  \label{fig_hydraulic_schematic}    
\end{figure}

 The servo-mechanism (Fig.\,\ref{fig_hydraulic_schematic}) is made of
several main components: a servovalve, a hydraulic pump, and high
pressure and low pressure accumulators \cite{minginette}.   There are
three circuits containing liquid hydraulic oil: high pressure circuit
(220 bar), low pressure circuit (20 bar), and the control circuit 
(running at a nominal pressure of 160 bar). All circuits are supplied by
a hydraulic pump delivering an outlet pressure of 350 bar and a nominal
flow rate of 20 l/min for a rotation speed of 1500 rpm. The power
consumption is about 26 kW. The high pressure circuit feeds the high
pressure accumulator (10 l capacity), which serves as an energy tank for
the expansion. A low pressure accumulator (which is normally not present
in these systems) is also included to avoid any shocks in the circuits
or back effects at the end of the piston travel. Each of the circuits is
equipped with pressure relief valves, with recuperation into a 100 l
tank, and flow rate regulating valves.

\begin{figure}[tbp]
 \vspace{-10mm}
  \begin{center}
      \makebox{\epsfig{file=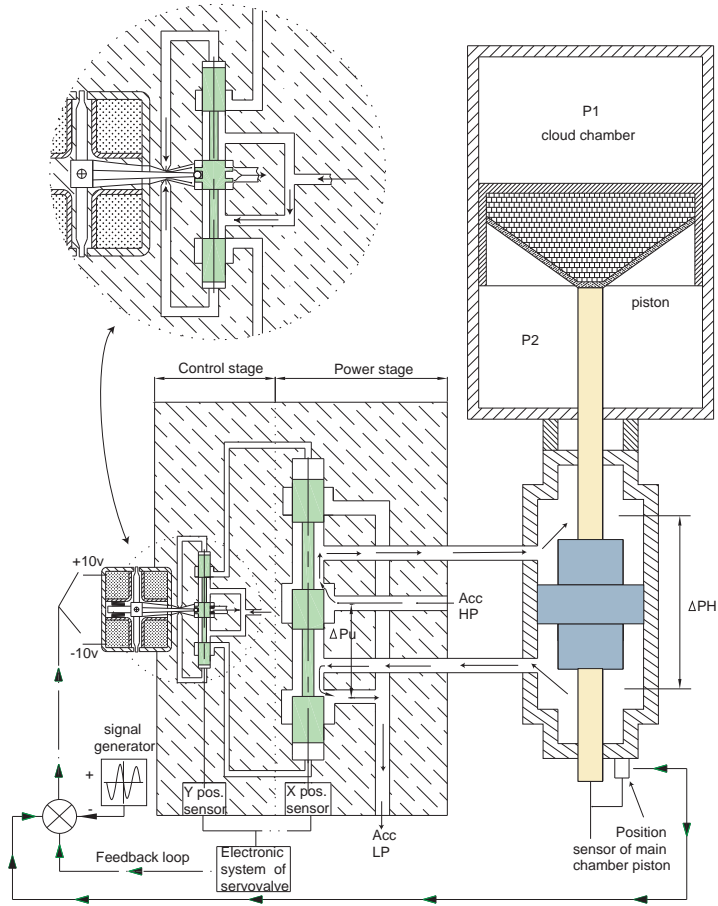,height=120mm}}
  \end{center}
   \vspace{-5mm}
  \caption{Principle of the servovalve for controlling the piston
expansions and re-compressions.}
  \label{fig_servovalve}    
\end{figure}

The control circuit drives the most critical component of the hydraulic
system: the servovalve (Fig.\,\ref{fig_servovalve}).  Depending on the
electrical signal ($\pm$10 V) delivered by a computer, the spool slides
inside the bushing of the control stage. The power stage spool is then
moved, and the inlet/outlet connect the expansion chamber of the
hydraulic piston with the high and low pressure accumulators. Any
movement of the latter has then a direct effect on the main piston rod.
The nominal control flow rate of the servovalve (control stage) is 28
l/min, and the opening time is about 6 ms. The power flow rate through
the servovalve can reach 1500 l/min, which leads to an expansion time of
about 150 ms for the maximum stroke of 200 mm.  Shorter piston strokes
are more rapid; for example, the expansion time is below 10 ms for a 5
mm expansion (which produces a supersaturation of about 6\%). The
limiting parameter is the ability of the main piston head to withstand 
acceleration.  The hydraulic system provides complete flexibility for
the choice of expansion and re-compression cycles, which are determined
by the analogue voltage pulse shape delivered by the control computer. 
The system also delivers an extremely precise ($\pm$5~$\mu$m) and
reproducible (better than
$\pm$5~$\mu$m) movement of the piston. 

The hydraulic piston (100 mm diameter) has no seal and its fit within
the cylinder is of prime importance. A prototype of the main piston,
hydraulic piston and servovalve system is required in order to validate
the study and fix the parameters for the final installation, such as the
pressure loss in the circuits and the leaks to be compensated. The
maximal allowable acceleration of the piston is another important  issue
that requires evaluation with the prototype.

\subsubsection{Liquid cooling system} \label{sec_cooling_system}

\begin{figure}[tbp]
  \begin{center}
      \makebox{\epsfig{file=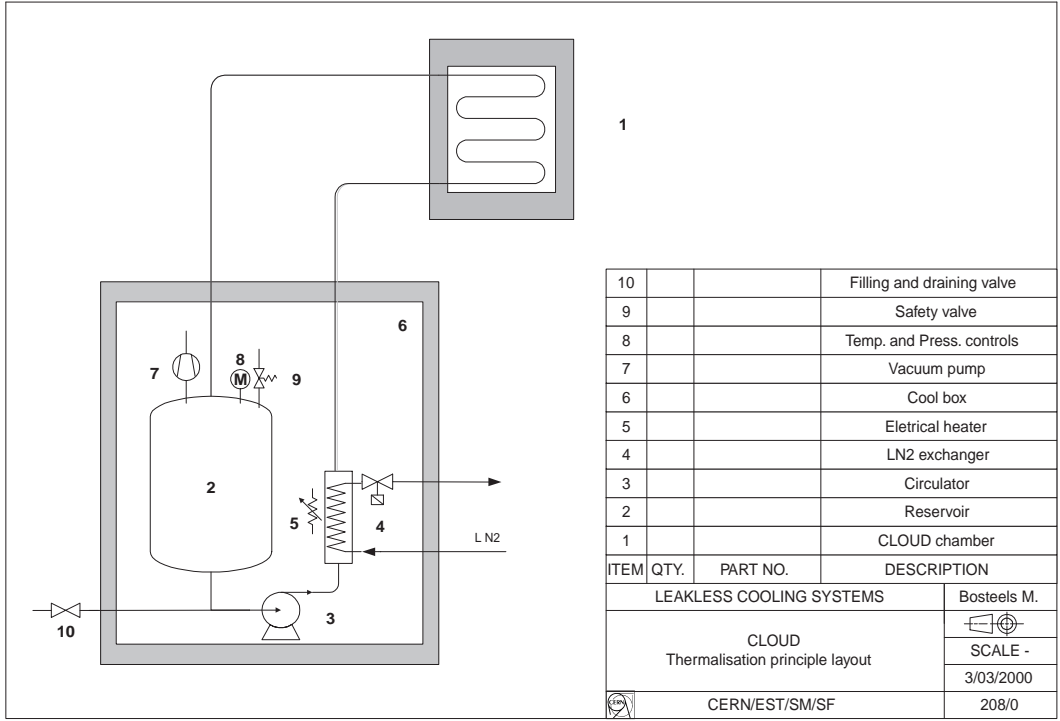,width=140mm}}
  \end{center}
  \caption{The liquid cooling and temperature control system.}
  \label{fig_cooling_system}    
\end{figure}

The liquid cooling and temperature control system for CLOUD  
(Fig.\,\ref{fig_cooling_system}) involves a closed circuit system,
insulated throughout by vacuum. The liquid coolant flows through a
jacket surrounding the cloud chamber and maintains the inner wall and
piston at a precisely-controlled temperature.  The gas inside the cloud
chamber is allowed to reach thermal equilibrium with the walls before
taking any measurements.   A fluorocarbon liquid is used, such as FC87
(C$_5$F$_{12}$), which is liquid in the range 163~K--303~K at
atmospheric  pressure.  The fluorocarbon is transparent to visible and
UV wavelengths down to about 180 nm.  Consequently the liquid can
circulate between the windows to obtain the optimum thermal jacket,
without compromising the optical measurements or the option of UV
irradiation of the active volume.

A coolant reservoir, circulation pump and heat exchanger are placed in a
cold box.  The temperature of the circulating fluorocarbon liquid is
adjustable in the range from 173~K--293~K.  Temperatures are measured
with Pt1000 platinum resistance thermometers which provide measurements
stable to 0.001~K.  The fluorocarbon is cooled by liquid nitrogen
evaporating at a constant pressure, and re-heated by an electrical
heater.  This allows fine temperature control and higher stability than
can be obtained with a refrigerator unit. Our experience of similar
liquid cooling systems for particle physics detectors shows that
temperature stabilities of 0.1~K can be achieved for relatively large,
non-insulated detectors involving internal heat sources and located in
experimental halls without air-conditioning.  In comparison, the CLOUD
system is relatively compact, insulated throughout by a vacuum layer,
and without any internal parasitic heat loads.  This indicates that,
with careful design, a temperature stability of 0.01~K can be achieved.

\subsubsection{Field cage} \label{sec_field_cage}

The walls of the cloud chamber are equipped with a field cage which
provides an electric field inside the active volume.   The functions of
the electric field are as follows:
\begin{itemize}
\item To clear ions and charged aerosols from the active volume.
\item To allow measurements at \emph{below} atmospheric ion
concentrations.  The typical ion concentration at the top of the
troposphere is a few 1000 cm$^{-3}$ 
(Section~\ref{sec_atmospheric_ions}); the clearing field can lower the
small ion concentration in the cloud chamber to near zero in about 2 s
(Section~\ref{sec_electric_field}).
\item To separate positive and negative ions and charged aerosols in
order to measure their properties separately.  In addition, this will
allow a technique for slowing charge neutralisation by +/- ion
recombination.  
\item To allow an approximate measurement of the aerosol size/charge
distributions by measurement of the electrical mobility before
activation. This is done by allowing the ions and charged aerosols to
drift in a uniform field for a known time before the expansion pulse. 
The resulting droplets effectively freeze the final locations of the
charged particles and allow their drift paths to be estimated and hence
also their electrical mobilities.  This provides a technique to check
for consistency with the more precise mobility measurements provided by
external detectors. 
\end{itemize}

\begin{figure*}[tbp]
  \begin{center}
      \makebox{\epsfig{file=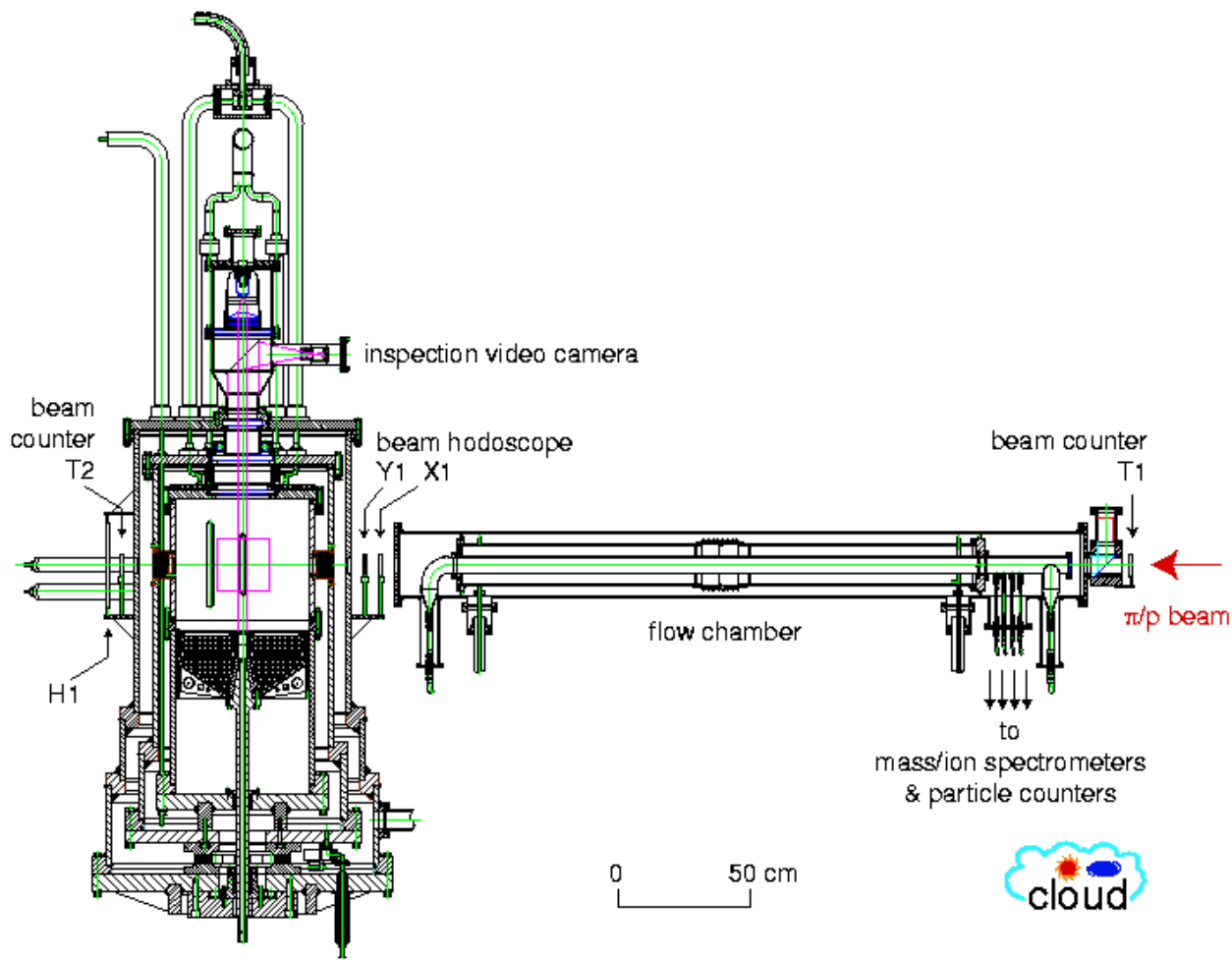,width=165mm}}
  \end{center}
\vspace{-5mm}
  \caption{Vertical section showing the flow chamber and beam counter
system.}
  \label{fig_flow_chamber}    
\end{figure*}

The field cage involves circular electrodes laid down on an  800
$\mu$m-thick ceramic insulating layer which lines the inside of the
cloud chamber. The electrodes are continued across the chamber windows
by means of thin, vacuum-deposited traces; these result in a negligible
loss of light intensity for the optical readout systems.   A layer of
electrodes at the top of the cylinder completes the field cage. Except
in the region of the windows, the electrodes are covered with a layer of
black teflon, which is in contact with the gas in the active volume. The
teflon has a small conductivity in order to avoid charge buildup and
consequent field distortion. The cloud chamber walls and piston are
grounded, and the field wire potentials are appropriately set relative
to this ground, up to a maximum of 500~V.   The performance of the field
cage is summarised in Section~\ref{sec_electric_field}.

\subsubsection{Flow chamber} \label{sec_flow_chamber}

A flow chamber of dimensions  2 m length $\times$ 6cm-diameter is
integrated with the cloud chamber assembly
(Fig.\,\ref{fig_flow_chamber}).  It is filled with the same gas/aerosol
mixture as the cloud chamber, and exposed to the same particle beam and,
where required, to the same UV irradiation. Furthermore it is operated
at the same temperature and pressure, but without the need for the same
high precision on temperature and pressure stability.  The exhaust gas
from the flow chamber is analysed using condensation particle counters
(CPC), differential mobility particle sizers (DMPS), mass spectrometers
and ion mobility detectors (see Fig.\,\ref {fig_gas_schematic} and
Sections\,\ref{sec_gas_and_aerosol_systems} and
\ref{sec_gas_ion_analysis}).

\subsection{Optical readout} \label{sec_optical_readout}

\subsubsection{Constant angle Mie scattering (CAMS) detector}
\label{sec_cams}

\begin{figure*}[htbp]
  \begin{center}
      \makebox{\epsfig{file=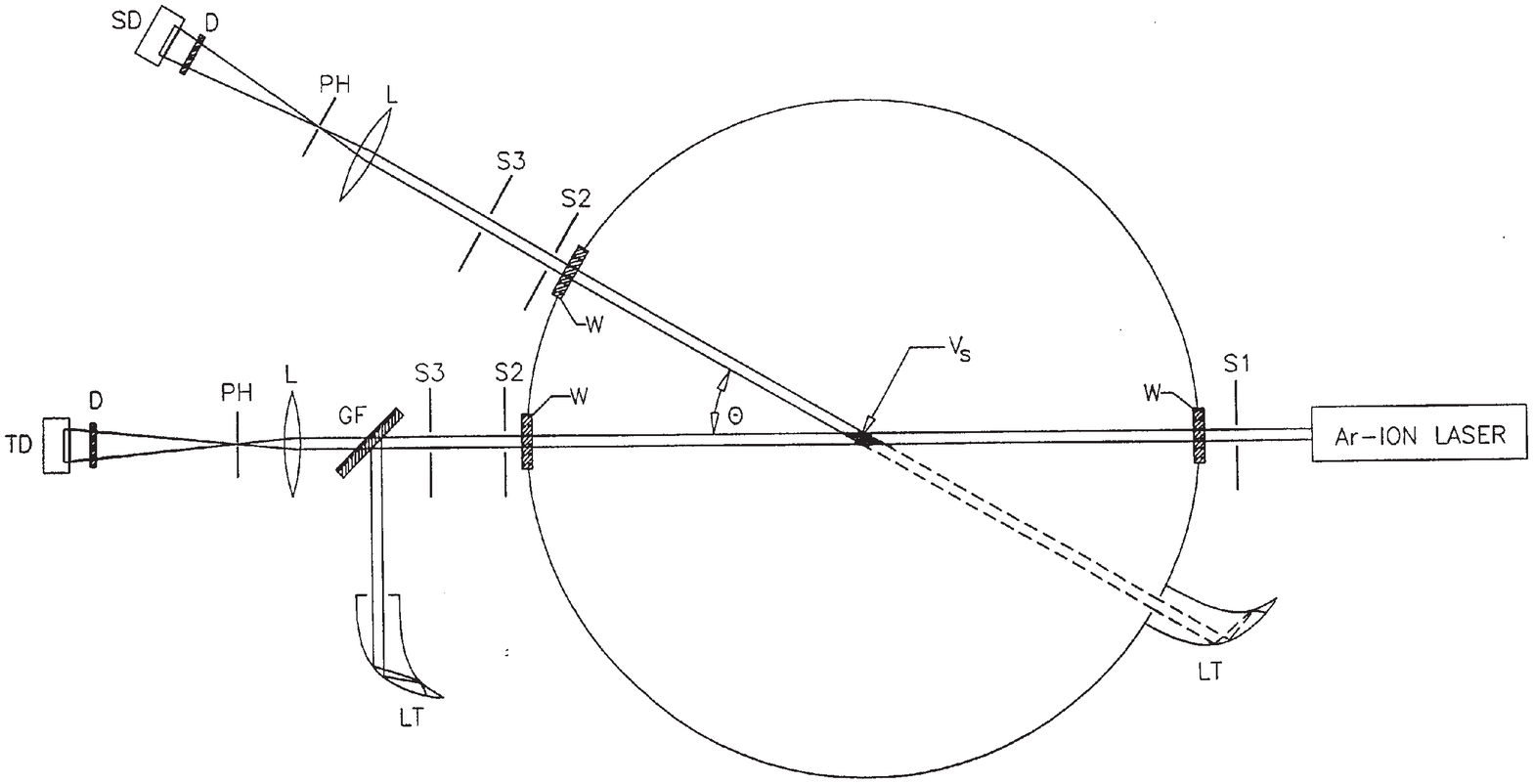,width=120mm}}
  \end{center}
  \vspace{-10mm}
  \caption{Schematic diagram showing the experimental arrangement for
the CAMS detection method.}
  \label{fig_cams_layout}   
\end{figure*}

The CAMS detection method allows simultaneous  measurement of number
concentration and size of growing droplets. This method has been
successfully applied in various experimental studies of nucleation and
condensation processes \cite{strey,rudolf,viisanen}. A detailed
description of theoretical and experimental aspects of the CAMS
detection method can be found elsewhere \cite{wagner, szymanski90}. 
The general schematic layout of a CAMS system is show in
Fig.\,\ref{fig_cams_layout} and the geometry for CLOUD is shown in
Fig.\,\ref{fig_cloud_detector_h}. In the following the main features of
CAMS are presented as they are relevant for the proposed research
project.

The CAMS detection method is based on light scattering. Growing
spherical particles are illuminated by a monochromatic, parallel light
beam, e.g.\,a laser beam (Fig.\,\ref{fig_cams_layout}). The light flux
$\Phi_{trans}$ transmitted through the measuring chamber is monitored
during the particle growth process by means of photodetector TD.
Simultaneously, the light flux $\Phi_{sca}$ scattered at a selectable
constant scattering angle $\Theta$ is monitored by means of
photodetector SD. The observation cones of the detectors are defined by
corresponding lens-pinhole arrangements L-PH. The actual sensitive
volume $V_S$ inside the measuring chamber is determined by the
intersection of laser beam and observation cone of detector SD for
scattered light.
 
\begin{figure*}[tbp]
  \begin{center}
      \makebox{\epsfig{file=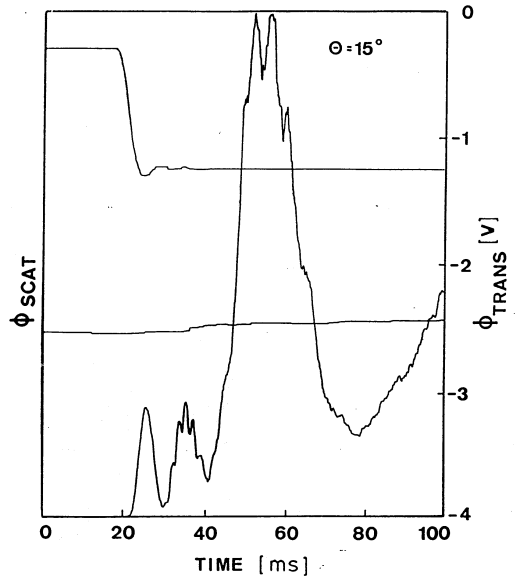,height=80mm}}
  \end{center}
  \vspace{-5mm}
  \caption{Result of a typical single CAMS experiments at Vienna for a
forward scattering angle $\Theta=15^\circ$.  The uppermost curve shows
the pressure step in the expansion cloud chamber.  The middle curve
shows the (inverted) transmitted light flux, and the lowest curve shows
the scattered light flux.}
  \label{fig_cams_data_1}   
\end{figure*}

Our measurements from a typical CAMS experiment are shown in
Fig.\,\ref{fig_cams_data_1}. Here simultaneous growth of liquid drops in
supersaturated vapour was observed. The vapour supersaturation was
achieved by means of an adiabatic pressure jump in an initially
saturated system. This pressure step is indicated by the upper curve in
Fig.\,\ref{fig_cams_data_1}. The middle curve in
Fig.\,\ref{fig_cams_data_1} shows (inverted) the transmitted light flux
$\Phi_{trans}$ as a function of time during the droplet growth process.
The lowest curve shows the scattered light flux $\Phi_{sca}$ as a
function  of time. In the experiment considered a forward scattering
angle $\Theta=15^\circ$ was chosen, the incident laser beam was linearly
polarized perpendicular to the plane of observation.

As can be seen from Fig.\,\ref{fig_cams_data_1}, the experimental
scattered light flux vs.\,time curve shows a series of maxima and
minima. These extrema are connected to light diffraction and can be
uniquely identified, as will be shown below. Thereby the particle size
can be determined at each of the experimental extrema. In the absence of
multiple light scattering the experimental scattered light flux will be
proportional to the number of particles inside the scattering volume.
Accordingly, the droplet number concentration can be evaluated from the
height of the experimental light scattering maxima independent of the
determination of particle  size.

The middle curve in Fig.\,\ref{fig_cams_data_1} (inverted in the figure)
shows that the transmitted light flux
$\Phi_{trans}$ decreases significantly during the droplet growth
process. Thus it can be concluded that considerable light extinction can
occur inside the measuring chamber, particularly at later stages of the
particle growth process. This light extinction will be increasingly
pronounced for increasing particle number concentrations. Obviously the
experimental scattered light flux will be reduced due to light
extinction in the measuring chamber as well. Thereby the values of the
particle number concentrations, as determined directly from the height
of the experimental light scattering maxima, will generally be somewhat
smaller than the actual concentrations. A correction for this light
extinction effect would require the knowledge of the actual particle
number concentration. Fortunately, the reduction of the scattered light
flux due to light extinction can be taken into account in the following
way. As seen from Fig.\,\ref{fig_cams_data_1}, the scattered light flux
$\Phi_{sca}$ will suffer approximately the same light extinction as the
transmitted light flux $\Phi_{trans}$, if the scattering volume
$V_S$ is restricted to the central part of the measuring chamber. Thus
the effect of light extinction on the scattered light flux can be
compensated by normalising the scattered relative to the transmitted
light flux. Accordingly, a linear relation between the {\em normalised}
scattered light flux $\Phi_{sca}/\Phi_{trans}$ and the particle number
concentration can be expected over a wide concentration range.

\begin{figure}[ht]
  \setlength{\unitlength}{1mm}
\begin{minipage}[t]{70mm}  
\noindent
  \begin{center}
      \makebox{\epsfig{file=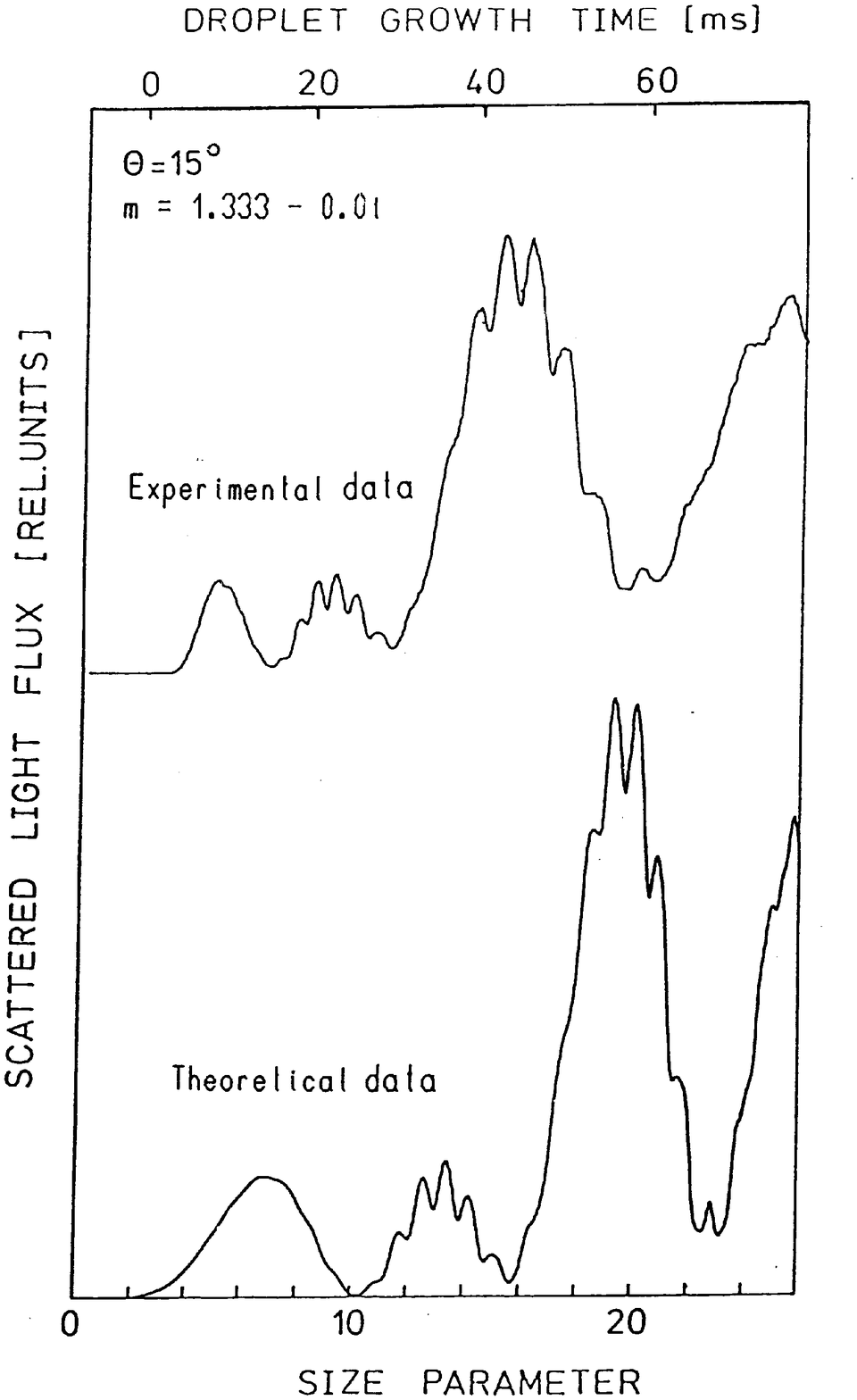,height=102mm}}
  \end{center}
   \vspace*{-5mm}
  \caption{Comparison of experimental scattered light flux vs.\,time
curves to theoretical scattered light flux vs.\,size curves for forward
scattering angle $\Theta=15^\circ$.}
  \label{fig_cams_data_2}    
\end{minipage}
\hfill
\begin{minipage}[t]{70mm}  
\noindent
  \begin{center}
      \makebox{\epsfig{file=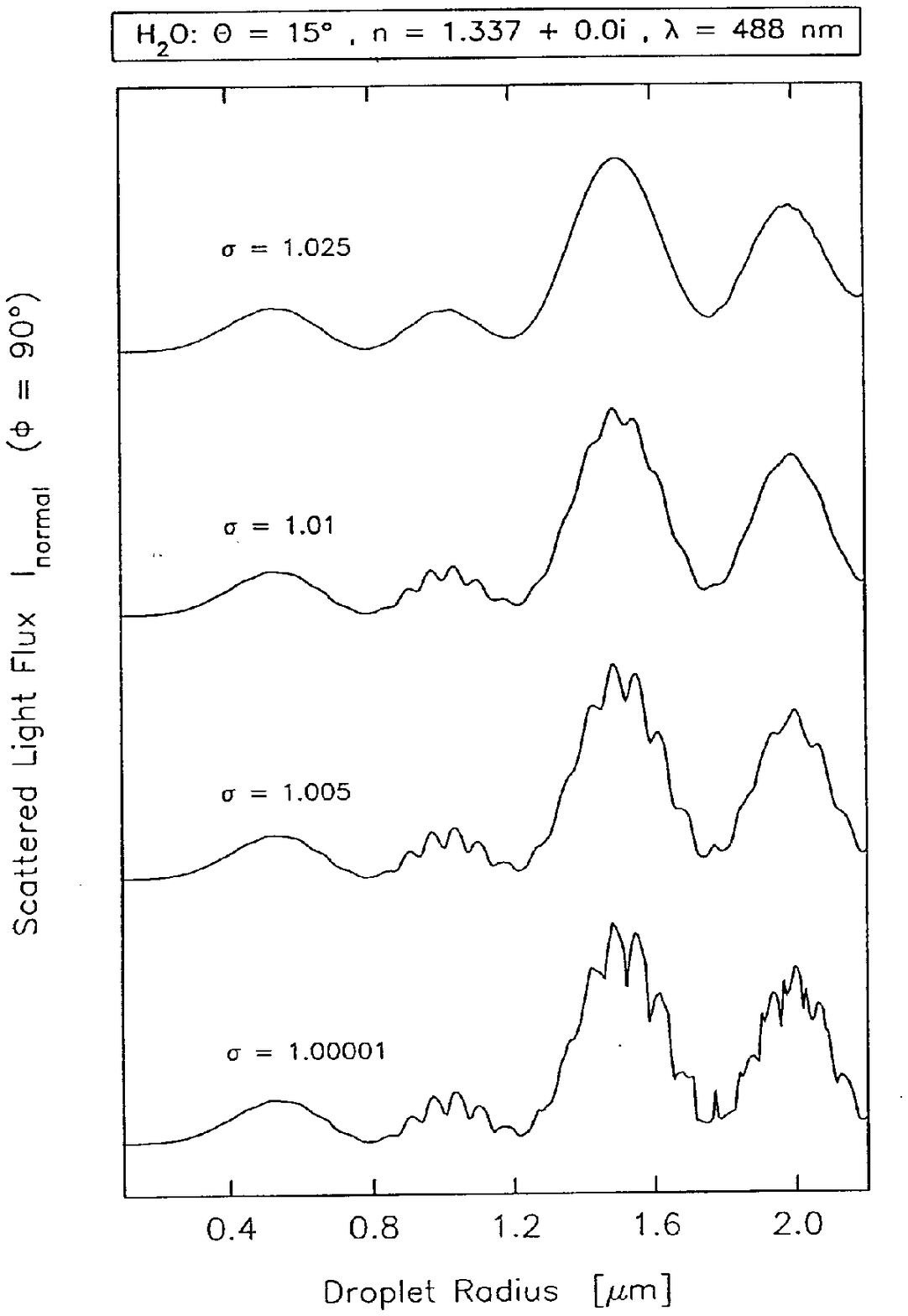,height=102mm}}
  \end{center}
   \vspace*{-5mm}
  \caption{Influence of drop size distribution on scattered light flux
for forward scattering angle $\Theta=15^\circ$.  The curves are
labelled according to the relative spread, $\sigma$, of the droplet
radii.}
  \label{fig_cams_data_3}  
\end{minipage}
\end{figure}

In order to achieve a unique quantitative interpretation of the light
scattering curves as obtained from CAMS experiments, it is important to
calculate the normalised scattered light flux
$\Phi_{sca}/\Phi_{trans}$ as a function of droplet size for various
constant scattering angles. The theory of light scattering by spherical
particles has been derived by Mie \cite{mie} and Debye
\cite{debye}; a recent treatise by Bohren and Huffmann
\cite{bohren} includes efficient numerical methods for evaluation of the
relevant light scattering functions. Fig.\,\ref{fig_cams_data_2} shows
a comparison of experimental scattered light flux vs.\,time curves
with  theoretical scattered light flux vs.\,size curves for forward
scattering angle $\Theta=15^\circ$. Satisfactory agreement between
experimental and theoretical data can be observed allowing to establish
a one-to-one correspondence of experimental and theoretical light
scattering extrema.

The morphology of the light scattering curves depends on the size
distribution of the growing droplets. As can be seen from
Fig.\,\ref{fig_cams_data_3}, the resonant ripple structure of the light
scattering curves tends to disappear with increasing width of the drop
size distribution. Thus CAMS measurements require sufficiently narrow
drop size distributions to identify unambiguously a single peak in
isolation.  However it can be seen that the first scattering maximum at
the forward scattering angle
$\Theta=15^\circ$ is not very sensitive to changes of the width of the
drop size distribution.

When a unique correspondence between experimental and theoretical light
scattering extrema is established, the particle sizes at specific times
during the growth process can be obtained from the positions of the
extrema of the experimental normalised light scattering curves relative
to the time axis. Independently, the droplet number concentrations at
specific times during the growth process can be determined from  the
heights of the maxima of the experimental normalised light scattering
curve by comparison to the corresponding theoretical curve. In this
connection it is important to note that, as mentioned above, the
normalised scattered light flux $\Phi_{sca}/\Phi_{trans}$ is linearly
related to the droplet number concentration. Determination of the
droplet number concentration can thus be performed by measurement of the
ratio $\Phi_{sca}/\Phi_{trans}$ of scattered and transmitted light
flux. To this end it is necessary to perform a mutual calibration of the
detectors TD and SD for transmitted and scattered light fluxes,
respectively. This calibration can be avoided by quantitatively
considering the light extinction inside the measuring chamber. From
measurements of the light extinction at certain droplet sizes the
droplet number concentration can be determined independently.

The choice of appropriate scattering angles depends on the actual
measurements to be performed. For determination of particle
concentrations it is advantageous to use comparatively small forward
scattering angles, because the corresponding light scattering curves
exhibit a rather simple structure with few broad maxima, which can
easily be identified even for comparatively broad droplet size
distributions. Furthermore, comparatively high scattered light fluxes
occur, allowing accurate measurements of the heights of the observed
light scattering maxima. For measurements of drop growth rates, however,
somewhat larger scattering angles are advantageous, because the
corresponding light scattering curves show a somewhat more complex
structure, providing detailed information on drop growth.

In summary the CAMS detection method is applicable to the observation of
growing spherical particles (droplets) with known refractive index. The
droplet size distribution needs to be sufficiently narrow so that
distinct extrema can be observed in the experimental light scattering
curves. Measurements over a comparatively wide range of droplet number
concentrations can be performed. The lower limit of the concentration
measuring range depends on the actual sensitive volume $V_S$ as
determined by the incident laser beam and the observation cone of the
detector SD for scattered light. In order to allow a unique evaluation
of the experimental scattered light curves, generally more than about 5
particles must be simultaneously present inside the sensitive volume
during particle growth. The upper limit of the particle concentration
range will depend primarily on the amount of multiple scattering inside
the measuring chamber \cite{szymanski90}. As shown recently
\cite{filipovicova}, CAMS measurements show a linear concentration
response over a range of concentrations up to as much as several
$10^7$ particles per $\mbox{cm}^3$. Finally it should be noted that the
CAMS detection method is a non-invasive method for absolute measurement
of drop size and number concentration. No empirical calibration
referring to external reference standards is required.

\subsubsection{CCD cameras and optics} \label{sec_ccd}

\paragraph{CCD cameras and readout:} 

The Rutherford Appleton Laboratory has designed and built a very fast,
twin channel CCD imaging system for the August 1999 total solar eclipse,
which is available for CLOUD. The SECIS (Solar Eclipse Coronal  Imaging
System) was successfully used to obtain in excess of 12,000  images
during totality (i.e.\,in a period of less than 3 minutes) from a site 
close to the Black Sea in Bulgaria. 

In essence, SECIS is an extremely fast, high-resolution imaging system.
It is  a considerable advance over previous systems in both frame rate
and photometric accuracy. To meet the requirements of the solar eclipse
observation,  consecutive large area images needed to be stored at a
very fast rate that  was not until recently possible in any existing
system but now is  technically feasible using new developments in CCD
cameras and memory  storage units.  Currently, SECIS uses two small-area
CCDs to obtain fast images and large on-line processing arrays to handle
the image data.   The computer system, interfaces and camera power
supplies fit in a  medium-sized, portable PC enclosure. Extra cooling is
built into the  enclosure to allow operation in hotter environments than
normal. An  uninterruptible power supply provides power for the whole
system.  Each frame of data is time-tagged for subsequent individual
access from  the PC.

The solar eclipse experiment required the capability of storing a large
number (several  thousand) consecutive frames of data without
interruption. The two CCD  cameras used are state-of-the-art digital
cameras. Each has a $512 \times 512$  format with $15 \times 15$
$\mu$m$^2$ pixels. The data is digitised to 12 bits and clocked  out at
a frame rate of 55 Hz (which is well above the 10 Hz rate required for
CLOUD).  These images are ``grabbed'' by a specially-adapted PC, with
dual  Pentium processors.  The PC has two PCI cards which provide the
two  camera interfaces; the cards  are linked to ensure that there is no
variation in timing between the two cameras. Exposure time and frequency
are derived from programmable  crystal-based  oscillators. The operating
system is Microsoft Windows NT Workstation.  The computer system
captures two synchronised video streams from the two  CCD cameras. It 
then reconstitutes the video images and stores  them on a computer
network for detailed off-line analysis.  The data is buffered and sent
to the computer via differential parallel  cables.  These also carry
camera control signals generated by the computer  to start and stop the
light collection on the CCDs and to trigger the  data  transfer. The
exposure time and frequency can therefore be selected by the operator.

The video data is packed by the PCI cards and temporarily stored in two 
buffer sets in main memory. As the buffers become full, the data is 
transferred to four SCSI disk drives, each 9 Gb in size. This is
sufficient for about 36 min continuous operation at the full rate of
1~Gbyte/min (50 Hz).  Each image is divided into 4 and each quarter
image is stored on a separate hard drive. For data analysis, the reverse
process takes the quarter images from each hard drive in the correct
sequence and rebuilds the original full  frame. These frames are then
written out to the network disk as FITS images for compatibility with
many different software packages. During normal operations, the 
workload  is shared by the two processors. Two small live images can be
displayed  together  with the buffer-filling status to allow the
computer operator to monitor the  performance of the system. 

This system currently exists and is available for the CLOUD experiment.
However, we are planning to upgrade the cameras to use newer CCD chips
with a smaller pixel size and more pixels per chip.

\paragraph{Optics for the cloud chamber:} 

For the performance figures presented in this section, we will assume
that each CCD camera has an upgraded chip of
$1024 \;\mathrm{(h)} \times 2048 \;\mathrm{(v)}$ pixels, with a pixel
size of $8 \times 8$
$\mu$m$^2$ and an active area of $8.2 \;\mathrm{(h)} \times 16.2
\;\mathrm{(v)}$~mm$^2$. 
 Since many CCD chips are already on the market with higher
specifications,\footnote{An example is the Kodak KAF-6303 which has
$3072 \times 2048$ pixels, with a pixel size of 
 $9 \times 9$ $\mu$m$^2$ and an active area of 
 $27.6 \times 18.5$~mm$^2$.} the figures presented here should be
considered as conservative.

\begin{figure}[htbp]
  \begin{center}
      \makebox{\epsfig{file=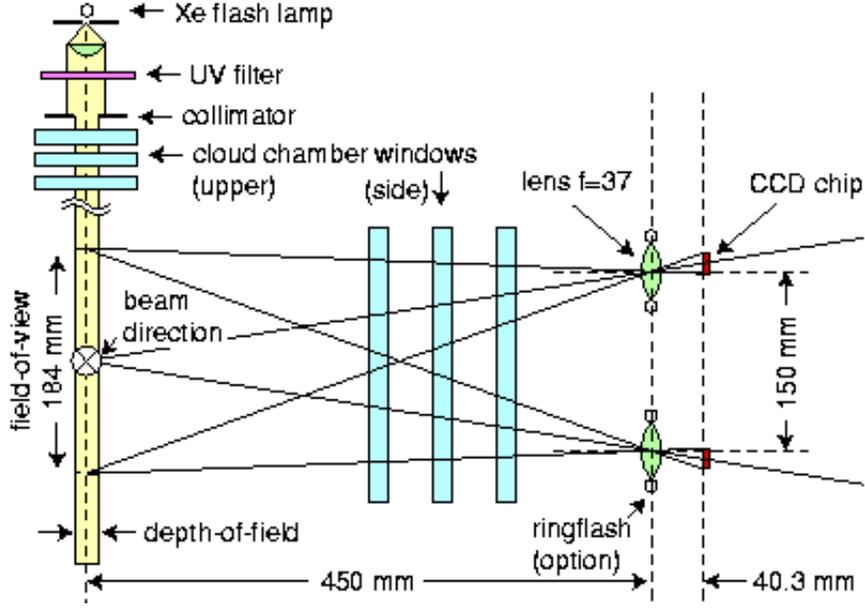,height=80mm}}
  \end{center}
  \vspace{-5mm}
  \caption{CCD camera optics and xenon flash illumination for the wide
field-of-view.}
  \label{fig_ccd_optics}    
\end{figure}

The CCD cameras are arranged in a stereo pair as indicated in
Fig.\,\ref{fig_ccd_optics}. The optics are adjustable between a wide
field-of-view and a narrow one. The former can be used to simultaneously
measure regions of the chamber exposed to beam and not exposed. This
requires a wide illumination field, which is provided by a xenon flash
lamps mounted at the top window.  This produces dark-field-illumination,
i.e bright droplet images on a dark background.   The flashlamps 
provide 100 J/pulse at up to 10 Hz; the power source is a 100~$\mu$F
capacitor-bank charged to 1500 V.  Experience with the 38 cm
Missouri-Rolla chamber has shown, for xenon flash illumination, the
scattered light intensity from a droplet is directly proportional to its
radius in the range 2--20~$\mu$m. (This was the limit of the
measurement; the proportionality probably continues either side of this
range.)  This indicates that the pulse height information from the CCD
will provide measurements of the radii of individual droplets.

\begin{table}[htbp]
  \begin{center}
  \caption{Optical parameters of the CCD camera system. The figures
assume a  CCD chip of 
$1024 \;\mathrm{(h)} \times 2048 \;\mathrm{(v)}$ pixels, with a pixel
size of $8 \times 8$
$\mu$m$^2$ and an active area of $8.2 \;\mathrm{(h)} \times 16.2
\;\mathrm{(v)}$~mm$^2$.}  
  \label{tab_ccd_optical_parameters}
  \vspace{5mm}
  \begin{tabular}{| l  l | c | c | c|}
  \hline
  \multicolumn{2}{|c|}{\textbf{Parameter}} &
\multicolumn{2}{|c|}{\textbf{Wide}} & 
 {\textbf{Narrow}}  \\
  \multicolumn{2}{|c|}{} &
\multicolumn{2}{|c|}{\textbf{field-of-view}} & 
 {\textbf{field-of-view}}  \\
 \hline
 \hline
 Lens focal length &  [mm]   & \multicolumn{2}{|c|}{37}   & 200  
\\
 De-magnification (object/image distance)  &    &
\multicolumn{2}{|c|}{11.2}   &  1.25   \\
 Transverse field-of-view in chamber (h $\times$ v) & [mm$^2$]  &
\multicolumn{2}{|c|}{$92 \times 184$}   &
 $10 \times 20$  \\
 \hline
   Lens aperture (f-number)  &   &  \hspace{1.5mm} 3.5 \hspace{1.5mm}  &
5.6 &  5.6
\\
   Lens useful diameter  & [mm]  &  11  & 6.6 &  36
\\
  Depth-of-field in object space (chamber) & [mm]  & 10 & 26 & 0.8 \\
  Diffraction resolution limit of image$^\S$ & [$\mu$m]  & 4.2 & 6.8 &
6.8 \\
  Diffraction resolution limit in object space$^\dagger$ & [$\mu$m]  &
90 & 
  90 & 10 \\
  2-droplet resolution$^\star$ & [$\mu$m]  & 180 & 
  180 & 20 \\
 Maximum droplet number concentration$^\ddagger$ & [cm$^{-3}$]  & 1250 &
500 &
 $2 \times 10^5$ \\[0.5ex]
  \hline
  \end{tabular}
  \end{center}
$^\S$  Full width of diffraction spot,  at wavelength $\lambda$ = 500
nm.
\\
$^\dagger$ Limited by pixel size, not image diffraction.\\
$^\star$ Assuming a centre-to-centre droplet separation of two pixels.
\\
$^\ddagger$ Assuming 10\% pixel occupancy. For the wide field-of-view
optics, we assume the xenon flashlamp is collimated to illuminate the 
depth-of-field region. For the narrow field-of-view optics we assume
CAMS laser beam illumination, with 20 mm of beam path subtended by the
observation cone. 
\end{table}

An alternative illumination scheme is provided by xenon flashlamps along
the viewing direction (``ringflash'' in Fig.\,\ref{fig_ccd_optics}).  
UV-absorbing filters ($\lambda_{min} = 450$ nm) are used to avoid
photo-ionisation reactions inside the chamber. To obtain the best
sensitivity at high droplet number densities,  only the shallow region
corresponding to the depth-of-focus of the CCD cameras  is illuminated
(see Table \ref {tab_ccd_optical_parameters}).  This is achieved by
focusing light from the  upper xenon lamp  with a cylindrical lens and
collimating the beam with slits. 
 The narrow field-of-view optics is optimised for measurements recorded
at the same time as the CAMS detector, where a small volume within the
acceptance of the cameras is illuminated by a narrow laser beam. 

Some representative choices of the optical parameters are indicated in
Table \ref{tab_ccd_optical_parameters}. For all lens apertures larger
than F5.6, the 8 $\mu$m pixel determines the size of a droplet image;
this is larger than the diffraction limit of lens and the physical
droplet size, for diameters  below 10 $\mu$m.  The wide field-of-view
has a transverse size of about 9.2 (h) $\times$ 18.4 (v)~cm$^2$ and a
depth-of-focus of 11--26 mm.  A large field-of-view limits the maximum
droplet number density to values of about $10^3$ cm$^{-3}$, which is
adequate for most studies.  If necessary, slightly higher values can be
achieved by restricting the illumination to a shallower depth-of-field.
In contrast, a narrow field-of-view allows measurements up to much
higher droplet number densities of about  $10^5$ cm$^{-3}$.

\subsection{Gas and aerosol systems}
\label{sec_gas_and_aerosol_systems}

\subsubsection{Overview} \label{sec_gas_system_overview}

\begin{figure}[htbp]
  \begin{center}
      \makebox{\epsfig{file=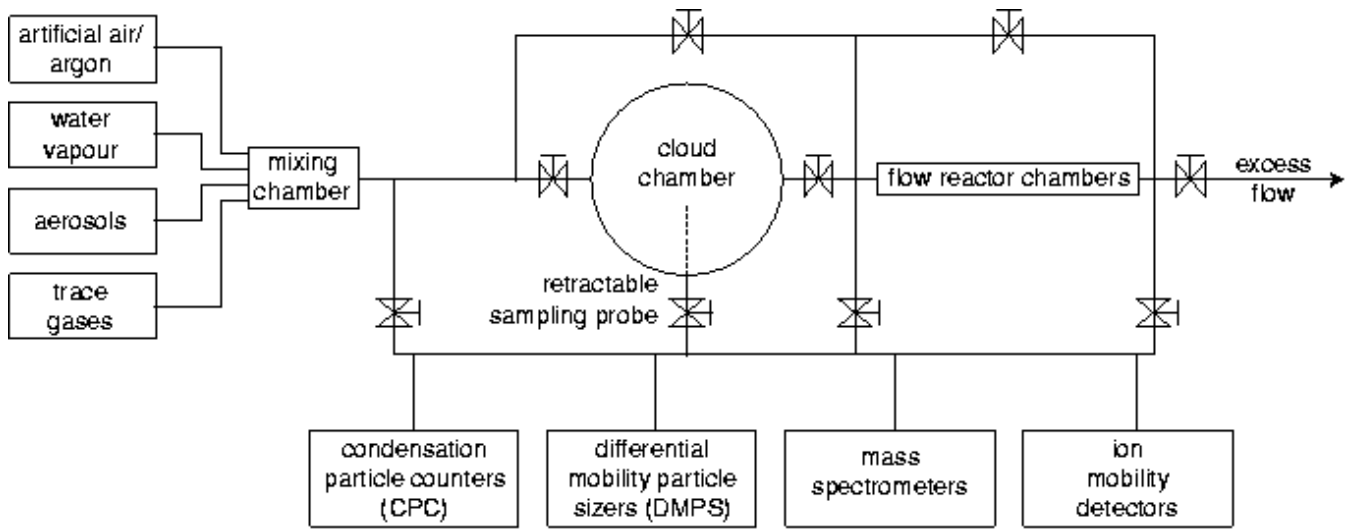,width=160mm}}
  \end{center}
  \caption{Schematic showing the elements of the gas and aerosol supply
systems and the physical and chemical analysis systems. } 
  \label{fig_gas_schematic}    
\end{figure}

A schematic of the gas and aerosol supply and analysis systems is shown
in Fig.\,\ref{fig_gas_schematic}.  The supply system involves four
components: carrier gas, water vapour, aerosols and trace gases.  The
carrier gas is either pure artificial air (80\% N$_2$, 20\% O$_2$) or
argon.  Water vapour, aerosols and trace gases are mixed into this
stream at the desired levels (see Table
\ref{tab_experimental_conditions}).  The water vapour and aerosol
systems are described in more detail below.

\subsubsection{Water vapour system} \label{sec_water_vapour_system}

The water vapour content in the chamber will be set by two techniques:
a) a liquid or ice film on the top of the piston and b) vapour and
carrier gas introduced from an external humidifier.   For most
experiments, the saturation ratio  before expansion will be
 $ p/p_0 = (100.00\pm0.05)\%$,  where $p$ is the partial pressure of the
water vapour and $p_0$ is the saturated vapour pressure over a plane
surface of water (or ice) at this temperature. (The error of 0.05\%
corresponds to a temperature variation of 0.01~K.) The vapour will be
provided by a liquid or ice film covering most of the top of the piston
and maintained at the precise ambient temperature. This design has
several important advantages: a) the amount of liquid needed is low and
thus can be high purity and also be changed easily if contamination of
the liquid film is suspected, b) it allows the introduction of liquids
other than water if needed, and c) the film will cover most of the
piston top but will not touch the walls, thus eliminating problems that
might occur when the liquid film (water) freezes. 

For experiments that require initial saturation ratios  in the range 
$50\% < p/p_0 < 100\%$, an external humidifier will be used.   A
straightforward method to produce a very precise water vapour content
from an external source is to use the saturated vapour and carrier gas
from a liquid pool that is very well temperature regulated, as
indicated above. The vapour and carrier gas can then be accurately
diluted to produce a known (water) vapour concentration.  In order to
avoid any problems due to contamination of the surface of the water with
surfactants, the surface is continually renewed by flowing the water
over many glass rods in a tower structure through which the carrier gas
flows.  

\subsubsection{Aerosol system} \label{sec_aerosol_system}

\paragraph{Particle generation:} Aerosol particle generation for the
project will depend on the particle size required.  To generate
submicron particles for laboratory experiments, two standard techniques
are available.   Nebulizers are used to generate  particles  in the size
range of 10--1000 nm.  In a nebulizer, a small liquid flow is introduced
into a pressurised air flow through  a nozzle. The liquid flow is
distributed into small droplets, which can be made to  evaporate  after
the nozzle by adjusting the fraction of the volatile compound in the
liquid.  The final particle size is set by the fraction of residual
non-volatile component  in each droplet \cite{hinds,willeke}. When a
slightly smaller size  range is needed e.g.\,particle diameter 3--200
nm, a technique based on evaporation and  condensation of either salts
or metals can be used \cite{scheibel}. 

\paragraph{Determination of number concentration of particles:}
Detection of submicron particles is usually carried out by use of
condensation particle  counters (CPC) or faraday cup electrometers
(FCE). The most appropriate instruments are  commercial CPCs by TSI.
One  model is used to determine number concentration of   particles with
diameters exceeding 10 nm \cite{mertes}.  Another instrument, 
especially suitable for ultrafine particles can be used for particles
with diameters   exceeding 3 nm \cite{stolzenburg}. A CPC connected with
a pulse height analyser can also, in certain conditions,  measure the
size distribution in a  very narrow size range of 3--10 nm
\cite{wiedensohler94}.  Smaller sizes can only be measured in the main
cloud chamber. 

\paragraph{Size classification of particles:} The above particle
counters (CPC 3010 and CPC 3010) can be used in combination with a size 
classification method, based on the electrical mobility of particles, to
provide a differential mobility  particle sizer, DMPS
\cite{knutsen,winklmayr}.  Alternatively,  as a more novel technique,
they can be used in a scanning mode as a scanning mobility particle
sizer  (SMPS).  High performance of these instruments requires a well
controlled flow arrangement \cite{jokinen}, determination of size
resolution of the size classification \cite{reischl97} and control of
the artificial electrical charging of particles
\cite{wiedensohler88,reischl96}. The mobility spectrum obtained from 
these measurements can be inverted to obtain the number size
distribution of the particles \cite{kousaka} with a time resolution of
1-10 min.

\paragraph{Transmission efficiency:} In general,  submicron particles
are quite efficiently transported in small  tubes by standard carrier
gas flows. The main mechanisms for transport losses are  deposition onto
tubing walls by diffusion,  gravitational settling of particles  in the
tubing, as well as inertial losses of particles at sharp tube bends.
Diffusion losses are only  significant for particles smaller than
approximately 20 nm \cite{gormley}.  These remaining processes are
usually  significant only for particles sizes exceeding about 100  nm
\cite{hinds,willeke}. For particle size range  of 10-1000 nm, usually no
other processes affect the transmission in flow tubes.  However if high
electrical fields or large thermal  gradients are present in the
tubing,  then electrical precipitation and thermophoretic effects may
become  important.

\subsection{Analysis of trace gases and ions} 
\label{sec_gas_ion_analysis}

\subsubsection{Overview}  \label{sec_gas_ion_overview}

We will analyse the trace chemical species and ions expected from the
flow chamber, as well as any new chemicals that may be formed. Trace gas
analysis will be performed by two mass spectrometers. A third mass
spectrometer (PITMAS) will determine the chemical composition of ions,
and ion mass and mobility spectrometers will detect growth processes,
which are crucial for the observation of ion-assisted particle
production.

\begin{figure}[htbp]
  \begin{center}
      \makebox{\epsfig{file=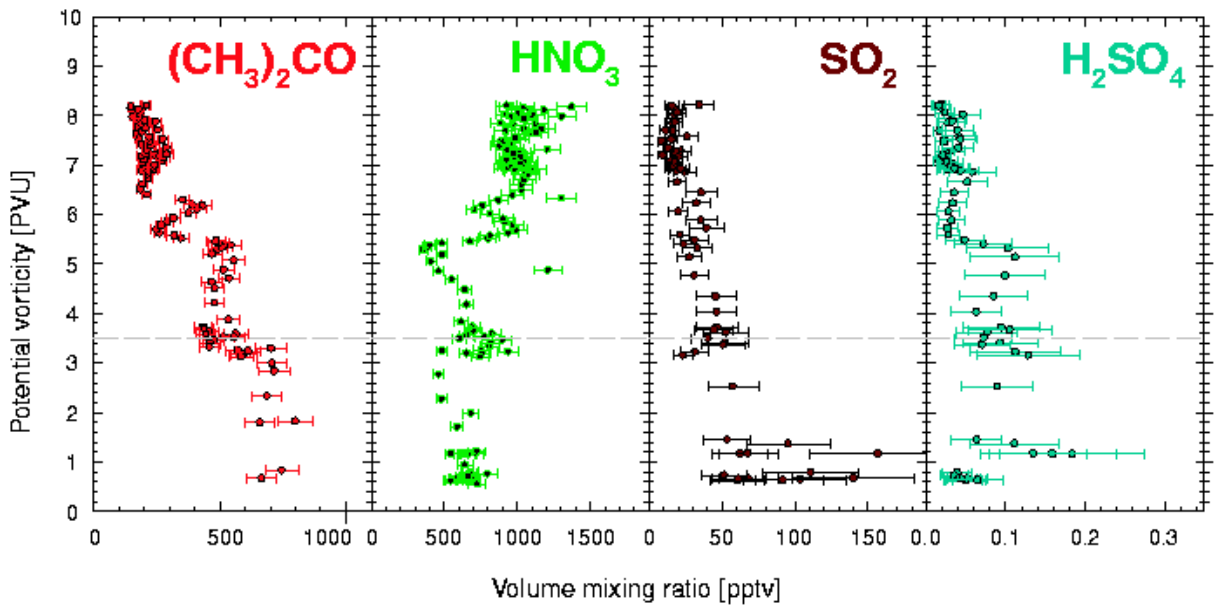,width=160mm}}
  \end{center}
  \caption {Measurements made with an aircraft-based Chemical
Ionisation Mass Spectrometer (CIMS) by MPIK-Heidelberg. The data show
vertical profiles of the mixing ratios of (CH$_3$)$_2$CO (acetone),
HNO$_3$, SO$_2$ and gaseous H$_2$SO$_4$ over central Europe in December
1995. The altitude is indicated as potential vorticity (PV)
\cite{holton}. Typically the tropopause is located around 3.5 PVU.}
  \label{fig_mpik_cims}    
\end{figure}

\subsubsection{Chemical ionisation mass spectrometer}
\label{sec_cims}

The MPIK-Heidelberg Chemical Ionisation Mass Spectrometer (CIMS) 
 measures trace gases with high sensitivity (typical detection limit
$\sim$ 1 pptv\footnote{\emph{pptv} denotes parts per trillion by 
volume.}) and fast time response ($\sim$ 10 s). The CIMS-instrument
consists of an ion flow reactor coupled to an efficiently-pumped linear
quadrupole mass spectrometer. The instruments used have been developed
and successfully employed by the MPIK-Heidelberg group for atmospheric
measurements at ground level as well as on research balloons, rockets,
and aircraft \cite{moehler91}--\cite{wohlfrom99}. A typical measurement
is shown in Fig.\,\ref{fig_mpik_cims}. In a recent development, the
linear quadrupole mass spectrometer has been replaced by a quadrupole
ion trap mass spectrometer (PITMAS) which has been modified for the
application to measurements of trace gases in the atmosphere
\cite{kiendler99}. The high duty cycle which is offered by this
instrument is of advantage for the measurements of trace gases down to
the sub-pptv range. In the CLOUD experiment, CIMS will be used for
measurements of the following trace gases: H$_2$SO$_4$, SO$_2$, HNO$_3$
and NH$_3$. These are potentially important with respect to ion-induced
aerosol formation under tropospheric conditions. Typical volume mixing
ratios of these gases in the ``clean'' upper troposphere are the
following: 0.05--0.5\,pptv (H$_2$SO$_4$), 10-1000\,pptv (SO$_2$), and
100--1000\,pptv (HNO$_3$). For upper tropospheric conditions the
CIMS-detection limits for measurement of these trace gases are currently
estimated to be 0.05 pptv (H$_2$SO$_4$, 1 minute time resolution),
1\,pptv (SO$_2$, 30\,s time resolution), and 50\,pptv (HNO$_3$, 10\,s
time resolution).

\subsubsection{Time-of-flight (ToF) mass spectrometer} 
\label{sec_tof}

The ToF ion mass spectrometer developed by St. Petersburg-Aarhus uses
pulsed electron impact ionisation of a gas sample, extraction by an
electric field of the produced ions, and subsequent drift of the ions to
a detector that counts the ions as a function of the time of arrival. 
The instrument will be used to measure changes of the chemical
composition of the gas mixture from the flow reactor chambers or the
cloud chamber (extracted with the sampling probes).  The main advantage
of ToF mass/charge analysis in comparison with other methods is the
possibility to register simultaneously ions over a very wide range of
masses and with relatively high mass/charge resolution.  The
measurements are made on-line and the results are available in
real-time.  The ToF analysis includes the following steps:

\begin{enumerate}
\item {\it Taking a gas sample:} A large reduction of pressure is
necessary because a pressure of less than $10^{-4}$ Torr is required by
the ToF spectrometer.  Two techniques are planned for use.  The first is
short sampling, synchronised and delayed relative to the particle beam
for studies of the time dependence of chemical processes.  The second is
sampling into a buffer volume with long-term analysis to increase
sensitivity to trace gases.  In both cases there is a problem of
maintaining the initial composition of gas mixture.  Therefore teflon or
proper plastic with non-nucleating surface can be chosen in the valves. 

\item {\it Ionisation of gas sample:}  Electron impact ionisation at
single collision conditions is planned.  In many cases non-dissociative
ionisation prevails or dissociation fractions are known.  In some
specific cases the identification  of the initial composition of
molecules will be verified by calibration experiments with mixtures
having known composition.

\item {\it Registration of ion signals:}  Ions are registered in current
or in counting mode.  Both will be used to provide at the same time high
sensitivity for weak lines (counting mode) and the ability to deal with
the large number of ions in the strong lines (current mode).  This is
facilitated by the fact that the intense lines correspond to masses less
than 40 amu and the interesting trace lines to larger masses.
\end{enumerate}

The ToF spectrometer has the following performance:
\begin{itemize}
\item Ion mass range 1-2000 a.m.u.
\item Mass/charge resolution power $M/\Delta M = 200$.
\item Maximum ratio of ion line intensities at simultaneous
registration: $10^{4}$.
\item Minimum ion line signal intensity (equal to background counting
rate): 0.01 Hz.
\end{itemize}

\begin{figure}[htbp]
  \begin{center}
      \makebox{\epsfig{file=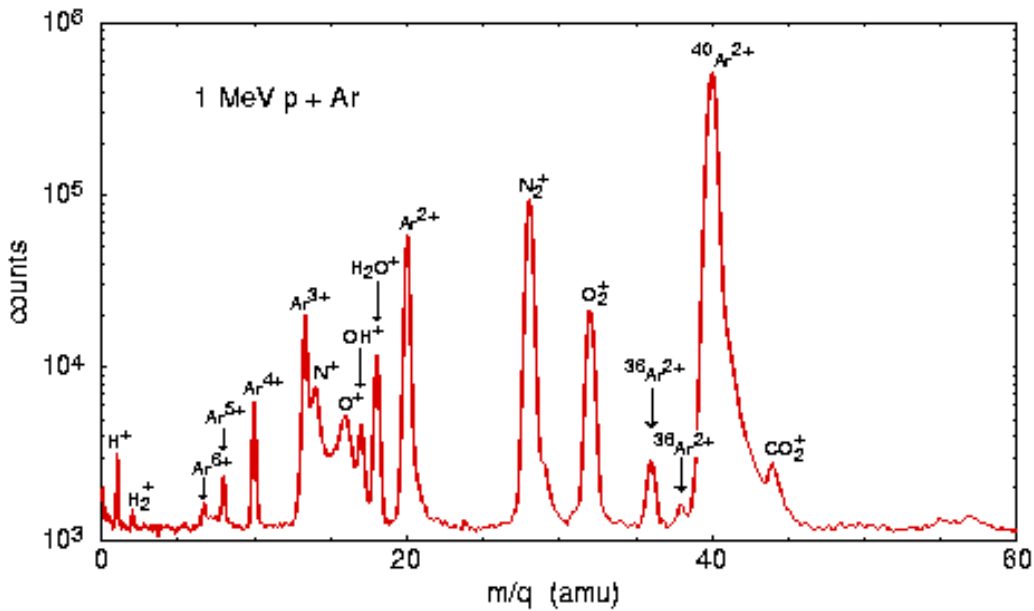,height=80mm}}
  \end{center}
  \caption{Time-of-flight mass spectrometer measurements of the ions
formed in Ar and residual gas by  1 MeV protons  \cite{fastrup}.}
  \label{fig_tof1}    
\end{figure}

The mass range and the resolving power can be improved if necessary. A
typical experimental spectrum is shown in Fig.\,\ref{fig_tof1} 
\cite{fastrup}. It shows on semi-logarithmic scale the composition of
ions formed by ionisation of ambient air and Ar.  As can be seen in
Fig.\,\ref{fig_tof1} there are lines corresponding to formation of
singly and multiply charged ions of different gases, molecular ions and
their dissociation fragment and also to isotopes of $^{36}$Ar and
$^{38}$Ar having low relative concentrations (about 0.34\%  and 0.06\% 
respectively).

\subsubsection{Ion mass spectrometers} 
\label{sec_iomass} 

\begin{figure}[tbp]
  \begin{center}
      \makebox{\epsfig{file=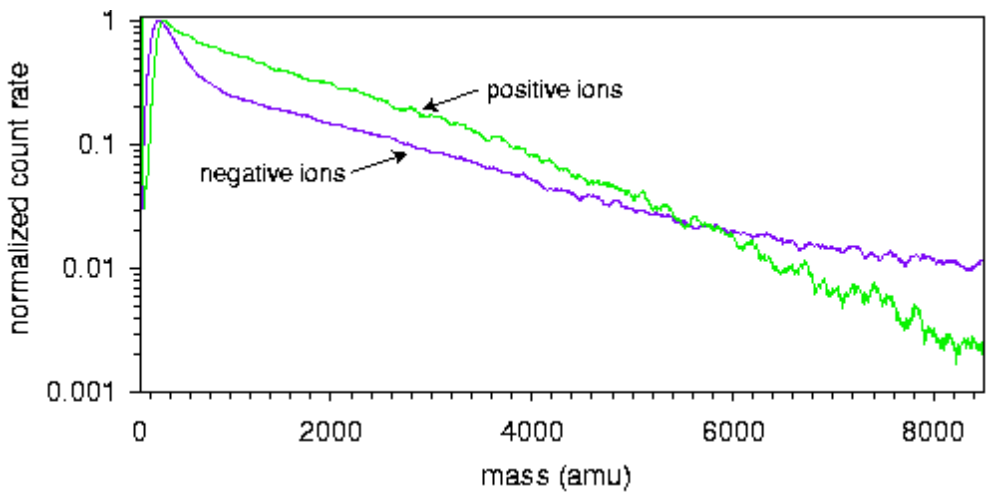,width=130mm}}
  \end{center}
  \vspace{-5mm}
  \caption{Measurements made with an Ion Mass Spectrometer (IOMAS) by
MPIK-Heidelberg. The data show high-pass-mode mass spectra of positive
and negative ions measured in the exhaust of a jet aircraft on the
ground. The plume age is around 0.1\,s, with 2\,mg/kg fuel sulphur
content, 30\% engine power setting and 2.2\,m distance between the exit
of the engine and the sampling point. The initial rise of the ion signal
is a well known instrumental effect. The spectra indicate the presence
of very massive ions, above even 8500 amu. The fractions of ions with
masses exceeding 8500 are 0.0023 (for positive ions) and 0.01 (for
negative ions).}
  \label{fig_mpik_iomas}    
\end{figure}

The MPIK-Heidelberg IOMAS (Ion Mass Spectrometer) and  quadrupole ion
trap mass spectrometer (PITMAS) instruments measure the mass
distribution of positive and negative ions with mass numbers $m\leq$\,10
000 (IOMAS) and $m\leq$ 4000 (PITMAS). IOMAS is a linear quadrupole mass
spectrometer which is usually operated in a high-pass-mode (HPM). The
HPM is very sensitive and particularly well suited for situations with
numerous ion species each having only a small fractional abundance. The
present mass range of IOMAS extends from gaseous molecules having mass
numbers of several hundred up to very massive ions having diameters in
the range of several nm. Therefore it covers the mass range of small
clusters which are formed subsequently during the initial steps of
nucleation. These clusters, however, are too small to be detectable by
condensation nucleus counters (CPCs)---although they are of course
readily detected in the CLOUD expansion chamber when operated with a
relatively large expansion. The performance of IOMAS has been
demonstrated recently in flow reactor experiments made at
MPIK-Heidelberg \cite{wohlfrom00a}. In these experiments large cluster
ions have been produced by adding of sulphur dioxide and water vapour
into a flow reactor where ions and OH radicals were produced. By varying
the amounts of sulphur dioxide and water vapour, efficient growth of
cluster ions consisting of numerous H$_2$SO$_4$/H$_2$O molecules could
be observed. In addition, our measurements of the exhaust from jet
aircraft made both on the ground and in flight show that very massive
ions with mass numbers exceeding 10000 are present
\cite{wohlfrom00b,wohlfrom00c} (Fig.\,\ref{fig_mpik_iomas}). These
measurements confirm the importance of ions in the formation of volatile
aerosols in the exhaust from jet aircraft, which has been suggested
recently \cite{yu97,yu98}. Evidence for the existence of ``chemiions''
in the exhaust of jet aircraft also comes from measurements of the total
ion concentration in the exhaust of a jet aircraft recently made by
MPIK-Heidelberg \cite{arnold99}, and from mass spectrometer
measurements of negative chemiions with mass numbers larger than 450
\cite{arnold00}.

In addition, the MPIK-Heidelberg PITMAS instrument would be available
for chemical identification of complex ions produced in the CLOUD
experiment. PITMAS offers a particularly attractive addition on the
analytical potential, namely fragmentation studies of mass-selected
ions. PITMAS can be operated in two modes. The first of these is a
standard mode that produces ion mass spectra with a very high mass
resolution and a large mass range (up to $m = 4000$). Provided that the
intensity of an ion signal is sufficient, the second mode allows for
specific fragmentation studies of mass-selected ions with mass number up
to 2000. In this mode only ions of a single mass are trapped. The
kinetic energy of the trapped ions is increased and leads to
collision-induced dissociation (CID) through collisions with He atoms
that are introduced into the spectrometer. For complex ions or cluster
ions, usually several different fragment ion species are produced. The
fragment ion pattern is, in most cases, characteristic of a parent ion
species. In a second CID step, one of the different fragment ion species
can be trapped and a CID study can be made leading to second generation
fragment ions. In the same way even higher order generations of fragment
ions can be produced \cite{kiendler00}. Consider, for example, an ion
with $m = 2000$\,amu. If measured by a conventional ion mass
spectrometer giving only $m$, an identification of the ion will be
extremely difficult, if not impossible. By contrast PITMAS has the
potential to pin down the identity of the ion. Another application of
PITMAS involves trapping of ions with one $m$ and subsequent interaction
of the trapped ions with a reagent gas which is introduced into the
trap. Thereby ions may be identified due to their different chemical
reactivity leading to characteristic product ions.

\begin{figure}[htbp]
  \begin{center}
      \makebox{\epsfig{file=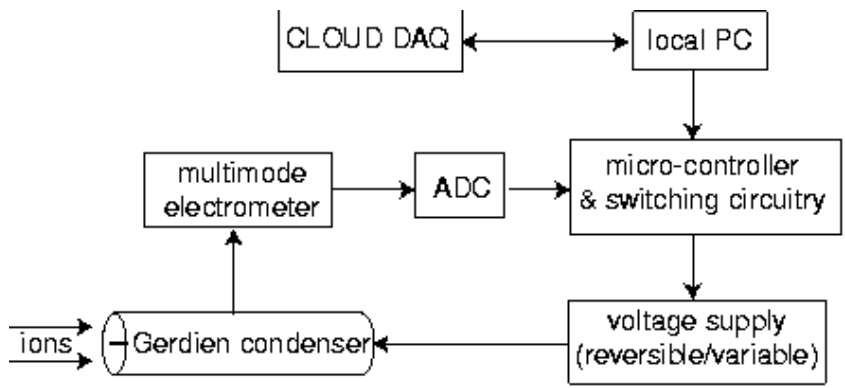,height=50mm}}
  \end{center}
  \caption{Schematic of the programmable ion mobility spectrometer
(PIMS) using a Gerdien condenser.}
  \label {fig_ion_mobility}    
\end{figure}

\subsubsection{Ion mobility spectrometer} 
\label{sec_pims}

Mobility-selective ion measurements are required in order to follow the
time evolution of small ions through complex clusters into aerosol
particles. A flow-driven programmable ion mobility spectrometer (PIMS)
\cite{harrison99} based on a classical mobility-selective detector
(Gerdien condenser) will be employed (Fig.\,\ref{fig_ion_mobility}). The
Gerdien condenser is a metal cylinder that is open at each end and
contains an axial electrode set at either a positive or negative
potential.  The cylinder is enclosed in another metal cylinder which
provides a shield.  Air containing the ions under analysis is flowed
through the cylinder at a known rate and the current is recorded with a
femtoammeter.

\begin{figure}[tbp]
  \begin{center}
      \makebox{\epsfig{file=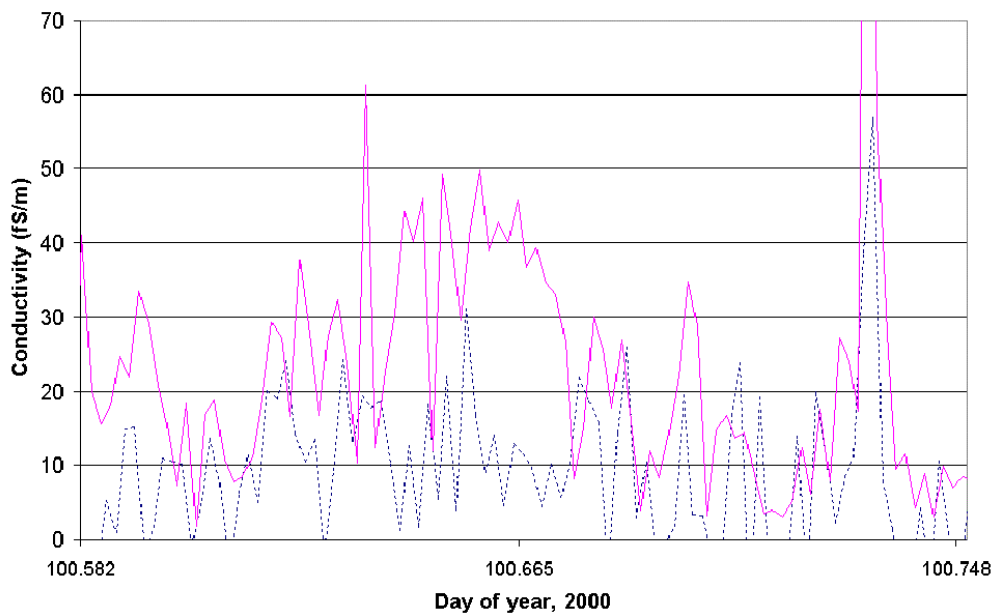,height=80mm}}
  \end{center}
  \vspace{-3mm}
 \caption{A comparison of air conductivity (which is directly
proportional to the ion concentration) measured sequentially using the
two modes of a PIMS under fair weather conditions at Reading between
1400 and 1800 BST on 9 April 2000. The solid trace is total conductivity
(from positive and negative ions) and the dotted trace is the negative
ion conductivity (which is approximately half the total).}
 \label{fig_air_conductivity} 
\end{figure}

The PIMS uses technology developed at Reading to measure the low ion
concentrations found in the atmosphere at ground level \cite{aplin00}.
The Reading ion spectrometer is a fully programmable device which can
measure ions self-consistently in two independent modes, as shown in
Fig.\,\ref{fig_air_conductivity}. The  PIMS can also be operated
dynamically to utilise mobility cut-offs appropriate to the ions
produced, which will allow the evolution of the whole ion spectrum to be
examined, before the particle measurement system. In addition, the PIMS
can measure the total ion concentration, which will permit calculation
of absolute concentrations from the ion mass spectrometers.

The ion counters for CLOUD are miniaturised versions of the recent
Reading design \cite{aplin00}, constructed in a cylindrical geometry
from stainless steel and PTFE, with a central platinum wire electrode.
In the flow chamber, a cross section of 1~cm will be used to reduce the
volume flow rate through the instrument. Adjustment of the bias voltage
depending on the flow rate is a simple procedure. Several PIMS detectors
measuring different ion mobilities will be arranged sequentially to
minimise the flow rate, and to ensure the same sample of air is
measured. Techniques have already been developed to reliably measure
the very small ion currents involved \cite{harrison97,harrison00}.

\subsection{Data acquisition  and offline analysis}
\label{sec_daq_and_offline_analysis}

\subsubsection{Data acquisition and slow control}
\label{sec_daq_and_sc}

\begin{figure}[tbp]
  \begin{center}
      \makebox{\epsfig{file=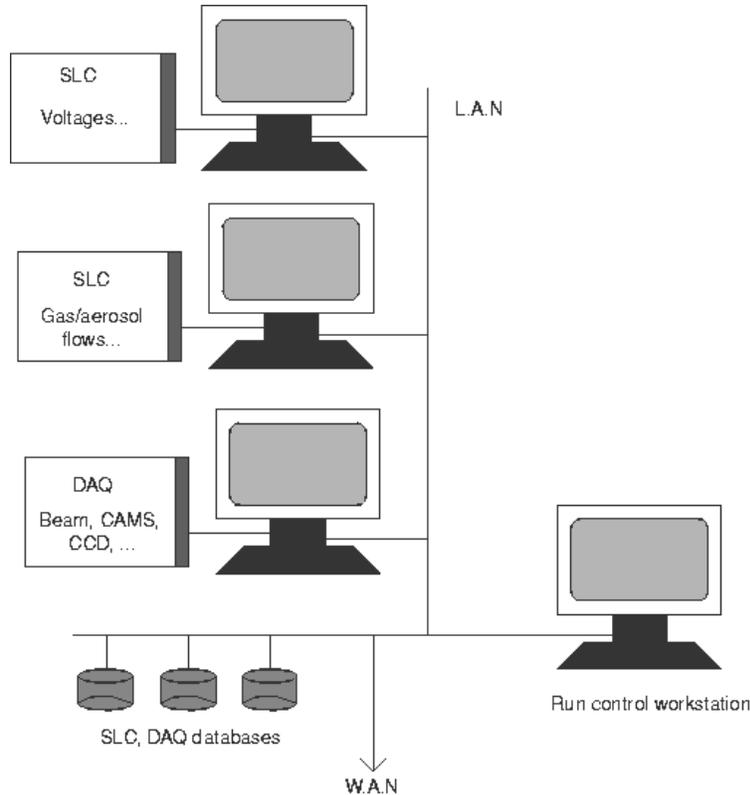,height=105mm}}
  \end{center}
   \vspace{-5mm}
  \caption{Schematic diagram of the data acquisition (DAQ) and slow
control (SLC) systems for CLOUD.}
  \label{fig_daq}    
\end{figure}

The CLOUD data acquisition (DAQ) system (Fig.\,\ref{fig_daq}) is
designed to collect and record a) the operating parameters of the cloud
chamber and other detector systems and b) the data from the optical
readout systems and other equipment during the periods of beam exposure
and droplet growth. It must be be highly flexible in order to
accommodate the many different run conditions necessary for the various
experiments.

The slow control (SLC) system\footnote{It is called \emph{slow control}
in particle physics terminology since it typically involves timescales
of  a few seconds rather than the nanosecond or microsecond timescales
characteristic of fast logic and data acquisition.} is responsible for
the software control of all  operating conditions for the experiment. 
It includes defining the timing and pressure cycle for the piston
expansion/compression, operating the gas valves, controlling the gas and
aerosol flow rates, setting the voltages for the field cage electrodes,
and setting voltages and other parameters for each of the detector
systems.

\paragraph{Functions:} The functions of the DAQ and SLC systems are to:
\begin{itemize}
\item Provide a user-friendly control of the run conditions,      
      (beam on/off, beam energy, etc.) and of the parameters for the
cloud 
        chamber and other detectors.
\item Implement a readout selection according to the run
conditions.     
\item Digitise, read out and assemble the data of the different
      detectors into sub-data blocks.
\item Assemble the sub-data blocks into full events.
\item Monitor the data quality, detect malfunctions and warn the
operator.
\item Record the events onto permanent storage.
\item Monitor and record the parameters of the cloud chamber and other
detectors.
\end{itemize}

\paragraph{Data components:}

The cloud chamber and detectors produce data which are divided into two
different data streams according to their readout period.

\begin{itemize}
\item The DAQ data with a readout period $\sim$msec are generated by the
CAMS detector,  CCD cameras, aerosol counters, mass spectrometers, ion
mobility detector and the beam counter systems.  This stream will also
include the temperature and pressure sensors for the active volume of
the cloud chamber, and the piston hydraulic control and position
monitor.

\item The SLC data with a readout period $\sim$sec/min are generated by
equipment such as gas flow monitors and the voltage sensors for the
field cage. 

\end{itemize}

For each data stream an event builder collects the data and transfers
them to disks for permanent storage. Data reduction and formatting can
be applied during transfer in order to reduce the event size and
therefore the amount of disk space required.

\paragraph{Central control and monitoring:} 

All the control functions of the CLOUD detector will be integrated in
order to:
\begin{itemize}
\item Support distributed and heterogeneous components.
\item Store, retrieve and modify the experimental configuration.
\item Store, retrieve and modify the CLOUD detector monitoring
      parameters.
\end{itemize} 

The CLOUD DAQ and SLC requirements are modest compared with a typical 
particle physics experiment and can be implemented with standard
commercial products. Standard CERN software is available to perform the
system integration.

\subsubsection{Offline analysis}
\label{sec_offline_anlysis}

\begin{figure}[htbp]
  \begin{center}
      \makebox{\epsfig{file=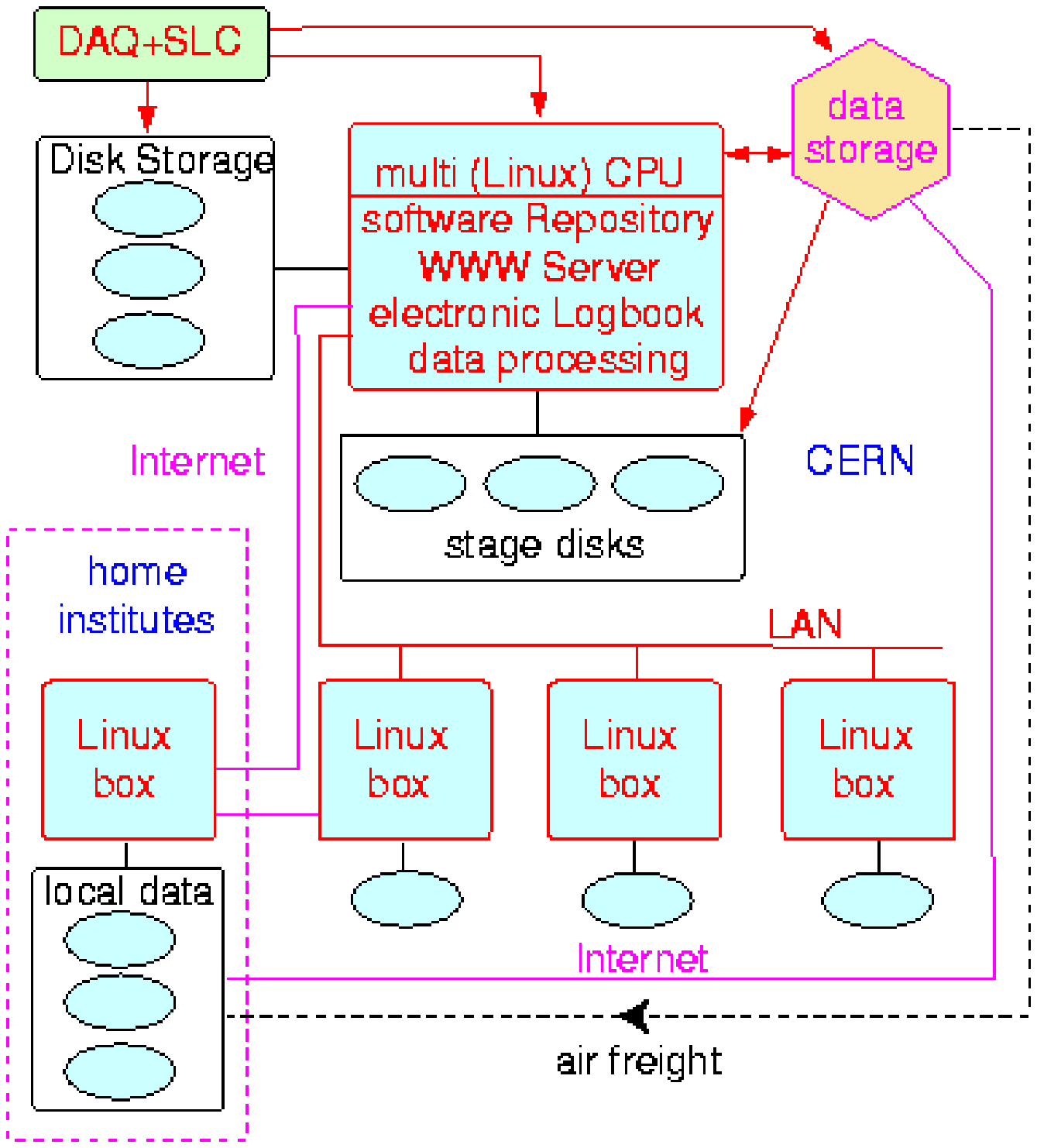,height=105mm}}
  \end{center}
   \vspace{-5mm}
  \caption{Schematic diagram of the offline computing environment.}
  \label{fig_offline}    
\end{figure}

Implementation of the offline computing environment for CLOUD
(Fig.\,\ref{fig_offline})  also poses no special difficulties in
comparison with present particle physics experiments.

The central part of the environment is a multi-CPU machine, or
alternatively a small cluster of several single/double-CPU machines.  
The platform of choice will be a PC running Linux in view of the high 
performance and low cost trends.  This machine is responsible for:
\begin{itemize}
\item Data processing.
\item Maintaining the database of the detector (``electronic logbook").
\item Providing a powerful web server which will be linked with the data
storage and the database.
\end{itemize}

The data flowing from the DAQ and SLC  will be stored in a mass storage
system and also on temporary disks attached to the central machine. This
will execute the necessary processing steps, such as reconstruction of
information from single sub-detectors and then combined event
reconstruction from the whole detector system. The central machine will
also manage storage of the processed data onto the same mass storage
system. 

Stage disks attached to the central machine will provide processed data
and, if needed, raw data for the end-user. End-user analysis will be
carried out either with a fraction of the CPU of the central machine or
with local desktop Linux machines. Home institutions will be equipped by
scalable Linux facilities, with a sufficient amount of disk space to
fetch---either over the network or by air freight---a subset or else the
full data sample for local analyses.

We plan to benefit as much as possible from the ongoing effort of the
HEP community in general and of the LHC community in particular towards
modern computing solutions, even given the relative smallness of our
data sample. As an example the integration with the currently proposed
solution for persistency and software architecture will be among the
first software development tasks. End-user data analyses will also
centre  around the software tools of the new generation of experiments,
such as ROOT\cite{root} and LHC++\cite{lhc++}.

\section{Detector performance} \label{sec_performance}

\subsection{Droplet growth time} 
\label{sec_droplet_growth_time}

\subsubsection{Principles of droplet growth} 
\label{sec_growth_principles}

An aerosol  will grow by condensation if the water vapour
supersaturation exceeds a critical value, which depends on the size and
nature of the aerosol (Appendix \ref
{sec_general_properties_of_clouds}).  Once the growth process has
started, it continues since the equilibrium vapour pressure decreases
with increasing droplet size.  Eventually the growth stops when
sufficient water vapour has been removed that the supersaturation is
reduced.

The driving force that determines the rate at which liquid mass
condenses onto the droplet is the supersaturation relative to the
equilibrium value, $SS(r)$:
\begin{eqnarray}
 SS(r) & = & (S-1)-{a \over r}+{b \over {r^3}} 
 \label{eq_ss}
\end{eqnarray} where $S$ is the water vapour saturation ratio with
respect to the vapour pressure over a plane surface of water, and $r$ is
the radius of the droplet.  The $a / r$ term corresponds to the effect
of curvature (Kelvin's equation) and the $b / {r^3}$ term corresponds to
the effect of dissolved solute (Raoult's law).  This is the form of the
K\"{o}hler equation (Fig.\,\ref{fig_kohler}).

The growth process of a droplet occurs by diffusion of water molecules
from the vapour onto its surface.  Associated with the condensation is
the release of latent heat, which tends to raise the temperature of the
droplet above its surroundings.  This increases the rate of evaporation
and decreases the rate of growth.  The size of this effect is determined
by the rate at which heat is conducted away from the droplet by the
surrounding gas. The resulting growth rate is given by \cite{rogers}:
\begin{eqnarray}  r{{dr} \over {dt}} & =  & {SS(r)}  \over {[F_k+F_d]}
\label{eq_rdr/dt} 
\end{eqnarray}  where $F_k$ represents the latent heat term and $F_d$
the vapour diffusion term.  When the droplet becomes sufficiently large,
the  $a / r$  and $b / {r^3}$ terms in Eq.\,\ref{eq_ss} are negligible
compared with $(S-1)$, so that $SS(r) = SS$.  Integrating 
Eq.\,\ref{eq_rdr/dt} then gives
\begin{eqnarray}  r_t^2 & = & r_0^2+2\xi t  \label{eq_rt2} \\
\mathrm{where }\; \; \xi & = & {{SS} \over {[F_K+F_d]}} 
\label{eq_xi}
\end{eqnarray} 
 Equations \ref{eq_rt2} and \ref{eq_xi}  illustrate the essential
feature that the time for a droplet to grow to a radius $r_t$ is $t
\propto r_t^2 / SS$.  A more rigorous calculation of droplet growth is
presented below. 

\subsubsection{Simulation of droplet growth}
\label{sec_growth_calculation} 

We have studied the formation and growth of the cloud droplets using a
simple convection model describing a rising cloud parcel. We have
assumed a constant updraft velocity without entrainment, since we are
primarily interested in droplet formation. The air parcel contains 
aerosol particles, water and the other condensable vapours (nitric acid,
hydrochloric acid and ammonia). The aerosol particle distribution is
described using four log-normal size distributions. The first
distribution consists of less hygroscopic Aitken mode\footnote{Aitken
mode aerosols are in the  20--100 nm diameter range whereas accumulation
mode aerosols are larger, with diameters of 100--1000 nm.} particles,
the second of more hygroscopic Aitken mode particles, the third of less
hygroscopic accumulation mode particles  and the fourth of more
hygroscopic accumulation mode particles. 

In the beginning of a simulation (at the cloud base) dry particles are
allowed to take up water and the other condensing  vapours and reach
equilibrium with the ambient air. The air parcel starts to rise and cool
adiabatically, the saturation ratio increases, and the vapours begin to
condense into the haze particles.  The saturation ratio of water vapour
rises typically slightly over unity, after which it decreases to a value
near unity. Although the water vapour saturation ratio decreases, some
particles continue to grow, because they have been activated and a 
cloud droplet population has formed.

\paragraph{Condensation model:} 

In the model the condensation of the vapours is simulated  by
numerically integrating a set of ordinary differential equations,  
based on the model used by Kulmala {\it et al.} \cite{kulmala93}. In the
present study we have slightly changed the differential equations
describing the change of the masses of the  species in the droplet.  The
rate of change of the mass of the condensing species $j$ in the droplet
size class $p$ is obtained from the equation
\begin{eqnarray*}
 \frac{dm_{j,p}}{dt} \equiv I_{j,p} ,
\end{eqnarray*}  
 where the mass flux  $I_{j,p}$ is the rate at which the mass of species
$j$  transports through  a surface area of a droplet in the size-class
$p$.  The mass fluxes can be calculated from the matrix equation
\begin{eqnarray*}
       \left(
       \begin{array}{cccc}
        B_{11}-1 & B_{12} & \cdots & B_{15} \\
        B_{21}   & B_{22}-1 &      & \vdots  \\
        \vdots   &        &\ddots  &         \\
        B_{51}   & \cdots &        &  B_{55} -1
       \end{array}
        \right)
       \left(
       \begin{array}{c}
        I_{1} \\
        I_{2} \\
         \vdots  \\
         I_{5}
       \end{array}
        \right)
       =
       \left(
       \begin{array}{c}
       A_{1}(S_{1a}-S_{1}) \\
        A_{2}(S_{2a}-S_{2}) \\
        \vdots  \\
        A_{5}(S_{5a}-S_{n})
       \end{array}
       \right),
\label{n6}
\end{eqnarray*} 
 where 
\begin{eqnarray*} 
 A_{j} =  \frac{4\pi  a_{p} M_{j} \beta_{Mj} D_{Ij} }{RT_{\infty}}
p_{s,j}(T_{\infty}) \nonumber \\ B_{ji} = -
S_{ja}\frac{M_{j}L_{j}}{RT_{\infty}^{2}}
\frac{ L_{i} }{4 \pi a_{p} \beta_{T} k } A_{j} .
\label{n5}
\end{eqnarray*}  
 In the preceding expressions $M_{j}$ is the molecular mass of species
$j$, $L_{j}$ is the latent heat of vapourisation, $a_{p}$ is the droplet
radius in the size class $p$, $D_{Ij}$ is the  binary vapour diffusion
coefficient of species $j$ in the  carrier gas, $S_{j}$ is the gas phase
activity, $S_{ja}$ is  the activity over droplet surface (i.e. liquid
phase activity $\times$ Kelvin effect), $p_{s,j}$ is the saturation
vapour pressure at the ambient temperature $T_{\infty}$ and $k$ is
thermal conductivity of gas mixture.  We have adopted the transition
coefficients for mass transfer $\beta_{Mj}$ and for heat transfer
$\beta_{T}$ from ref.\,\cite{fuchs}. The preceding equation may be
solved by computer using  the Gauss elimination method or LU
factorisation
\cite{press}. The equation  couples heat and mass transfer of water and
other condensing vapours \cite{mattila}.

\paragraph{Droplet growth results in the cloud chamber:} 

Using the condensation models described above we have performed several
model studies which have been verified using data measured in  Vienna
\cite{rudolf94}.  We have assumed initial aerosol particles of radius
about 25 nm. To grow these into droplets of radius 1
$\mu$m  when the saturation ratio, $S$, is 1.2 takes less than 100 ms. 
If $S = 1.02$ it takes about 1 s and if $S = 1.002$ it takes about
10~s.  These values can be scaled to other droplet radii and other
supersaturations since the growth time $t \propto r_t^2 / SS$ (Eqs.
\ref{eq_rt2} and \ref{eq_xi}), where $SS = S - 1$.

\subsection{Sensitive time of the cloud chamber} 
\label{sec_sensitive_time}

After an expansion has taken place, heat gradually leaks into the
chamber, the supersaturation diminishes and after a certain time (termed
the {\it sensitive time}) condensation can no longer take place on the
activated droplets.  It is therefore important that the sensitive time
is sufficiently long to allow activated droplets to grow at least to the
minimum detectable size (about 1 $\mu$m diameter). 

Immediately after the expansion, the gas temperature has fallen but the
walls of the chamber remains at the equilibrium temperature value  prior
to the expansion. The gas layer close to the walls then begins to heat
up, expand and rise by convection.  The expansion of the boundary layer
causes the gas inside to re-compress adiabatically, which warms it and 
causes the supersaturation to fall.  Because of the low  thermal
conductivity of air, heat conduction is an insignificant process for 
re-heating of the central gas volume.

This sensitive time of a cloud chamber, $t_s$, due to this
re-compression  process is calculated \cite{das_gupta} to have the
following dependencies:
 \begin{eqnarray} 
 t_s \propto {L^2 \cdot P \over  (\Delta V /V)^2 \cdot  T^2 } 
\label{eq_ts}
 \end{eqnarray}
  where  $L$ is the linear dimension of the cloud chamber, $ \Delta V
/V$ is the volume expansion ratio, $P$  the pressure and $T$ the
temperature.  This expression shows that long sensitive times require
large chambers since the sensitive time increases as the square of the
linear dimension (i.e.  as the square of the ratio of volume/surface
area, $(V/A)^2$).  

The sensitive time is expected to be proportional to
$P/T^2$ ( Eq.\,\ref{eq_ts}). Since low pressures and low temperatures
occur together in the troposphere, these effects partially compensate. 
However the reduction in pressure dominates and so $t_s$ decreases, e.g.
it is reduced by a factor of about two for the conditions corresponding
to 10 km altitude.

\begin{table}[htbp]
  \begin{center}
  \caption{Values of the temperature drop, $\Delta T$, and water vapour
supersaturation, $SS$, resulting from a small pressure expansion
ratio,   $\Delta P /P$.  The volume expansion ratio $\Delta V / V = 0.71
\cdot \Delta P /P$.  The initial conditions are P = 101 kPa and T =
298~K and $SS$~=~0\% (100\% relative humidity).}
  \label{tab_ss}
  \vspace{1mm}
  \begin{tabular}{| c c c | | c c c |}
  \hline
 $\Delta P / P$ & $\Delta T$ & $SS$  & $\Delta P / P$ & $\Delta T$ &
$SS$ \\
 \mbox{[$\times 10^{-3}$]}  & [K]  &  [\%]  
  & [$\times 10^{-3}$]  & [K]  &  [\%]  \\
  \hline
  \hline
    &  & &  &  &   \\[-2ex]
    0.1  &  0.009  &  0.04  &    1.0  &  0.085  &  0.40  \\
    0.2  &  0.017  &  0.08  &    2.0  &  0.170  &  0.81  \\
    0.4  &  0.034  &  0.16  &    3.0  &  0.255  &  1.22  \\
    0.6  &  0.051  &  0.24  &    4.0  &  0.341  &  1.63  \\
    0.8  &  0.068  &  0.32  &    5.0  &  0.426  &  2.04  \\[0.5ex]
  \hline
  \end{tabular}
  \end{center}
\end{table}

Equation \ref{eq_ts} indicates that the sensitive time increases for
smaller expansions (which give rise to smaller temperature differences
between the walls and the gas).  For small expansion ratios, there is a
linear relationship with the resultant supersaturation, e.g.\,at 298 K
(Table \ref{tab_ss}):  
  \begin{eqnarray} 
  \Delta P / P = \gamma \cdot \Delta V / V 
 = [\gamma/(\gamma-1)]\cdot \Delta T / T = 0.25 \cdot SS, 
 \label{eq_dp/p}
 \end{eqnarray}
 where $\gamma = 1.4$ is the ratio of the ratio of the specific heats
for an air-water vapour mixture, and the supersaturation 
$SS$ is expressed as a fraction. Therefore  $t_s \propto 1/SS^2$. 
Since the droplet growth time $\propto 1/SS$, this increase in the
sensitive time at low supersaturations more than compensates for the
longer   growth time.

Our experience with the 25 cm Vienna chamber indicates a sensitive time
of at least 10~s at  2\% supersaturation and stp.  Droplet number
concentrations typically did not exceed 5000 cm$^{-3}$.  Vapour
depletion and latent heat production during growth, however, must be
quantitatively accounted for, even for growth times less than 10~s.
From Eq.\,\ref{eq_ts}, the 50~cm CLOUD expansion chamber should
therefore have about a factor 4 longer sensitive time under the same
conditions. At lower supersaturations, the sensitive time should
further increase, proportionally to $1/SS^2$, e.g. by a factor 100 at
0.2\% supersaturation. Under these very low supersaturations, the
expected sensitive time is therefore above ten minutes. In conclusion,
we expect that the sensitive time of the  50~cm CLOUD expansion chamber
is sufficient to observe droplet growth at all thermodynamic and water
vapour conditions of interest.

\begin{figure}[tbp]
  \begin{center}
      \makebox{\epsfig{file=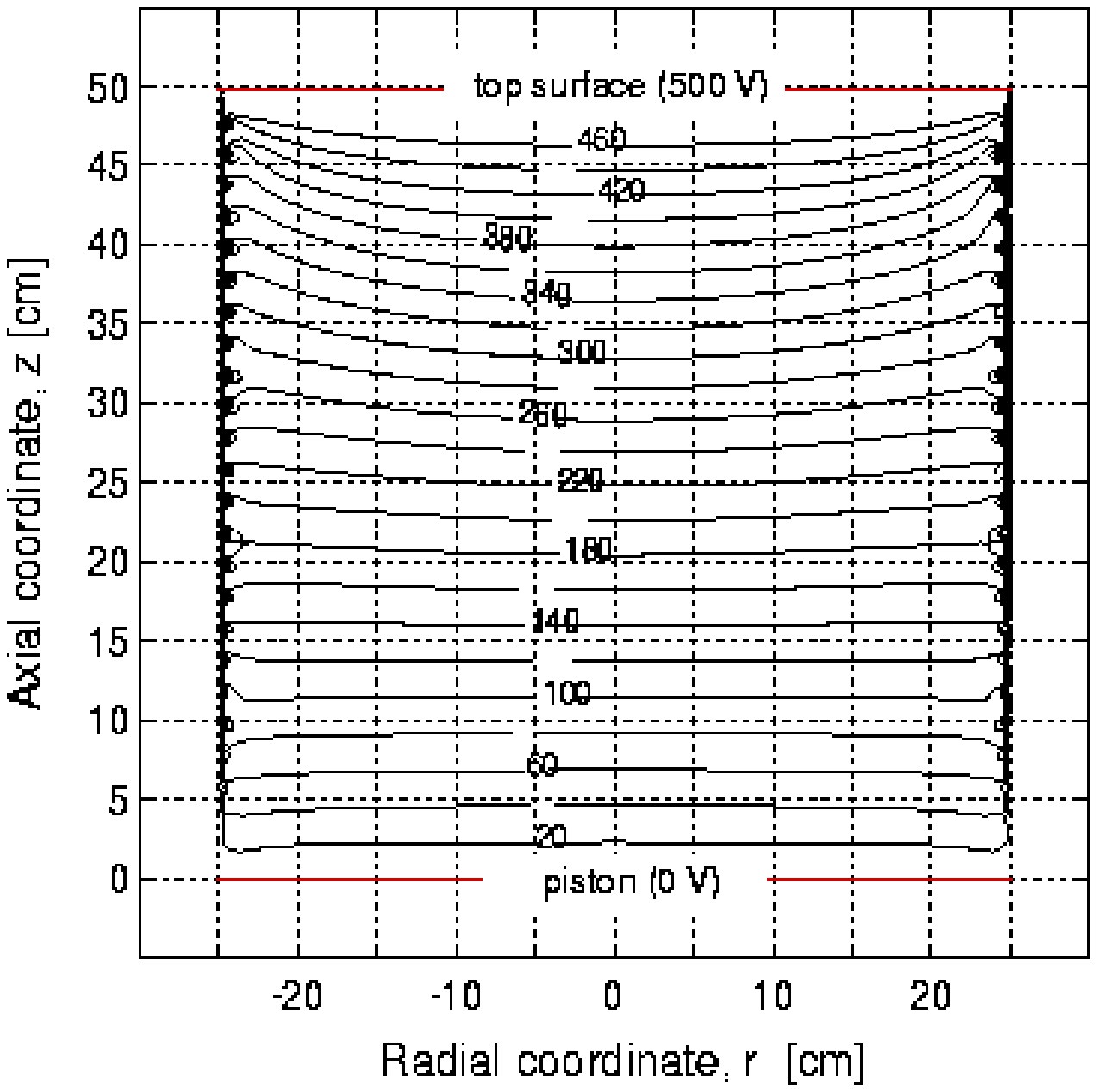,height=95mm}}
  \end{center}
   \vspace{-5mm}
  \caption{Equipotential contours inside the cloud chamber for linear
voltage  gradient on the electrodes, and the uppermost electrode at 500
V.}
  \label{fig_equipotentials}    
  \begin{center}
      \makebox{\epsfig{file=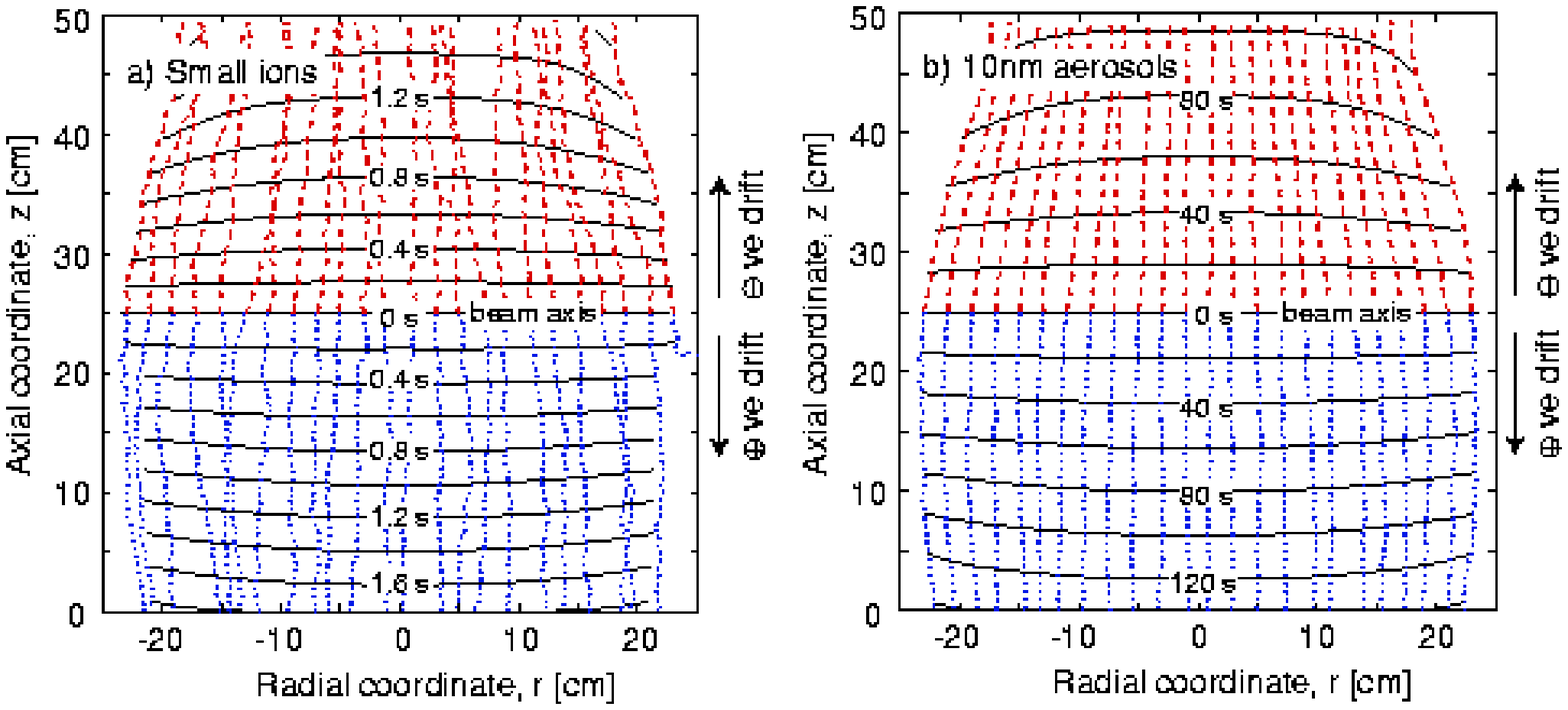,width=160mm}}
  \end{center}
  \caption{Drift paths and isochronous contours for a) small ions,  and
b) 10 nm diameter aerosols in air at 293 K and 101 kPa pressure.  A
total of 24  pairs of particles with unit positive and negative electric
charge, respectively, are uniformly generated along the beam axis ($z$ =
25 cm) at $t$ = 0 s.  The electric field is shown in
Fig.\,\ref{fig_equipotentials}.  The scattering of the drift paths is
due to Brownian diffusion.}
  \label{fig_drift_paths}    
\end{figure}

\subsection{Electric field and charged particle drift}
\label{sec_electric_field}

We have made a preliminary study with the GARFIELD program
\cite{veenhof}  of the electric field produced by of the field cage. 
The equipotentials are shown in Fig.\,\ref{fig_equipotentials} when the
uppermost electrode is set at +500 V and the piston at 0 V, and a linear
voltage division is provided  for the intermediate electrodes.  An
electric field of about 1000 Vm$^{-1}$ is produced within the chamber
volume, and is sufficiently uniform to perform the required functions,
as described in Section~\ref{sec_field_cage}.

The drift velocity, $v_D$ [ms$^{-1}$], of a particle with charge $ne$
[C] in and electric field $E$ [Vm$^{-1}$] is given by 
\begin{eqnarray*}
  v_D & = & \mu E,  \\
  \mathrm{and} \; \; \mu & = & neB 
\end{eqnarray*}   where  $\mu$ [m$^2$V$^{-1}$s$^{-1}$] is the electrical
mobility and $B$ [mN$^{-1}$s$^{-1}$] is the  mechanical mobility.  Some
values of these quantities for particles of various sizes are given in
Table \ref{tab_mobilities}.

The drift characteristics of charged particles in this electric field
are shown in Fig.\ref{fig_drift_paths}.  Small ions can be efficiently
cleared from the cloud chamber in about 2 s.  Charged aerosols, however,
drift much more slowly.  In the example shown, 10 nm-diameter aerosols
with unit electric charge are swept from the chamber in about 2
minutes.  These large differences in electrical mobility can be usefully
applied as an experimental tool to provide, for example, mobility
measurements inside the chamber, or selective removal of light charged
particles while leaving heavier particles. Because of their low
mobilities (Table~\ref{tab_mobilities}), aerosols larger than a few tens
of $\mu$m in diameter are most efficiently cleared from the cloud
chamber either by activation and sedimentation or else by vacuum
evacuation of the chamber and then refilling.

\begin{table}[tbp]
  \begin{center}
  \caption{Mobilities and diffusion parameters for ions and aerosol
particles at 293~K and 101 kPa \cite{hinds}. The electrical mobility
assumes unit charge. Since each of these quantities is approximately
inversely proportional to pressure, the values at another pressure P
[kPa] can be estimated by multiplying the numbers in the table by 101/P.
}
  \label{tab_mobilities}
  \vspace{5mm}
  \begin{tabular}{| l | l l l l|}
  \hline
  \textbf{Particle} & \textbf{Electrical} & \textbf{Mechanical}  
   & \textbf{Diffusion} & \textbf{Slip cor.}  \\[-0.5ex]
  \textbf{diameter}  & \textbf{mobility} & \textbf{mobility}
   & \textbf{coefficient}  & \textbf{factor}  \\
    $d$  & $\mu$ & $B$  & $D$ & $C_c$  \\
   \mbox{[$\mu$m]} & [m$^2$V$^{-1}$s$^{-1}$]  &  [mN$^{-1}$s$^{-1}$] 
   &  [m$^2$s$^{-1}$] & \\
  \hline
  \hline
    &  &  & & \\[-2ex]
   0.00037$^\dagger$ &  & $4.9 \cdot 10^{15}$ & $2.0 \cdot 10 ^{-5}$ &
\\
   $\ominus$ve air ion & $1.6 \cdot 10^{-4}$ &  & &  \\
   $\oplus$ve air ion & $1.4 \cdot 10^{-4}$ &  & &  \\
   0.001 & $1.6 \cdot 10^{-4}$ & $1.0 \cdot 10^{15}$ & $4.2 \cdot 10
^{-6}$ & 200 \\
   0.003 & $2.3 \cdot 10^{-5}$ & $1.4 \cdot 10^{14}$ & $5.8 \cdot 10
^{-7}$ & 73 \\
   0.010 & $2.1 \cdot 10^{-6}$ & $1.3 \cdot 10^{13}$ & $5.3 \cdot 10
^{-8}$ & 22 \\
   0.030 & $2.5 \cdot 10^{-7}$ & $1.5 \cdot 10^{12}$ & $6.3 \cdot 10
^{-9}$ & 7.9 \\
   0.100 & $2.7 \cdot 10^{-8}$ & $1.7 \cdot 10^{11}$ & $6.9 \cdot 10
^{-10}$ & 2.9 \\
   0.300 & $4.9 \cdot 10^{-9}$ & $3.1 \cdot 10^{10}$ & $1.2 \cdot 10
^{-10}$ & 1.6 \\
   1.000 & $1.1 \cdot 10^{-9}$ & $6.8 \cdot 10^{9}$ & $2.7 \cdot 10
^{-11}$ & 1.2 \\[0.5ex]
  \hline
  \end{tabular}
  \end{center}
   \hspace{10mm} $^\dagger$ Diameter of an air molecule.
\end{table}

\subsection{Diffusion effects}  \label{sec_diffusion_effects}

\subsubsection{Principles of diffusion}
\label{sec_diffusion_principles}

In the absence of an electric field, ions and aerosol particles migrate
through the cloud chamber by diffusion and sedimentation (gravitational
settling).  The latter can be completely neglected for the size of
aerosol particles  under consideration.  For example, a 0.1~$\mu$m
diameter aerosol has a settling velocity of about 1 $\mu$m s$^{-1}$ at
stp.  

The ions and aerosols will diffuse a projected distance 
\begin{eqnarray*}
  x_{rms} & = & \sqrt {2Dt}  
\end{eqnarray*}  
 in a time $t$, where $D$ [m$^2$s$^{-1}$] is the diffusion
coefficient.   The diffusion coefficient and  the mechanical mobility
$B$ are simply related:
\begin{eqnarray*}
  D & = & kTB,  \\
  \mathrm{and} \; \; B & = & {{C_c} \over {3\pi \eta d}}  
\end{eqnarray*} 
 where k [JK$^{-1}$] is Boltzmann's constant, T [K] is the temperature,
$\eta$ [Pa s] is the viscosity of air and $d$ [m] is the particle
diameter.  The quantity $C_c$ is a dimensionless slip correction factor
\cite{hinds} which is close to unity for particles above 1 $\mu$m
diameter but increases in magnitude for smaller particles. Values of
these  quantities for various sizes of particle are given in Table
\ref{tab_mobilities}.

\begin{figure}[htbp]
  \begin{center}
      \makebox{\epsfig{file=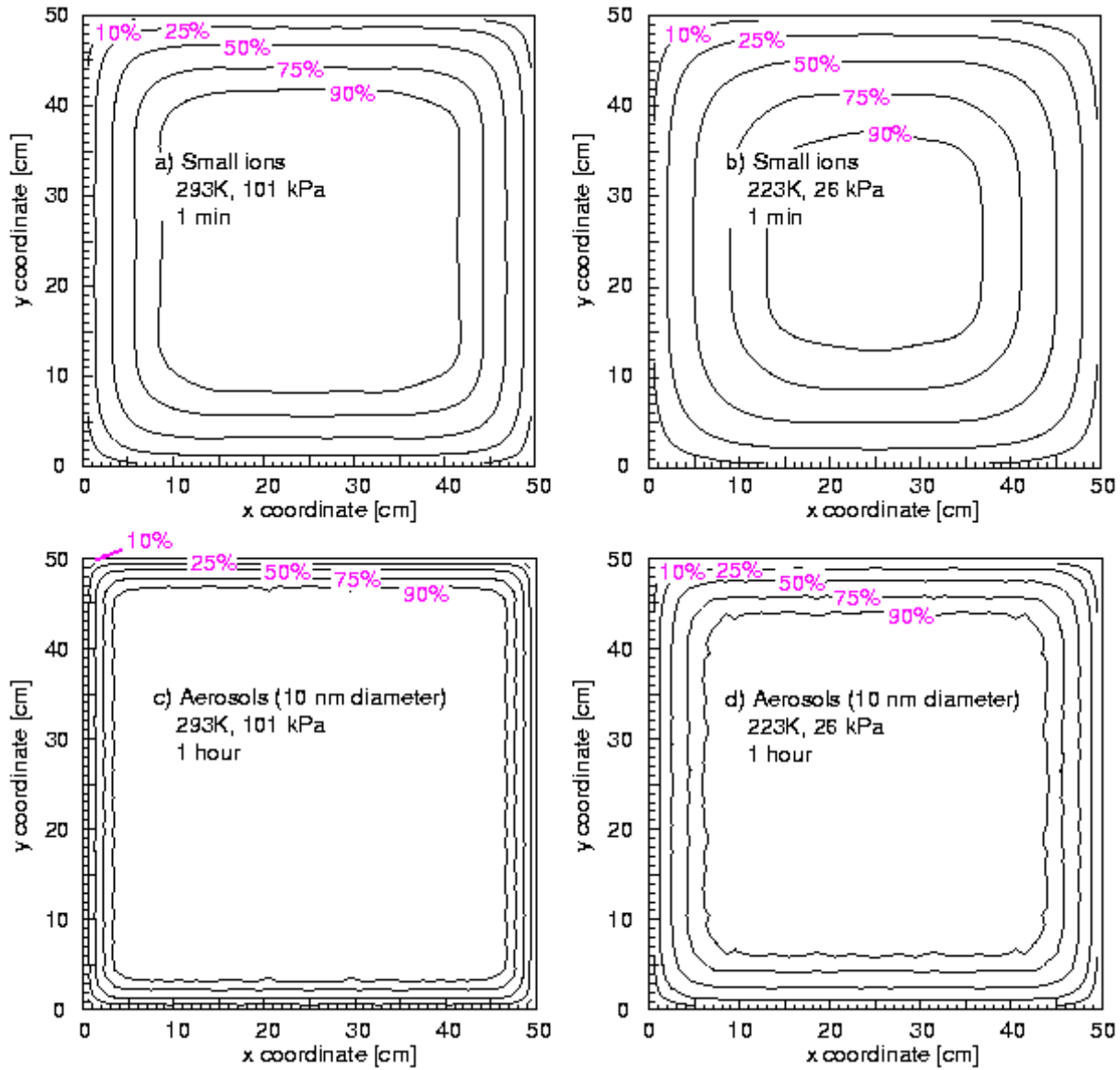,width=160mm}}
  \end{center}
  \caption{Wall losses of particles in the cloud chamber due to thermal 
diffusion. The upper plots show the number density of small ions after a
time $t = $  1 minute at a) 293~K and 101 kPa (standard conditions)  and
b) 223~K and 26 kPa (10 km altitude).  The lower plots show the number
density of 10 nm-diameter aerosols after a time of 1 hour at c)~293~K
and 101 kPa  and d) 223~K and 26 kPa. A particle is assumed to be lost
if it touches one of the walls (which are located at the boundaries of
the plots).  The initial charged particle distributions were generated
uniformly in $x$ and $y$ in the range $0 < x,y < 50$ cm.  The contours
indicate the fraction of the original number density of particles
remaining after the indicated times.  The losses can be estimated at
other times by scaling the distance between a contour and its nearby
wall as
$\sqrt{t}$.}  
  \label{fig_diffusion_losses} 
\end{figure}

\begin{figure}[htbp]
  \begin{center}
      \makebox{\epsfig{file=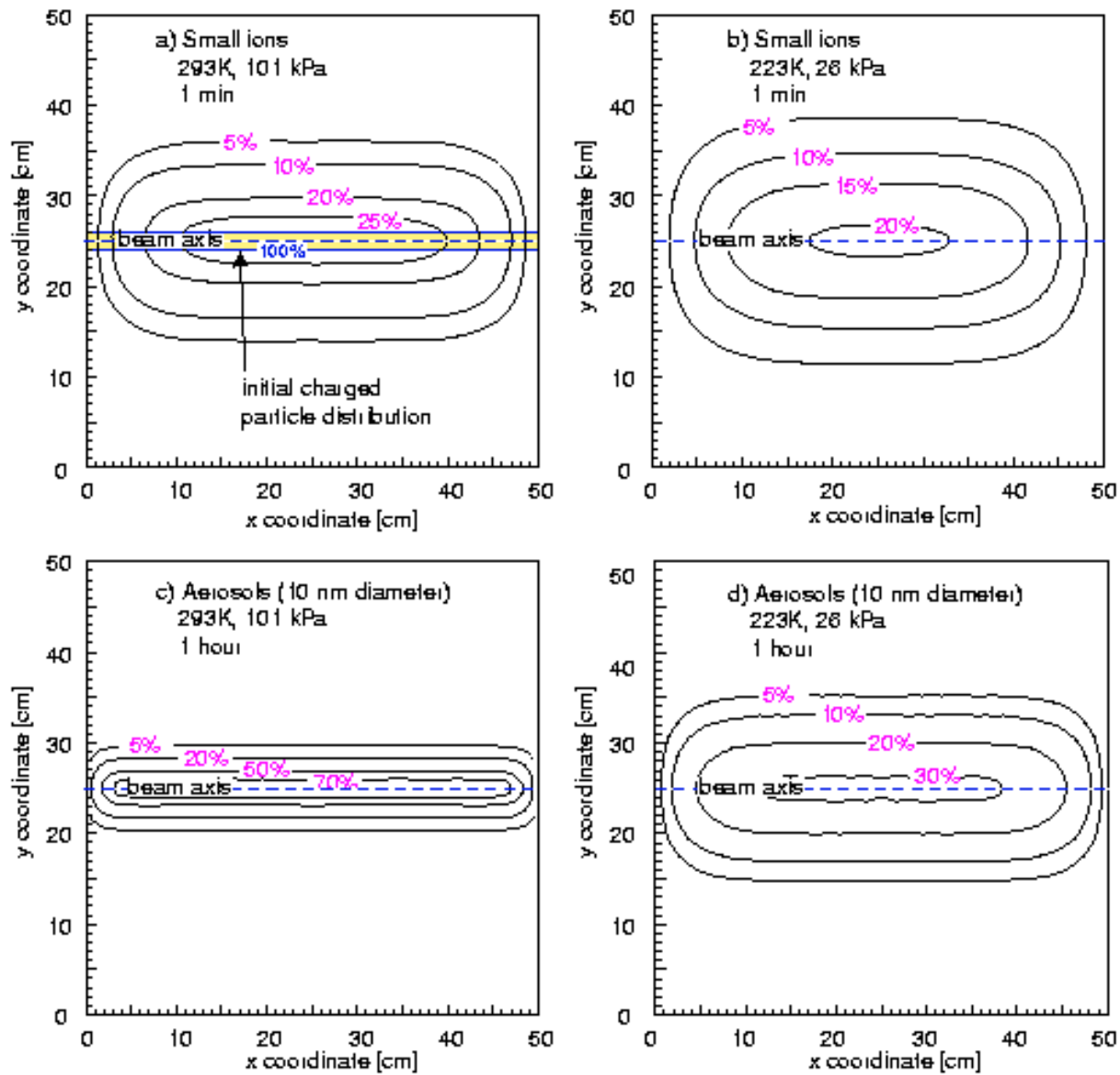,width=160mm}}
  \end{center}
  \caption{The spreading of charged particles from the beam region of  
the cloud chamber due to thermal  diffusion.  The upper plots show the
number density of small ions after a time $t$ = 1 minute at a) 293~K and
101 kPa (standard conditions)  and b) 223~K and 26 kPa (10 km
altitude).  The lower plots show the number density of 10~nm-diameter
charged aerosols after a time of 1 hour at c) 293~K and 101 kPa  and d)
223~K and 26 kPa.  The initial charged particle distributions were
generated according to the beam profile: a Gaussian distribution in the
y projection, centred on $y = 25$~cm and with $\sigma_y$ = 0.4~cm, and a
uniform distribution in the $x$ projection.  The contours indicate the
number density of charged particles after the indicated times.  A figure
of 100\% corresponds to the original particle density at $t$ = 0,
averaged over a 2-cm-wide bin centred on the beam axis (as shown in
panel a)).}
  \label{fig_beam_diffusion}    
\end{figure}

\begin{figure}[htbp]
  \begin{center}
      \makebox{\epsfig{file=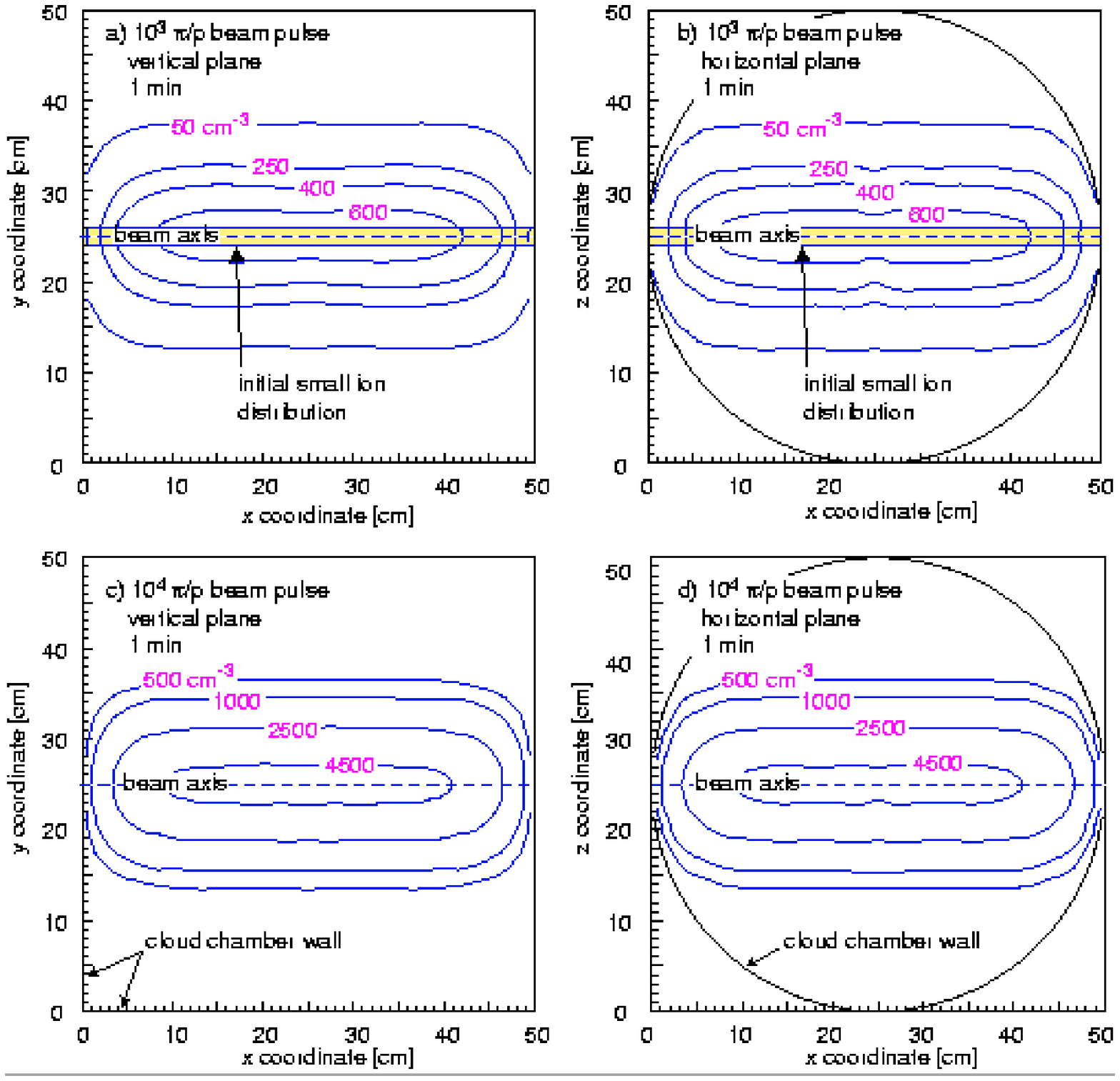,width=160mm}}
  \end{center}
  \caption{The small ion concentrations (ion pairs cm$^{-3}$) in the
cloud chamber one minute  after the passage of a beam pulse, at  293K
and 101 kPa (standard conditions).  The upper figures correspond to a
beam pulse of 
$10^3$ $\pi$/p and show the concentrations in a) the vertical ($xy$)
plane   and b) the horizontal  ($xz$) plane through the beam axis.  The
lower figures  correspond to a beam pulse of $10^4$ $\pi$/p in c) the
$xy$ plane and d) the
$xz$ plane. Both thermal diffusion and ion-ion recombination effects are
included.  The initial small ion distributions were generated according
to the beam profile as described in the caption for
Fig.\,\ref{fig_beam_diffusion}.}
  \label{fig_beam_ionisation}  
\end{figure}

\subsubsection{Ion and aerosol losses to the walls}
\label{sec_particle_losses}

The finite size of the cloud chamber will result in the loss of charged
particles and aerosols to the walls by diffusion.  When an aerosol
touches a wall, it attaches by van der Waals forces and is lost.  For
the purposes of the study described here we have also assumed that a
small ion is lost (or, equivalently, its charge is neutralised) if it
collides with a wall.

We have estimated diffusional losses in the cloud chamber by a
2-dimensional Monte Carlo simulation.  Two atmospheric conditions were
considered: a) standard conditions (T = 293~K and P = 101 kPa), and b)
10 km altitude (223~K and 26 kPa).  In the case of small ions
(Figs.\,\ref{fig_diffusion_losses}a and b), the fiducial region in the
centre of the chamber is unaffected for periods of about one minute for
the 10 km conditions, and for longer periods at ground level
conditions.  This is to be compared with ion-ion recombination lifetimes
in the atmosphere (i.e. for small ion concentrations of a few $\times
1000$  cm$^{-3}$) of about 5 minutes or less (Section 
\ref{sec_atmospheric_ions}).  Since copious small ions can be
replenished inside the cloud chamber by the particle beam each 14.4 s of
the PS accelerator cycle, these losses of small ions can easily be
accounted for and  compensated. We also note that these calculations
assume small ions, whereas the actual mobilities of the ions will be
reduced due to the rapid attachment of molecules and conversion to
complex ions (Appendix
\ref{sec_cosmic_rays}). 

Aerosols have much smaller mobilities than small ions (Table
\ref{tab_mobilities}) and so they are expected to be less affected by
wall losses.  This can be seen in  Figs.\,\ref{fig_diffusion_losses}c)
and d), which show the losses of 10 nm-diameter aerosols after a period
of 1 hr.  The losses are relatively modest, even for the 10 km
conditions (the 50\% loss region is a distance of about 2.5 cm from the
walls).  Since the width of the loss region scales quite slowly with
time as $\sqrt{t}$, the cloud chamber dimensions are sufficient for
measuring long aerosol growth processes that may last several hours,
without appreciable loss of aerosol particles.   In summary, therefore,
we conclude that the 50 cm dimension for the cloud chamber is a suitable
choice from considerations of diffusional losses of ions and charged
aerosols.

\subsubsection{Diffusion of beam ionisation} \label{sec_beam_diffusion}

We have also studied the diffusion of ionisation from the beam region of
the cloud chamber. The results of a 2-dimensional simulation
(Fig.\,\ref{fig_beam_diffusion}) indicate that distinct beam and no-beam
regions of the chamber exist only for exposure times less than about one
minute. Beyond that time, small ions will be spread throughout the
chamber (although, if required, they can be rapidly cleared in a few
seconds by application of the clearing field).  Therefore experiments
that involve beam exposure times of more than about one minute will
require separate runs for the beam/no-beam measurements. We note that
the charged particle and ion profiles inside the chamber can be measured
experimentally with deep piston expansions that activate all particles
and small ions into droplets. 

In order to determine the beam intensity required to reproduce natural
levels of ionisation, we have performed a 3-dimensional simulation
including the effects of ion-ion recombination. 
 Figures \ref{fig_beam_ionisation} c) and d) show that the ion-pair
concentration in the region around the beam about one minutes after a
single pulse of $10^4$ $\pi$/p is similar to the natural concentration
at 15 km altitude (i.e. a few $\times 10^3$ i.p.\,cm$^{-3}$, see
Fig.\,\ref{fig_cosmics}).  The effects of ion-ion recombination losses
can be seen by comparing the results at a factor 10 lower beam intensity
(Figs. \ref{fig_beam_ionisation} a) and b)), which indicate a smaller
reduction (factor 7.5) in ionisation concentration.

\section{Data interpretation and cloud modelling}
\label{sec_cloud_simulations}

The purpose of these investigations is twofold. Firstly, we will use the
laboratory results to incorporate ion-mediated aerosol and cloud
processes into  models. Secondly, we wish to establish whether the
experimental results have important consequences for the behaviour of
aerosols and clouds in the atmosphere.

Although it is clear that atmospheric ion formation has the potential to
influence aerosol and cloud processes, the significance of cosmic rays
must be evaluated within the natural variability of other parameters
that influence aerosols and clouds. Ultimately, modelling within the
CLOUD project must address these issues and, specifically, should:

\begin{enumerate}

\item {\it Obtain a quantitative understanding of the effects of 
ionisation on aerosols and cloud droplets, including ice formation.}
This will be achieved by comparing cloud chamber and aerosol flow
chamber observations with simulations using both an aerosol and a cloud
microphysical box model.

\end{enumerate}

With regard to aerosols:

\begin{enumerate}

\setcounter{enumi}{1}

\item {\it Test the importance of ion effects on aerosols under
atmospheric conditions} for different cosmic ray fluxes and compare the
magnitude of the ion effect on aerosols with the range of natural
variations in other parameters. This will be achieved by performing
aerosol box model simulations under atmospheric conditions. This work
will focus particularly on the aerosol properties that are important for
cloud formation.

\end{enumerate}

With regard to clouds:

\begin{enumerate}

\setcounter{enumi}{2}

\item {\it Develop parameterisations of droplet activation and ice
formation suitable for inclusion in cloud models.} This will be achieved
by comparing box model simulations of droplet activation and freezing
with cloud chamber results.

\item {\it Assess whether the direct effect of cosmic rays on cloud
droplets and ice nucleation has a discernible effect on cloud formation
and development.} This work will involve simulations of different cloud
types under atmospheric conditions for a range of cosmic ray fluxes
typical of their variation with altitude.

\item {\it Assess the relative importance of cosmic rays on cloud
development.} The effect of ion formation will be compared with natural
variations in other parameters.
\end{enumerate}

\subsection{Modelling of aerosol processes}

\subsubsection{Evaluation of experimental results}

Aerosol microphysical models will be used to simulate the formation,
growth and coagulation of aerosols measured by CLOUD.  Radius-resolved
models of \hso--\ho\ aerosols including the processes of nucleation,
coagulation, evaporation and condensation are appropriate for this task.
Such models exist at the University of Helsinki and the University of
Leeds. These models simulate the competition between condensation of
\hso\ on existing aerosols and on new aerosols, leading to formation of
stable clusters. A recent study by Yu and Turco \cite{yu00} has
demonstrated the importance of air ions in the formation of new aerosols
in the natural environment, although the potentially important process
of attachment of air ions onto existing aerosols \cite{clement92} was
not considered. The magnitude of such effects remains uncertain and
should be constrained in the models by comparison with experimental
results from CLOUD.

The models, including ion-mediated nucleation and coagulation, will be
compared with results from the cloud chamber and flow chamber, with and
without the simulated cosmic ray source. According to
Fig.~\ref{fig_gcr_to_cn}, there are three processes that should be
constrained in model simulations of the experimental results:

\paragraph{Charge-enhanced nucleation:} Experiments investigating ion
effects on nucleation of H$_2$SO$_4$/H$_2$O and
H$_2$SO$_4$/H$_2$O/NH$_3$ aerosols will be carried out in the flow
chamber and expansion chamber. The results will be compared with the
classical ion-induced nucleation theory. Possible enhancements with
respect to the classical nucleation theory will be parameterised in
terms of gas concentrations, temperatures, and ionisation rates.

\paragraph{Charge-enhanced growth of nucleated clusters:}  The growth
rate of freshly nucleated clusters can be affected by charges carried by
the nuclei. The Helsinki group is developing a theory for the
enhancement of the condensation rate of H$_2$SO$_4$ molecules on charged
clusters due to ion-dipole interactions. The theory will be tested
against growth experiments carried out with the CLOUD apparatus.

\paragraph{Charge-enhanced coagulation of nucleated clusters:}  The
freshly nucleated clusters will coagulate within themselves and with
larger (pre-existing) particles. The theory for charge-enhanced
coagulation is well established and can be tested by CLOUD. The CLOUD
results will be the first to observe directly the coagulation rates of
charged nanometre aerosols.

\subsubsection{Evaluation of atmospheric aerosol effects}

Simulation of the CLOUD observations will enable a model including
ion-mediated aerosol effects to be validated. The next element of the
work is to apply this model to atmospheric conditions. While the
experimental measurements will be performed under realistic
temperatures, pressures and atmospheric concentrations of
\hso\ and \ho, an important missing element in the initial proposed
programme is the influence of a larger background-aerosol mode. Such a
mode in reality will scavenge condensable vapours and nucleated aerosols
and suppress new particle formation.  These effects may require
experimental study by CLOUD. Using a model to scale the rates of new
aerosol formation (determined by the beam intensity) to atmospheric
conditions will be essential.

Idealised atmospheric box model simulations will be performed to study
the influence of several quantities that affect the conversion of
freshly nucleated particles into CCN (Fig.~\ref{fig_gcr_to_cn}):

\begin{description}

\item[1. H$_2$SO$_4$ concentrations.] The \hso\ concentration is
determined by production from oxidation of SO$_2$ by OH radicals and
loss due to scavenging and aerosol nucleation. Realistic \hso\
concentrations are therefore determined in a model simulation by
specifying the background aerosol surface area and SO$_2$ and OH mixing
ratios.

\item[2. Background aerosols.] A range of aerosol size distributions
ranging between those typical of the free troposphere, the pristine
marine boundary layer and polluted conditions will be used.

\item[3. Humidity and temperature] will be specified to be appropriate
for the different atmospheric regions in item 2.

\item[4. Ion source strengths] vary between about 2 and 20 ion
pairs/(cm$^3$s) between the boundary layer and the upper troposphere.
Fluctuations in these source strengths at each altitude will be tested.

\item[5. Sensitivity to  nucleation and coagulation rates]. These
quantities, determined from the laboratory experiments, are likely to be
uncertain and the effect of this on the calculations will need to be
assessed.
\end{description}

Model simulations taking account of the full range of natural variations
in the above quantities will be performed to establish whether
variations in the cosmic ray flux are likely to have a significant
effect on aerosol processes. In particular, we wish to establish whether
there is an effect on aerosols in the CCN size range of about 100 nm,
and which parts of the atmosphere are likely to be affected the most.

\subsection{Modelling of cloud processes}

The planned experiments in the cloud chamber will yield data that can be
applied directly to numerical models of clouds. Two elements to a cloud
modelling study based on these experiments can be identified:

\begin{enumerate}

\item Detailed simulation of the cloud chamber results to constrain
microphysical parameters affected by ionisation.

\item Simulation of real clouds incorporating the effects identified in
item 1. 

\end{enumerate}

The simulation of real clouds should take particular account of the
relative effects of cosmic rays against a background of other natural
variations.

\subsubsection{Simulation of cloud chamber results}

The cloud chamber experiments will enable the effect of ionisation on
cloud droplet formation and freezing to be studied under well controlled
conditions. It should be noted that the cloud formation processes in the
chamber (principally droplet formation and freezing) are considerably
simpler than the complex interacting processes that govern real cloud
development (which include also coagulation, riming, sedimentation,
droplet breakup etc). This relative simplicity will enable the cloud
chamber observations to be compared with results of a highly simplified
box model incorporating commonly used descriptions of droplet formation
and freezing. Comparisons with and without beam should enable the
effects of air ions to be identified and parameterised in the models.

\subsubsection{Simulation of real clouds} The purpose of this element of
the work is to determine whether the properties of natural clouds are
sensitive to the effects of cosmic ionisation. There are \emph{direct}
effects and \emph{indirect} effects that must be studied. The direct
effects include charge-enhanced CCN activation into cloud droplets
(Section~\ref{sec_activation}) and ice nucleation
(Section~\ref{sec_ice_formation}), while the indirect effect of interest
is the charge-enhanced conversion of CN into CCN
(Section~\ref{sec_nucleation}). 

In particular, we wish to examine the effect on the following cloud
properties:
\begin{enumerate}
\itemsep-1mm
\item Cloud droplet number concentration (CDNC);
\item Precipitation formation in non-glaciating and glaciating
clouds\footnote{\emph{Glaciating} clouds are those in which ice
formation occurs. Clouds in which this does not occur are often termed
``warm clouds''}, which could influence cloud lifetime;
\item Cloud reflectivity and cloud emissivity. 
\end{enumerate} There have been numerous studies of the factors
affecting these cloud parameters. We anticipate that the contribution of
small changes in the mean ionisation state of the atmosphere to the
variance in these cloud properties will be small, though perhaps
non-negligible. We also cannot exclude that a significant contribution
to the variance could arise due to large amplitude fluctuations in
ionisation rate.

We seek to understand whether ionisation variations are a {\it
statistically significant} contributor to variations in these cloud
properties. Our approach is to perform a Monte Carlo simulation using
two detailed cloud microphysical models, one simulating marine stratus
(the MISTRA model) and one simulating cumulus (described in the
Appendix~\ref{sec_cloud_models}).  This requires running the cloud
models a number of times with uncertain input parameters sampled at
random from within appropriate ranges. Such a global stochastic approach
also allows the synergistic effects between various input parameters to
be determined. It is also the most appropriate technique for examining
the significance of relatively minor variations in a parameter of
interest. The models we propose to use are suitable for the calculation
of these cloud properties but sufficiently simple to enable multiple
calculations.

The important model variables to include in such a Monte Carlo
simulation include, but are not restricted to, the following:
Atmospheric thermodynamic state, which determines updraft velocities in
clouds and entrainment rates and, in the case of stratus, the depth of
the mixed layer; temperature and specific humidity; time of day, which
affects radiative imbalance and entrainment at the top of marine
stratus. To test the effect of perturbations to the CCN abundance
(Section~\ref{sec_nucleation}) we will use the output of the aerosol
models incorporating the charge-enhanced processes of nucleation,
condensation and coagulation. 

Several important questions need to be addressed at the single cloud
level. The results of these detailed cloud resolving model studies could
be parameterised for inclusion in atmospheric general circulation models
to assess global changes.

\begin{enumerate}

\item What is the effect of aerosol changes on cloud droplet number
densities and what effect does this have on cloud reflectivity (at solar
wavelengths) and emissivity (in the infrared)? Can modelled changes in
cloud optical properties be seen in satellite data? Cloud  model studies
will be needed using, as input, aerosol distributions calculated from
off-line aerosol microphysical models.

\item How do changes in cloud droplet number affect the precipitation
process in clouds? Changes in the efficiency of precipitation  will
influence cloud lifetime and, hence, on a statistical basis, average
cloudiness. However, a direct relationship between changes in
precipitation rate and cloudiness will be difficult to establish. The
key point here is to establish the direction and approximate magnitude
of any change.

\item How do changes in the efficiency of ice formation affect cloud
thermodynamics (heat transport) and precipitation efficiency? Increased
efficiency of ice formation in an individual cloud caused by a greater
cosmic ray flux might be expected to lead to increased rainfall and
decreased cloudiness, which would be opposite to the correlation that
has been observed.

\end{enumerate}

\section{Accelerator and beam} \label{sec_accelerator_and_beam}

In principle there are alternatives to a particle accelerator beam for
providing a source of ionising radiation, such as UV or radioactive
sources.  However we consider that a particle accelerator beam has
several advantages and is the best choice.  Firstly a few-GeV hadron
beam comprising protons and pions most closely duplicates the
composition and energy of the cosmic rays and hadronic showers found in
the upper atmosphere. The known timing of the beam pulse is also
important for experiments such as ice nucleation where expansions are
necessary at a precise time either before or after the beam exposure.  
A particle beam has the advantages that it is well collimated and
easily traverses the material of the windows and gases of the cloud
chamber and flow chamber, with negligible scattering. Finally, the beam
intensity is easily adjusted over a broad range and can be measured
precisely.

\begin{figure}[htbp]
  \begin{center}
      \makebox{\epsfig{file=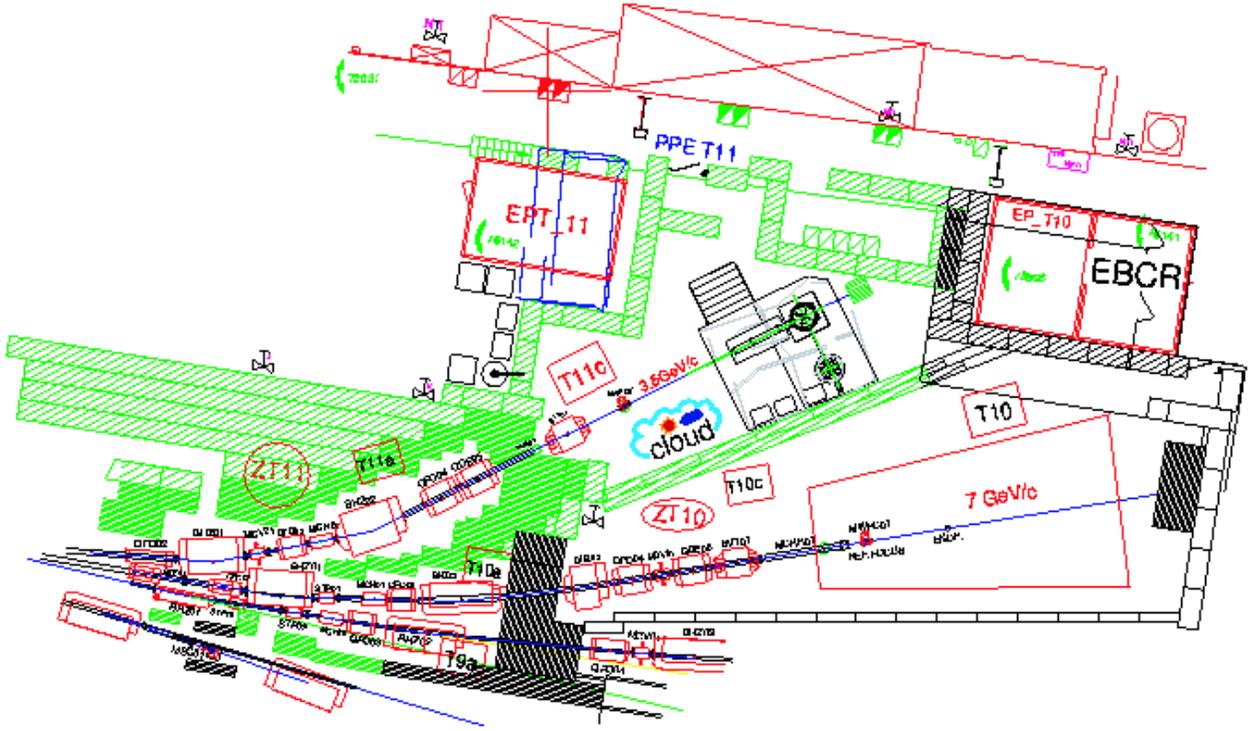,width=165mm}}
  \end{center}
  \vspace{-5mm}
  \caption{Experimental area layout of the CLOUD experiment at the CERN
PS in the T11 beamline of the East Hall.}
  \label{fig_experimental_area}    
\end{figure}

\subsection{Experimental area} \label{sec_experimental_area}

It is proposed to install CLOUD at the CERN PS (Proton Synchrotron) in
the T11 secondary beamline of the East Hall.   A preliminary study of
the experimental installation  has been carried out
(Fig.\,\ref{fig_experimental_area}).  The cloud chamber itself would be
permanently installed on the beamline for the duration of the
experiment.  However, to provide space for other experiments to test
detectors during periods when CLOUD is not running, we have designed the
flow  chamber to be dismountable from the beamline.  The associated gas
and aerosol generation and analysis systems, together with the liquid
cooling and hydraulic systems, would be permanently installed near the
apparatus.  A dedicated counting room for CLOUD is foreseen to be
installed above the current T11 counting room.

\subsection{Beam requirements} \label{sec_beam_requirements} 

The T11 beamline can deliver up to $\sim 10^6$
$\pi$/p per pulse (0.46 s duration). All lower intensities can be
delivered. We assume one pulse would be available for CLOUD per
supercycle (14.4 s). The T11  beam energy, which is selectable  in the
range $\sim$1--3.5 GeV, matches the main part of cosmic ray spectrum .
The transverse size of the beam at the detector (a minimum of about $7
\times 7$ mm$^2$ FWHM) is well-suited to the detector requirements. 

We can compare the T11 beam with cosmic ray intensities in the upper
troposphere as follows.  One pulse of $10^6$ $\pi$/p deposits $\sim 2
\cdot 10^7$ ion-pairs cm$^{-3}$ in air at a pressure corresponding to 15
km altitude.  At this altitude the molecular number concentration is
about $5 \cdot 10^{18}$~cm$^{-3}$, so a single beam pulse deposits about
4 pptv ion-pairs (and also produces about 6 pptv NO and OH molecules).
This beam flux is equivalent to  about $5 \cdot 10^5$~s  (5 days) of
cosmic rays at 15 km altitude, and the initial ion-pair concentration is
about a factor $10^4$ higher than at 15 km altitude.  The difference in
these two factors is due to the ion lifetime in the atmosphere before
recombination or scavenging, which is of order 100 s.

After beam passage, the ion concentration in the beam region will fall
due to diffusion and to losses from ion-ion recombination and from
scavenging by aerosols.   The effects of diffusion and recombination
have been simulated, as described in Section~\ref{sec_beam_diffusion}.
The results show that the ion-pair concentration in the region around
the beam about one minutes after a single pulse of $10^4$ $\pi$/p is
similar to the natural concentration at 15 km altitude (i.e. a few
$\times 10^3$ i.p.\,cm$^{-3}$, see Fig.\,\ref{fig_cosmics}).

In summary, the T11 beam will allow a  flexible exposure for CLOUD, 
with time-averaged ionisation cpncentrations ranging from below the
natural level of cosmic rays (with the use of the clearing field) up to
about a factor 100 times higher.

\subsection{Beam counter system} \label{sec_beam_counter_system}

The beam counter system (BCS) measures the particle beam intensity and
its transverse profile.  A telescope of plastic scintillation counters
(T1, X1, Y1 and T2 in Fig.\,\ref{fig_flow_chamber}) measures the beam
flux.  The dimensions of the counters are approximately 
$60 \times 60$~mm$^2$ transversely and 8 mm thickness.  A larger counter
(H1) at the downstream end monitors the beam halo.  The counters X1 and
Y1 are beam hodoscopes to measure the beam profile in the horizontal and
vertical projections, respectively.  They are constructed from $4
\times 4$ mm$^2$ scintillator strips (or perhaps from scintillating
fibres), read out with multianode phototubes, such as the Hamamatsu 
H6568/R5900-M16 with $4 \times 4$ photocathode pixels, each of size $4
\times 4$~mm$^2$.  Scintillation counters will also be mounted above the
cloud chamber to tag the presence of cosmic rays during data taking (not
shown in the figure).

\section{Planning}  \label{sec_planning}

\subsection{Cost estimates and responsibilities}
\label{sec_costs}

Table \ref{tab_cost_estimate} summarises the cost estimates for CLOUD 
and the proposed responsibilities of the collaborating partners. 
Following the standard CERN procedure, these financial responsibilities
are to be considered as preliminary until they are agreed by the
national funding agencies.  When finalised they will be enacted by
signing a memorandum of understanding between CERN and the collaborating
institutes. 

\begin{table*}[htbp]
  \begin{center}
  \caption{Cost estimates and proposed responsibilities.}
  \label{tab_cost_estimate}
 \vspace{5mm}
  \begin{tabular}{| l  l  | r |  l | } 
    \hline  
 &  \textbf{Item} & \textbf{Cost} &\textbf{Responsible}  \\  
 &   &  \textbf{\small [kCHF]} & \textbf{Institutes}   \\
  \hline
  \hline
\multicolumn{2}{|l|}{\textbf{Cloud chamber \& flow chambers:}}  &  &  
\\
 & Mechanical assembly \& field cage & 150  &  DSRI-Copenhagen \\
 & Piston \& hydraulic system & 95  &  CERN \\
 & Liquid cooling system & 75  &  CERN \\  
 \hline
\multicolumn{2}{|l|}{\textbf{Optical readout:}}  &  &   \\
 & CAMS detector \& laser$^\dagger$ & 200  &  Vienna \\
 & CCD camera system \& Xenon flasher & 75  &  RAL \\
 \hline
\multicolumn{2}{|l|}{\textbf{Gas \& aerosol generation systems:}}  & 
&   \\
 & Gas \& aerosol systems  & 50  &  \small Helsinki, FMI, Kuopio
\\
 & Water vapour system  & 20  &  \small UM-Rolla \\
\hline
\multicolumn{2}{|l|}{\textbf{Physical, chemical and ion analysis:}}  & 
&   \\
 & Aerosol physical analysers  & 100  &  \small Helsinki, FMI, Kuopio,
UM-Rolla \\ 
 & Linear quadrupole mass spectrometers & 1100  & MPIK-Heidelberg \\
 & ToF mass spectrometer & 175  & Aarhus, Ioffe \\
 & Ion mobility detector & 75  &  Reading \\
   \hline
\multicolumn{2}{|l|}{\textbf{Beam measurement:}}  &  &   \\
 & Beam counter system & 30  &  Lebedev \\
  \hline
\multicolumn{2}{|l|}{\textbf{DAQ \& computing:}}  &  &   \\
 & DAQ \& slow control system & 65  &  - \\
 & Atmospheric cloud simulations & 50  &  Leeds, Mainz \\[0.5ex]
  \hline
  \multicolumn{2}{|l|}{\textbf{Infrastructure:}}  &  &   \\
 & Support structure \& counting room & 40  &  CERN \\[0.5ex]
    \hline
   \multicolumn{2}{|r|}{\textbf{Total:}} & \textbf{2300} &   \\
	  \hline    
  \end{tabular}
  \end{center}
 $^\dagger$ Does not include contract manpower, estimated at 240 kCHF.
  \begin{center}
  \caption{Construction milestones.}
  \label{tab_milestones}
  \vspace{5mm}
  \begin{tabular}{| r  l | l|}
  \hline
  \multicolumn{2}{| c  |}{\textbf{Date}} & \textbf{Activity} \\
  \hline
  \hline
 Dec  & 2000  &  Finish detailed design \& prototyping  \\
 Jan  & 2001  &  Start construction \\
 Nov  & 2001  &  Begin installation at CERN PS \\
 Mar  & 2002  &  Checkout with beam   \\
 May  & 2002  &  Start data taking   \\[0.5ex]
  \hline
  \end{tabular}
  \end{center}
\end{table*}

In addition to PS beam-time, the collaboration requests CERN provide the
following:
\begin{enumerate}
\item Joint construction of the cloud chamber in collaboration with the
Danish Space Research Institute (DSRI).  This concerns:
\begin{enumerate}
\item Construction of subsystems for which CERN has special expertise,
namely the piston and hydraulic system (which duplicates the BEBC
design) and the liquid cooling system.
\item Engineering design support for the cloud chamber and associated
equipment. 
\end{enumerate}
\item Technical coordination of the overall experiment.
\item Experimental research group (research staff and fellows).
\item Infrastructure support (offices, lab space, experimental counting
room, computing support, etc.). 
\end{enumerate}

The present estimate of the total construction cost is 2300~kCHF (Table
\ref{tab_cost_estimate}), assuming an efficient use of existing
equipment and the recuperation of some materials and infrastructure
equipment from previous, terminated experiments. These estimates are
based on delivered value on-site and do not include the technical
infrastructure used in the home institutes.  Neither do the above
estimates include manpower costs since they are accounted for in the
home laboratory infrastructure costs.  Discussions are proceeding with
the national funding agencies in order to secure adequate home
laboratory infrastructure support for the experiment. 

Plans are underway to apply for complementary funding from the European
Union for equipment support as well as to fund a number of
Ph.D.\,students and research assistants to participate in the
experiment. 

The annual maintenance and operation costs of CLOUD are estimated at 100
kCHF.  These cost would be paid by the collaborating institutes,
including CERN, following the standard CERN procedure. Since the total
power consumption of the experiment is under 60 kW, the electricity
costs are  negligible.

\subsection{Technical coordination} \label{sec_technical_coordination}

Technical coordination covers the following tasks:

\begin{itemize}
\item Establishment of a master schedule linked to each institute's own
schedule, and monitoring of the progress.
\item  Management of all interfaces between the various components, and
resolution of any technical inconsistencies.
\item Management of the engineering data.
\end{itemize}

Concerning the last item, all relevant documents (technical
specifications, engineering and calculation notes, documents from
manufacturers, etc.) would be organised within the framework of the CERN
EDMS (Engineering Data Management System). In particular, the CDD (CERN
Drawings Directory) software will be used to provide a central library
of all relevant drawings produced by the  institutes or manufacturers.
Once stored and archived in the CDD, the drawings are available to all
registered collaborators through a web interface.  The documents can be
retrieved via the web and viewed under HPGL (Hewlett Packard Graphical
Language) format, which is the equivalent of Postscript for technical
drawings. Evolution of the detector design and the release of new
technical drawings can then be efficiently  managed in a controlled and
readily-accessible manner.

\subsection{Milestones} \label{sec_milestones}

Assuming the SPSC recommends approval of the experiment in May 2000, the
detailed design and prototyping will be finished by the end of 2000 and
the construction will take place in 2001.
 The initial checkout of the equipment with beam is expected to start in
March 2002.  A preliminary estimate of the time required to complete the
data taking is about 3 years, but at present this should be considered
as approximate.

\section{Conclusion} \label{sec_conclusion}

We have joined together in an unprecedented team of atmospheric
physicists, solar-terrestrial physicists and particle physicists, to
offer CERN the opportunity to make a major contribution to environmental
science. Clouds are the engines of the weather, yet they are only
sketchily understood at the microphysical level. Now satellite
observations give empirical evidence for an astonishing link between
high-energy physics and meteorology, namely that cosmic rays from the
Galaxy may influence cloud formation and behaviour. 

If this link between cosmic rays and clouds is real, it provides a major
mechanism for climate change. During the 20th Century the cosmic rays
reaching the Earth diminished by about 15\% as a result of increasing
vigour in the solar wind, which scatters the cosmic rays. The inferred
reduction in cloud cover could have warmed the Earth by a large fraction
of the amount currently estimated to be due to man-made carbon dioxide. 
In that case, the effect of carbon dioxide may have been overestimated. 
If, on the other hand, the link to cosmic rays proves to be illusory,
present diplomatic efforts to curb emissions of carbon dioxide will be
more strongly supported scientifically. Settling the issue, one way or
the other, is therefore an urgent task.  

To find out whether cosmic rays can affect cloud formation, and if so
how, we propose to simulate the cosmic rays with a beam of charged
particles from CERN's Proton Synchrotron, in the CLOUD experiment. The
beam will pass through a cloud chamber where the atmosphere is to be
represented realistically by moist air charged with condensation nuclei
and trace condensable vapours, and chilled by expansion.  We shall be
able to compare processes when the beam is present and when it is not.
    
Our team brings to the planning of the CLOUD experiment a thorough
knowledge of atmospheric and cloud science, derived from field and
laboratory experiments, airborne data-gathering, satellite
observations, and microphysical theories. We also possess considerable
experience with cloud chambers and their optical readout, and with the
mass spectrometers and aerosol particle detectors required for chemical
and physical analyses of aerosols and ions.  The design of the cloud
chamber draws on CERN's own experience, notably with the Big European
Bubble Chamber. 

Space research has shown how "big science" can make spectacular
contributions to knowledge of the environment by bringing together
experts from different disciplines. As an analogous multidisciplinary
team for particle physics, we do not claim total originality. More than
100 years ago C.T.R. Wilson invented the cloud chamber to investigate
weather phenomena. It evolved into a prime instrument for particle
physics. Now the wheel of history turns and we go back to Wilson's
concept to investigate the possibility that the Earth's atmosphere acts
like one big cloud chamber that echoes the whims of the Sun.  

\section*{Acknowledgements}

We would like to warmly thank for their advice and important
contributions to this proposal, J\"{u}rg~Beer, J.J.\,Blaising, Nigel
Calder, Luc Durieu, Gregory Hallewell, Alain~Herv\'{e},
\mbox{Markus~Nordberg,} Michael Price, David Ritson, Jean-Pierre
Riunaud, Thomas Ruf, Tom Taylor, Johann Tischhauser, Robert Veenhof and 
Alan Watson.

\newpage 
\section*{APPENDICES}  \vspace{2ex}

\addcontentsline{toc}{section}{APPENDICES}

\appendix

In the following we provide some general background information on:
cloud physics (Appendix \ref{sec_cloud_physics}), 
aerosol-cloud-climate interactions   (Appendix
\ref{sec_aerosol_cloud_climate}), classical operation of a Wilson cloud
chamber (Appendix \ref{sec_wilson_chamber}), cosmic rays in the
atmosphere (Appendix \ref{sec_cosmic_rays}), and cloud models (Appendix
\ref{sec_cloud_models}).

\section{Cloud physics}  \label{sec_cloud_physics} 

\subsection{General properties of clouds}
\label{sec_general_properties_of_clouds}

Clouds are principally composed of ice and water, and in some cases they
may contain electrified particles.  High and low altitude clouds of
limited vertical extent may contain only one principal phase of liquid
water or ice; for example, cirrus clouds in the upper troposphere
(altitudes greater than about 8 km) are composed principally of ice.
Clouds are often classified as stratiform or convective.  The bulk
properties of clouds have a marked influence on climate, although
whether clouds cause a net warming or cooling of the atmosphere depends
on the cloud properties and altitude.

Water vapour is effectively a gaseous constituent of atmospheric air,
and its concentration can be determined by its gaseous partial pressure.
At any given temperature there is an associated maximum value of partial
pressure due to water vapour, the {\it saturation vapour pressure}. Air
containing sufficient water vapour to generate the saturation vapour
pressure is said to be saturated, and has a relative humidity of 100\%.
Most natural clouds form when air becomes saturated upon cooling, either
adiabatically (cooling upon expansion of rising air) or isobarically
(e.g.\, by radiatively cooling near the ground, producing fog).
Occasionally it is possible for the saturation vapour pressure to be
exceeded in localised regions, known as {\it supersaturation} ($SS$). In
the atmosphere, this is never greater than an excess relative humidity
of few percent, due to the abundance of small particles (or aerosol) on
which the water can condense. Many different kinds of aerosol are
capable of acting as {\it condensation nuclei} (CN) but it is the subset
able to permit condensation (activation) at atmospheric $SS$ ({\it cloud
condensation nuclei}, CCN) which are of greatest interest for the
formation of atmospheric clouds.

If the temperature is below 0\degc, liquid water droplets may persist in
a thermodynamically unstable {\it supercooled} state from which freezing
may be readily initiated. Freezing can be initiated by heterogeneous or
homogeneous nucleation. In {\it heterogeneous nucleation}, the
supercooled water freezes in the presence of a suitable ice nucleus.
{\it Homogeneous nucleation} occurs if cooling is continued further, and
all supercooled water in atmospheric clouds becomes ice at about
-40\degc\  by this process. Suitable ice nuclei are very rare in the
atmosphere, and it is certainly the case that only a very small (but
variable) fraction of the atmospheric aerosol is able to initiate ice.
Consequently a significant fraction of atmospheric clouds contain
supercooled water. At temperatures between about -5 and -10\degc, ice
multiplication processes can occur. When a particle freezes at these
temperatures, mechanical stresses during freezing lead to the ejection
of small ice fragments, which in turn act as efficient ice nuclei.

The {\it droplet number density} in water clouds depends upon the
cooling rate of air as it enters the cloud (since this affects the  peak
$SS$ that is reached) and upon the concentration, size and chemical
composition of the CCN. Although highly variable, typical number
densities are a few $\times$ 100 cm$^{-3}$ in continental  clouds and a
few $\times$ \mbox{1--10~cm$^{-3}$} for maritime clouds. Number
densities are usually higher in convective clouds than in stratiform
clouds.

\begin{figure}[tbp]
  \begin{center}
      \makebox{\epsfig{file=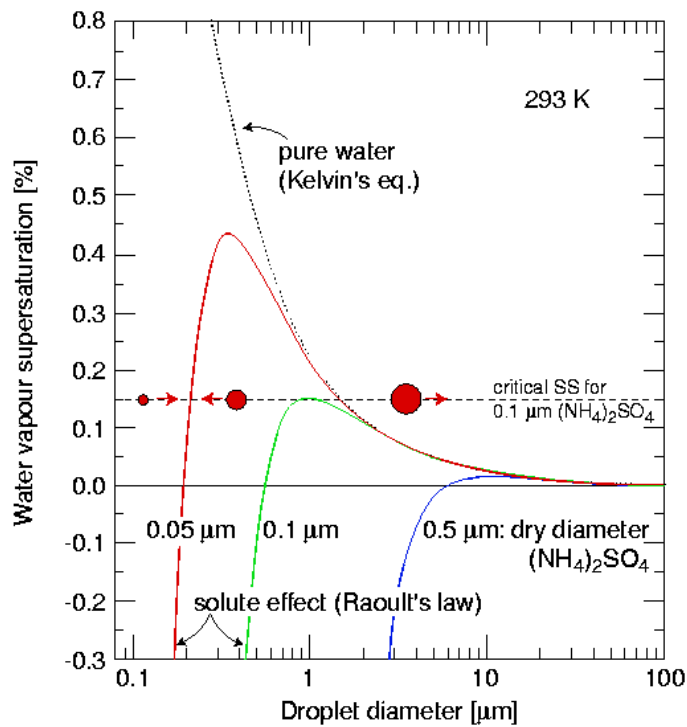,height=95mm}}
  \end{center}
   \vspace{-5mm}
  \caption{K\"{o}hler curves showing the equilibrium water vapour
supersaturation at 293~K for droplets of pure water (dotted curve) and
for droplets containing various masses of dissolved (NH$_4$)$_2$SO$_4$
(solid curves) vs.\,diameter of the droplet \cite{seinfeld}.  The water
vapour supersaturation, $SS$ (\%)  $= (p/p_0 - 1) \cdot 100$, where $p$
is the partial pressure of the water vapour and $p_0$ is the saturated
vapour pressure over a plane surface of water at this temperature. In
the indicated example, an ambient water vapour $SS$ of 0.15\% (dashed
line) exceeds the critical value for all ammonium sulphate aerosols with
dry diameter \mbox{$\ge$ 0.1 $\mu$m}. These aerosols will therefore
activate and grow into cloud droplets, whereas smaller aerosols remain
as unactivated haze particles. Droplets below their corresponding
equilibrium curve will shrink by evaporation whereas those above will
grow by condensation (the indicated droplets correspond, for example, to
a dry diameter of 0.05 $\mu$m).}
  \label{fig_kohler}    
\end{figure}

Activation of a CCN occurs when the water vapour $SS$ exceeds a 
critical value.  This can be seen from the K\"{o}hler curves
(Fig.\,\ref{fig_kohler}), which show the equilibrium $SS$ (and
therefore equilibrium vapour pressure) over droplets of various sizes
and containing various masses of dissolved salts. The equilibrium $SS$
of pure water droplets increases with decreasing radius  due to the
effect of  curvature (Kelvin's equation; ln~$(p/p_0) \propto 1/r$). 
However dissolved salts reduce the equilibrium $SS$ due to a reduction
of the molar concentration of the water (Raoult's law; $p/p_0 \propto -
1/r^3$). The latter effect dominates at small radii, i.e. at high solute
concentrations.

Once activated, droplets grow by diffusion of water vapour. {\it
Diffusional growth} is rather slow and it is unusual for droplet radii
to exceed 20--30 $\mu$m by this process.  Cloud droplets typically
attain sizes of 10 $\mu$m within a few minutes but take over an hour to
reach 100 $\mu$m (the growth time $\propto r^2/SS$). Droplet {\it
collision and coalescence} (which occurs when droplets collide while
falling under the influence of gravity) takes over as the principal
growth mechanism for radii greater than about 20 $\mu$m.

For clouds to generate {\it rainfall}, some drops must grow to
precipitable sizes of 1 mm or greater. This is achieved not by
diffusional growth of water droplets, but by collision and coalescence
of droplets or formation of ice ({\it glaciation}). Ice formation
usually occurs in only a small fraction of the cloud droplets, allowing
these to grow by vapour diffusion preferentially due to the lower vapour
pressure of ice crystals compared with water droplets. The ability of a
cloud to generate rain is an important factor in determining its
lifetime.

\begin{figure}[htbp]
  \begin{center}
      \makebox{\epsfig{file=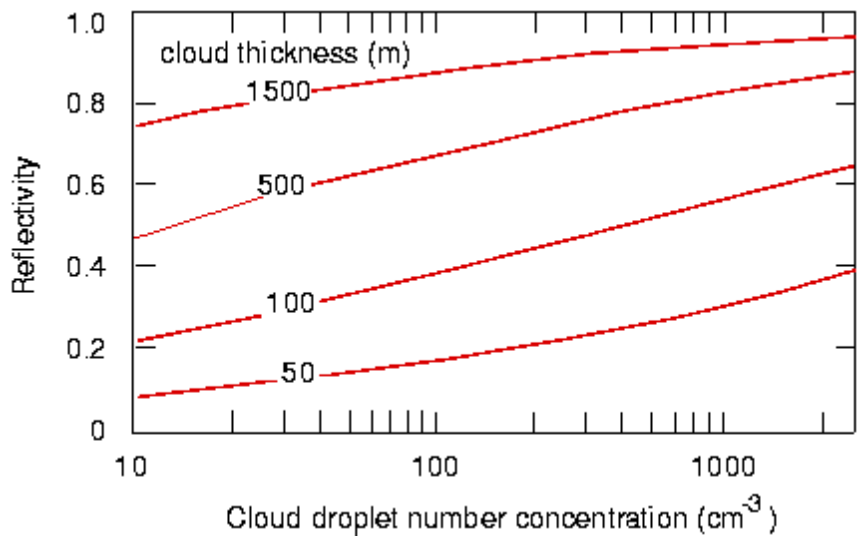,height=70mm}}
  \end{center}
  \caption{The variation of cloud albedo with cloud thickness and
droplet number concentration for a fixed liquid water content of
\mbox{0.3 g m$^{-3}$} \cite{schwartz}.}
  \label{fig_cloud_reflectivity}    
\end{figure}

Clouds have a high {\it reflectivity} at visible wavelengths and
contribute significantly to the net {\it albedo} of the planet, reducing
the net amount of radiation that is absorbed at the Earth's surface. Any
changes in cloud reflectivity would therefore have potential
implications for the radiative balance of the climate system. Cloud
reflectivity depends on the mass of condensed water (termed the {\it
liquid water content}), the depth of the cloud and the droplet number
density. Figure
\ref{fig_cloud_reflectivity}  shows the variation of cloud reflectivity
with cloud depth and cloud droplet number density for a fixed liquid
water content of \mbox{0.3 g m$^{-3}$} (most data confirm that there is
little or no dependence of liquid water content on cloud droplet number
density). Any perturbation of cloud droplet number density by cosmic
rays could lead to changes in cloud reflectivity.

\subsection{Aerosols and cloud condensation nuclei}
\label{sec_aerosols_and_ccn}

Atmospheric aerosols are liquid or solid particles or clusters of
molecules suspended in the air. The atmosphere contains significant
concentrations of aerosols, sometimes as high as 10$^6$ cm$^{-3}$, with
diameters spanning over four orders of magnitude from a few nm to a 100
$\mu$m or so. Aerosols are often classified as either {\it primary} or
{\it secondary}. Primary particles are those injected directly into the
air (e.g.\, by wind erosion, sea spray, pollen, etc). These may be
either of natural or anthropogenic origin. Secondary aerosols are those
created by gas-to-particle conversion of molecules ({\it nucleation}).

\begin{figure}[htbp]
  \begin{center}
      \makebox{\epsfig{file=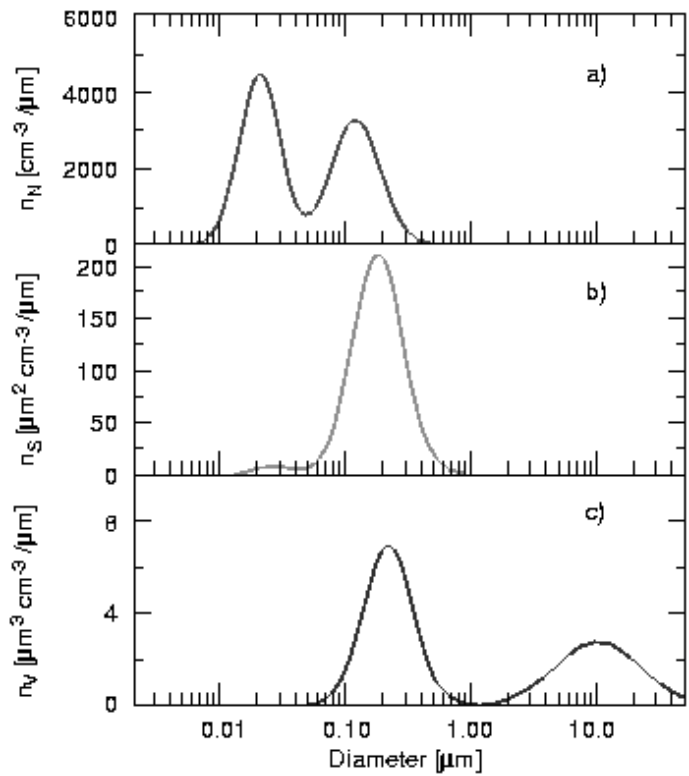,width=90mm}}
  \end{center}
  \caption{ Typical remote continental aerosol distributions: a) number,
b) surface area and c) volume \cite{seinfeld}.}
  \label{fig_aerosol_distributions}    
\end{figure}

Aerosol composition varies widely depending upon geographical location
and proximity to specific sources. It also varies significantly across
the size distribution, with the smallest aerosols often being clusters
of volatile species such as sulphuric acid and water (formed from
gas-to-particle conversion) and the largest often being inorganic salts
and dust particles. Although it is not possible to define a canonical
aerosol distribution, observations indicate that aerosol sizes can often
be described by quasi-distinct modes. For example, a typical remote
continental aerosol  (Fig.\,\ref{fig_aerosol_distributions}) is
composed of a {\it nucleation} mode (median radius $\simeq$ 0.01
$\mu$m), an {\it accumulation} mode (median radius $\simeq$ 0.1
$\mu$m) and a {\it coarse} mode (median radius $\simeq$ 10 $\mu$m).

The size and composition distribution are determined by many different
processes. They include the following: nucleation of new particles at
small sizes, formation of larger primary particles, coagulation,
deposition, condensation of soluble gases, phases changes (e.g.\,
crystallisation), in-cloud transformations and washout, and water
uptake. Inorganic aerosols are usually weakly acidic, with the most
common aqueous cationic components being H$^+$, ammonium (NH$_4^+$) and
sodium, and with anionic components sulphate, chloride, nitrate. Such
aerosols are  hygroscopic. Aerosols can also be partly or wholly
composed of organic compounds derived from plant waxes and combustion
sources. These aerosols may be either hygrophobic or hygroscopic.

Nucleation of new aerosols is an important way in which anthropogenic
emissions can perturb the properties of aerosols. An important source of
new aerosols is the oxidation of sulphur dioxide in the presence of
water vapour to give sulphuric acid, which readily combines with water
vapour to form new aerosols. This is believed to be primarily a {\it
homogeneous} nucleation process, although atmospheric ions from cosmic
rays may affect the rate of formation of new aerosols. The rate of
nucleation is extremely difficult to predict from theory and must be
measured in the laboratory under well-controlled conditions.

Hygroscopic aerosols serve as centres for the growth of cloud droplets.
Such {\it cloud condensation nuclei (CCN)} are normally aqueous acids or
dissolved salts. Not all aerosols grow into cloud droplets. The number
that do depends on their composition, the $SS$  attained in the cloud,
and the size distribution of aerosols competing for moisture. An
increased aerosol number density in general leads to an increased cloud
droplet number density.

\subsection{Atmospheric electricity}
\label{sec_atmospheric_electricity}

Electric fields present in the atmosphere vary between fair weather
values of typically 100~Vm$^{-1}$ at the surface to about 100 kVm$^{-1}$
in thunderstorms before a lightning discharge. The atmospheric electric
circuit, which involves a global current of 2000 A, is thought to be
sustained by thunderstorms continually active around the tropics. This
current between the ionosphere and the Earth's surface flows throughout
the atmosphere, in regions of disturbed and undisturbed weather, and is
carried by vertical migration of molecular ions.

The molecular ions (more traditionally known as small ions) are formed
continuously in the atmosphere by galactic cosmic rays (see Appendix
\ref{sec_atmospheric_ions}) and, close to the land surface, also by
natural radioactivity. The ions produced are rarely single species but
clusters of water molecules around a central ion. Typical atmospheric
ion concentrations at low altitudes in unpolluted air and fine weather
are about 500 ions cm$^{-3}$, and vertical charge fluxes caused by
conduction in fair weather conditions are typically 2 pA m$^{-2}$
($\sim$12~ions cm$^{-2}$s$^{-1}$). Ions are removed by
self-recombination and by attachment to atmospheric aerosols. Ion
concentrations are generally lower in clouds than in clear air
conditions due to scavenging by droplets. Chemical differences between
positive and negative ions leads to a natural asymmetry in which the
negative ions have a slightly higher (20\%) electrical mobility: this
asymmetry leads to a small average charge on atmospheric particles.

Electrification in disturbed weather (thunderstorms) is not completely
understood but is thought to result from ice-water interactions in
thunderclouds, in which vertical convective motions lead to differential
transport of rising ice crystals and falling soft hail. Their
interaction leads to electrification, the sign and magnitude of charge
exchange depending on the temperature at which they interact which, in
turn, is a function of height.  It has also been suggested that
negatively-charged aerosols have faster activation and growth than
positive aerosols and that this may contribute to the charge-separation
mechanism in thunderclouds.

\section{Aerosol-cloud-climate interactions}
\label{sec_aerosol_cloud_climate}

The increased concentration of greenhouse gases such as water vapour,
carbon dioxide and methane may have large effects on the Earth's climate
by enhancing the greenhouse effect. Much attention has also been paid to
the possibility that the increased concentrations of aerosol particles
can have a cooling effect on the global climate
\cite{charlson91,charlson92,charlson94}.

Clouds can affect on electromagnetic radiation fluxes in the Earth's
atmosphere  by scattering and absorption. Locally the net effect is not
very evident (e.g.\,because of different kinds of surfaces)
\cite{curry}. According to measurements, however,  the global net effect
of today's clouds on the climate is probably cooling
\cite{hartmann91}  and thus opposite to the effect of greenhouse gases.
The effects of clouds  on the radiation fluxes depend on their depth,
liquid water content and cloud droplet size distribution
\cite{twomey77} (Appendix \ref{sec_general_properties_of_clouds}). 

The formation and growth of cloud droplets occur on pre-existing aerosol
particles. The characteristics of the  pre-existing particle
distribution strongly affect the  developing cloud droplet distribution.
Field experiments show that the pre-existing aerosol particle
distribution is usually composed of mixed particles, i.e. particles
including both hygroscopic and insoluble components
\cite{zhang,svenningsson92, svenningsson94}.  According to recent model
studies by the Helsinki group \cite{korhonen96a,kulmala96}, the soluble
mass of pre-existing aerosol particles is the most important factor
(other than dynamics) in determining the developing cloud droplet 
distribution. Hygroscopic material decreases the saturation vapour 
pressure of water vapour above the surface of a solution droplet and
makes the formation of a cloud droplet easier. The condensation of
different gaseous substances on the aqueous particles during their
growth increases the hygroscopicity of the particles. When the amount of
the hygroscopic material in the particles increases it also decreases
the critical supersaturation needed for their activation to cloud
droplets
\cite{kulmala93,korhonen96b} (see Fig.\,\ref{fig_kohler}). It has been
recently shown that thin organic films can also affect the water vapour
pressure above the growing droplet's surface and thus influence cloud
droplet formation \cite{schulman}. 

The amount of water vapour available during the growth process  depends
on various dynamical aspects and on the particle size distribution. When
an air parcel exceeds 100\% relative humidity the most hygroscopic
particles activate first and start to consume water vapour. The maximum
supersaturation achieved is therefore smaller when there is a  larger
amount of soluble material in the nascent aerosols.  Condensation of
some strongly hygroscopic gaseous  substances on the particles
simultaneously with water vapour can allow a larger fraction of the
pre-existing particle distribution to grow to cloud droplets. 

Recently we have investigated the effects of changing hygroscopicity 
(the availability of condensable material in the  gas phase) on the
formation of cloud droplets \cite{kulmala98,laaksonen97}.  In the 
simulation, water (H$_{2}$O), nitric acid (HNO$_{3}$), hydrochloric acid
(HCl) and ammonia  (NH$_{3}$) vapours condense on mixed particles
composed of ammonium sulphate ((NH$_{4})_{2}$SO$_{4}$) and an insoluble
substance. In order to take into account the dynamical aspects, we used
an adiabatic air parcel 
 model \cite{kulmala93}. The initial mean radii, standard deviations and
number concentrations as well
 as the fraction of soluble salt in the particles and the concentration
of gaseous HNO$_{3}$, HCl and NH$_{3}$ in ambient air are varied about
their observed values in marine and continental air (Table
\ref{tab_gases}). The Helsinki group has also investigated how the
optical thickness of the simulated clouds  varies as a function of
initial ratio of acid and ammonia mixing ratios.

\begin{table}[htbp]
  \begin{center}
  \caption{Measured ranges of gaseous ammonia, nitric acid and
hydrochloric acid concentrations in marine and continental air.}
\label{tab_gases}
  \vspace{5mm}
\begin{tabular}{|l|c|c|c|}
\hline
\textbf{Trace gas}  & \textbf{Marine} & \textbf{Continental} 
  & \textbf{Continental} \\[-0.8ex]
 &  & \textbf{(rural or semi-rural)} & \textbf{(polluted)} \\ 
 & \textbf{[ppb]} & \textbf{[ppb]} & \textbf{[ppb]} \\  
\hline
\hline
 NH$_3$ & 0.1- 14 & 0.2 - 5.4 & 7 - 100   \\   
 HNO$_3$    & 0.05 - 0.8 & 0.2 - 1.8 & 4 - 8  \\ 
 HCl & $<$ 0.1 - 1.0 & 0.08 - 1.4 & 1  - 23   \\[0.5ex] 
  \hline
  \end{tabular}
  \end{center}
\end{table}

Our simulations confirm that gas phase hygroscopicity has a large effect
on the formation and properties of clouds 
\cite{kulmala98,laaksonen97}. The K\"ohler curves are also  modified 
\cite{kulmala97,laaksonen98}. In general, this study shows that the
increase in hygroscopicity of growing droplets implies an increasing
number of cloud droplets and increasing optical thickness of a  cloud.
The way in which the soluble mass is distributed among  aerosol
particles and between the size modes also seems to be important
\cite{ogren}. When condensed on aqueous solution droplets, nitric and
hydrochloric acids enhance the activation of cloud droplets.  This is
also the case when no other condensable trace gases present. In
contrast, ammonia alone cannot enhance the number of activated cloud
droplets significantly. But in real atmospheric situations where
particles also take up strong acids, ammonia has a clear increasing
influence on the activation of cloud droplets.  This is due to the fact
that larger amounts of free hydrogen ions enhance the formation of
ammonium ions significantly. On the other hand condensed ammonia
neutralises the solution droplets  efficiently, which allows larger
amounts of condensable acids to be taken up.

We have also studied how the optical properties of the resulting cloud
droplet distribution vary with changing hygroscopicity. The change of
optical thickness is usually very large (the maximum  $\Delta \tau
/\tau$ is 1.6, where $\tau$ is the optical thickness).  This is the case
even at  moderate concentrations, for example 
$\Delta \tau / \tau$ is around 0.5 when the sum of acids is 1~ppbv and
ammonia is 1 ppbv. This change in optical thickness means that the
albedo of an individual cloud may change significantly. In summary,
therefore, it is important to estimate the  global indirect forcing due
to the effect of condensable gases on cloud droplet formation, and to
understand the possible influence of galactic cosmic rays on this
process.

\section{Classical operation of a Wilson cloud chamber}
\label{sec_wilson_chamber}
 
\begin{figure*}[htbp]
  \begin{center}
      \makebox{\epsfig{file=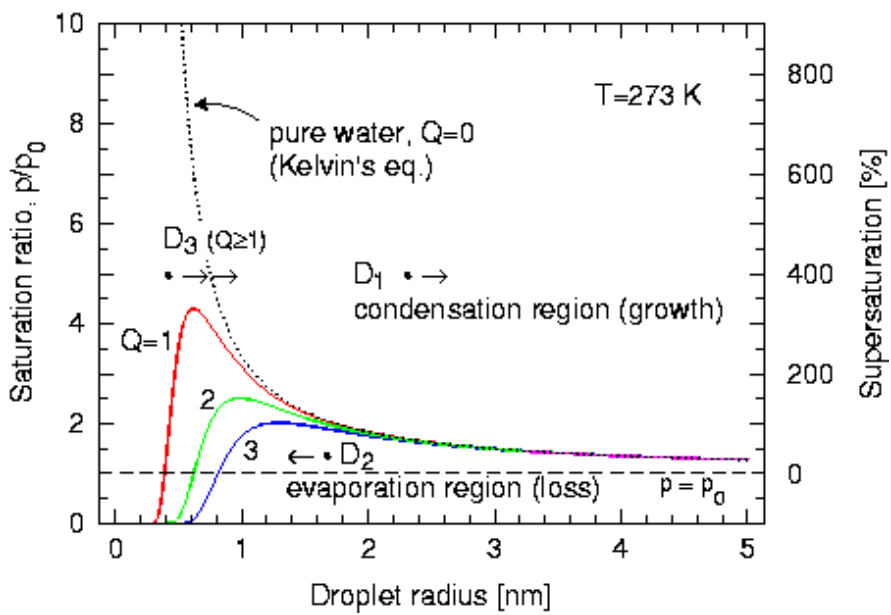,height=80mm}}
  \end{center}
 \vspace{-5mm}
  \caption{Thomson-Kelvin curves showing the equilibrium saturation
ratio for charged droplets of pure water at 273 K. The curves are
labelled according to the droplet electronic charge, Q$e$.  The
saturated vapour pressure, p$_0$, corresponds to equilibrium with a
plane water surface (0\% supersaturation).  Droplets formed in the
region above the curves will grow (condensation region) while those
below will shrink (evaporation region).}
  \label{fig_droplet}    
\end{figure*}

The principle of classical operation of a Wilson cloud chamber
\cite{wilson,segre} for the detection of charged particles can be
understood from Fig.~\ref{fig_droplet} which shows the water vapour
pressure equilibrium curves for small droplets carrying a charge $Qe$:
\[\log _e\left( {{p \over {p_0}}} \right)={M \over {RT\rho }}\left[ {{{2
\gamma } \over r}-\left( {{{Q^2e^2} \over {4\pi \epsilon _0r^2}}\cdot {1
\over {8\pi r^2}}} \right)} \right],\] 
 where $p$ is the vapour pressure, $p_0$ the saturated  vapour pressure
at a plane water surface,  $R$ the gas constant, $T$ the absolute
temperature,
$\gamma$ the surface tension, $M$ the molecular weight,
$\rho$ the density, $\epsilon _0$ the permittivity of free space  and
$r$ the radius of the droplet.  The curves divide an upper region of
vapour pressure in which water droplets grow by condensation (e.g.
droplet D$_1$ in Fig.~\ref{fig_droplet}) from a lower region where they
shrink by evaporation (e.g.\,D$_2$).  

The curves are similar in shape to the K\"{o}hler curves for the
equilibrium vapour pressure over aerosol-nucleated droplets, but much
higher supersaturations are required for activation of charged droplets
of pure water (compare Fig.\,\ref{fig_kohler} with the right-hand scale
of  Fig.\,\ref{fig_droplet}). The (purely electrostatic) influence of
droplet charge is significant only for very small radii $\sim$1 nm.  To
set this scale, the effective radius of one water  molecule is
$\sim$0.2 nm and a droplet of radius 0.5 nm contains about 18
molecules.  The charge reduces the equilibrium water vapour pressure by
an increased attraction of the polar water molecules.   Evidence
suggests that negative ions form larger condensation nuclei than do
positive ions, and hence the former activate more rapidly and at lower
water vapour supersaturations. 

The necessary supersaturation in a cloud chamber is generated by fast
adiabatic expansion.  For an ideal gas and an adiabatic expansion,
\begin{eqnarray*}
  P_1V_1^\gamma      & = & P_2V_2^\gamma, \; \; \; \mbox{and}  \\
  T_1V_1^{\gamma-1} & = & T_2V_2^{\gamma-1},
\end{eqnarray*} 
 where $\gamma$ is the ratio of specific heats
$C_P/C_V$ (1.40 for  air and saturated water vapour).  For example, when
a mixture of air and saturated water vapour at 1 atmosphere and 293~K is
adiabatically expanded by a large volume ratio
$V_2/V_1 =1.3$, the temperature falls to 264~K and the resultant 
saturation ratio is 5.9 (i.e. 490\% supersaturation).
 Under these conditions all small ions ($Q \geq1$) are activated and
rapidly grow to visible droplets.  

In contrast, very small adiabatic expansions are required to generate
the low supersaturations of a few $\times$ 0.1\% found in the atmosphere
(Fig.\,\ref{fig_kohler}). For example, a 100~$\mu$m piston movement of
the CLOUD expansion chamber at one atmosphere and 293~K gives a volume
expansion ratio $dV/V = 2.0 \cdot 10^{-4}$ and a pressure expansion
ratio $dP/P = 2.8 \cdot 10^{-4}$.  This reduces the gas temperature by
0.023 K and produces a supersaturation of 0.12\%.  These considerations
set the required temperature stability of 0.01 K for the CLOUD expansion
chamber.

\begin{figure*}[htbp]
  \begin{center}
      \makebox{\epsfig{file=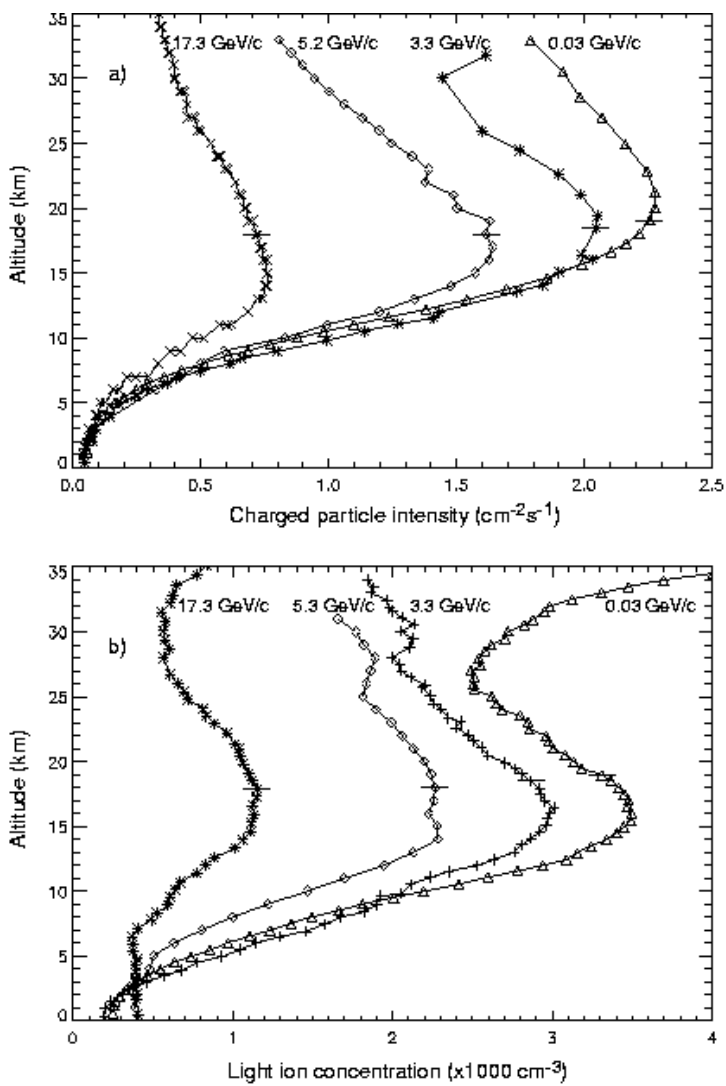,width=95mm}}
  \end{center}
   \vspace{-5mm}
 \caption{ a) The  charged  particle intensity 
  and b) the  light ion concentration  vs.\,altitude, measured at
several latitudes with cutoff rigidities as indicated
\cite{ermakov97}. The intensity measurements in a) were made with the
omnidirectional Geiger counter described in
Appendix~\ref{sec_atmospheric_cr_meaurements}. In b) the five-point
running  averages are shown. The data were recorded in or near 1990,
corresponding to a sunspot maximum (but without solar proton events),
i.e during a cosmic ray
\emph{minimum}. The horizontal  bars  show the typical experimental
statistical errors.}
  \label{fig_cosmics}    
\end{figure*}

\section{Cosmic rays in the atmosphere}
\label{sec_cosmic_rays}

\subsection{General characteristics of atmospheric ions}
\label{sec_atmospheric_ions}

The maximum cosmic ray fluxes occur at altitudes of 15--20 km, where the
charged particle intensities vary between about 0.8 and 2.3
cm$^{-2}$s$^{-1}$, depending on geomagnetic latitude 
(Fig.\,\ref{fig_cosmics}a)
\cite{ermakov97}.  In air at 101 kPa pressure ($1.22 \times
10^{-3}$~gm~cm$^{-3}$), the mean ionisation density for minimum-ionising
charged particles is 68 ion-pairs cm$^{-1}$. At high altitudes the
fraction of heavily-ionising non-relativistic particles becomes
significant and the mean ionisation density is about 110 ion-pairs
cm$^{-1}$, corrected to 1 atmosphere pressure
\cite{ermakov97}.  At 15 km the density of air is
$0.20 \times 10^{-3}$~gm~cm$^{-3}$ and the mean ionisation density is
therefore about 18  ion-pairs cm$^{-1}$ per charged particle. Therefore
the ion-pair production rate by cosmic rays at 15 km altitude is
$q =18 \times$ (0.8--2.3) = (14--41) cm$^{-3}$s$^{-1}$, depending on
geomagnetic latitude.  Free radicals are also created by galactic cosmic
rays and these may lead to  chemically-reactive molecules.  As examples,
about 1--2 OH radicals \cite{crutzen86} and 1.5 NO molecules
\cite{nicolet,jackman,crutzen75} are estimated to be produced per
ion-pair. Mixing ratios of about 0.7 pptv OH or NO are therefore
generated  by cosmic rays per day in the upper troposphere.  

Once created, the charged particles will interact with atmospheric gas
molecules and thereby become converted to complex positive and negative
cluster ions \cite{viggiano95}. Free electrons will rapidly ($\tau
\sim$ 200~ns)  attach to O$_2$, leading to O$_2^-$ as the most important
primary negative ion. Primary positive ions are mostly N$_2^+$, O$_2^+$,
N$^+$, and O$^+$. Both positive and negative primary ions experience
rapid ion-molecule reactions with relatively abundant atmospheric gases
leading to the cluster ions H$^+$(H$_2$O)$_n$ and CO$_3^-$(H$_2$O)$_n$.
The former react further with basic molecules B possessing proton
affinities larger than that of H$_2$O leading to H$^+$B(H$_2$O)$_n$. An
important example for B is acetone (CH$_3$)$_2$CO (proton affinity:
194\,kcal/mol). Negative ions react with acidic molecules, particularly
HNO$_3$ and H$_2$SO$_4$, leading to NO$_3^-$(HNO$_3$)$_m$ and
HSO$_4^-$(H$_2$SO$_4$)$_l$(HNO$_3$)$_k$. The above species have been
observed in the upper troposphere and lower stratosphere by
aircraft-based ion mass spectrometers 
\cite{heitmann83,krieger94}.

The variation of ion concentration with time is given by
\begin{eqnarray} 
 {{dn} \over {dt}}=q-\alpha n^2-\beta n N_a -\gamma n^p
\label{eq_dn/dt}
\end{eqnarray}
 where $q$ [cm$^{-3}$s$^{-1}$] is the ion-pair production rate by cosmic
rays, $n$ = $n_+$ = $n_-$ [cm$^{-3}$] are the positive and negative
small  ion concentrations, $\alpha$ [cm$^3$s$^{-1}$] is the ion-ion
recombination coefficient (about $1.5 \times 10^{-6}$ cm$^3$s$^{-1}$),
$\beta$ [cm$^3$s$^{-1}$] is the ion-aerosol attachment coefficient
(which varies with aerosol size and charge) and 
$N_a$ [cm$^{-3}$] is the aerosol number concentration.  The final term
represents the unknown contribution of ion induced nucleation; $\gamma$
is the ion induced  nucleation coefficient, and the power $p$ may lie
between 1 and 2, depending on the mechanism.  If we assume for the
moment that the principal removal mechanism is ion-ion recombination,
then equilibrium is reached when the ion-pair production and
recombination rates are equal, i.e. when 
 $q = \alpha n^2$.  This implies an equilibrium ion-pair density,
$n = \sqrt {{{q} \mathord{\left/ {\vphantom {{r_{ip}} \alpha }}
\right. \kern-\nulldelimiterspace} \alpha }} = \sqrt{(14-41)/(1.5 \times
10^{-6})} = (3.1-5.2) \times 10^3$~cm$^{-3}$.   Therefore the total
light ion concentration (both + and -) is about  $(6-10) \times
10^3$~cm$^{-3}$.   The measured light ion concentrations
(Fig.\,\ref{fig_cosmics}b) are between a factor 3--5 smaller than this
estimate, indicating extra losses due to ion-scavenging by aerosols and
perhaps also to ion-induced nucleation.

Figure\,\ref{fig_cosmics}b) shows that the maximum light ion
concentration occurs at an altitude of about 17 km, and the
concentration at 8--9~km is about a factor of two lower.  Since the
troposphere has a depth of about 18~km over the tropics, decreasing to
about 8 km over the poles, this indicates that between one third and one
half of the total ionisation from galactic cosmic rays is deposited
directly in the troposphere, depending on latitude.  

From Eq.\,\ref{eq_dn/dt}, the recombination lifetime of an ion is 
$\tau \simeq 1 / (\alpha n)$.  This implies ion lifetimes of about 3--9
min. Additional loss mechanisms, such as aerosol attachment, will
reduce the actual lifetime. The ions drift vertically in the electric
field created by the negatively-charged Earth and the positively-charged
ionosphere.  The field strength is $E
\sim 100$~V/m at sea level, producing an drift velocity for small ions 
of about \mbox{1.5 cm s$^{-1}$}. At an altitude of 15 km, however,  $E
\sim 2$~V/m due to the higher conductivity of the air, and the drift
velocity is only 
\mbox{0.1 cm s$^{-1}$} This results in an  ion drift distance at 15~km
altitude  of only 0.4--1.0~m before recombination.  The ions and free
radicals produced by cosmic radiation will in general be transported
substantially further by tropospheric dynamics, both vertically and
horizontally.

\subsection{Atmospheric measurements of cosmic rays}
\label{sec_atmospheric_cr_meaurements}

We present here a summary of the regular balloon cosmic ray (CR)
observations started by Lebedev Physical Institute in the former Soviet
Union in 1957. The cosmic ray monitoring in the atmosphere consists of
the launching of small rubber balloons, each carrying a special
radiosonde and a charged particle detector
\cite{charakhchyan,bazilevskaya}. 

The charged particle detector comprises two cylindrical Geiger counters,
each of  9.8~cm length and 1.8 cm  diameter, and with steel walls of
0.05~g~cm$^{-2}$  thickness.  The counters are arranged in a vertical
telescope with their axes separated by 2.7 cm and a  7~mm
(2~g~cm$^{-2}$) Al filter inserted between them.   A single counter
provides  the omnidirectional flux of charged particles: electrons with
energy  E$_e >$ 0.2 MeV and protons with  E$_p >$ 5 MeV. A coincidence
of the two counters records the vertical flux of charged particles
within a solid angle of about 1 sr: electrons with E$_e >$ 5 MeV and
protons with E$_p >$~30~MeV.  Both the omnidirectional and vertical
fluxes of charged particles in the atmosphere are measured
simultaneously. A radio pulse caused by a charged particle passing
through one or both counters is transmitted to a ground-level receiver. 
In addition, the air pressure (atmospheric depth) is measured by a
special barometric sensor. In this way the charged particle counting
rates versus atmospheric depth are recorded.  Measurements are made
continuously from ground level up to altitudes of 30--35 km. 

\begin{figure*}[htbp]
  \begin{center}
      \makebox{\epsfig{file=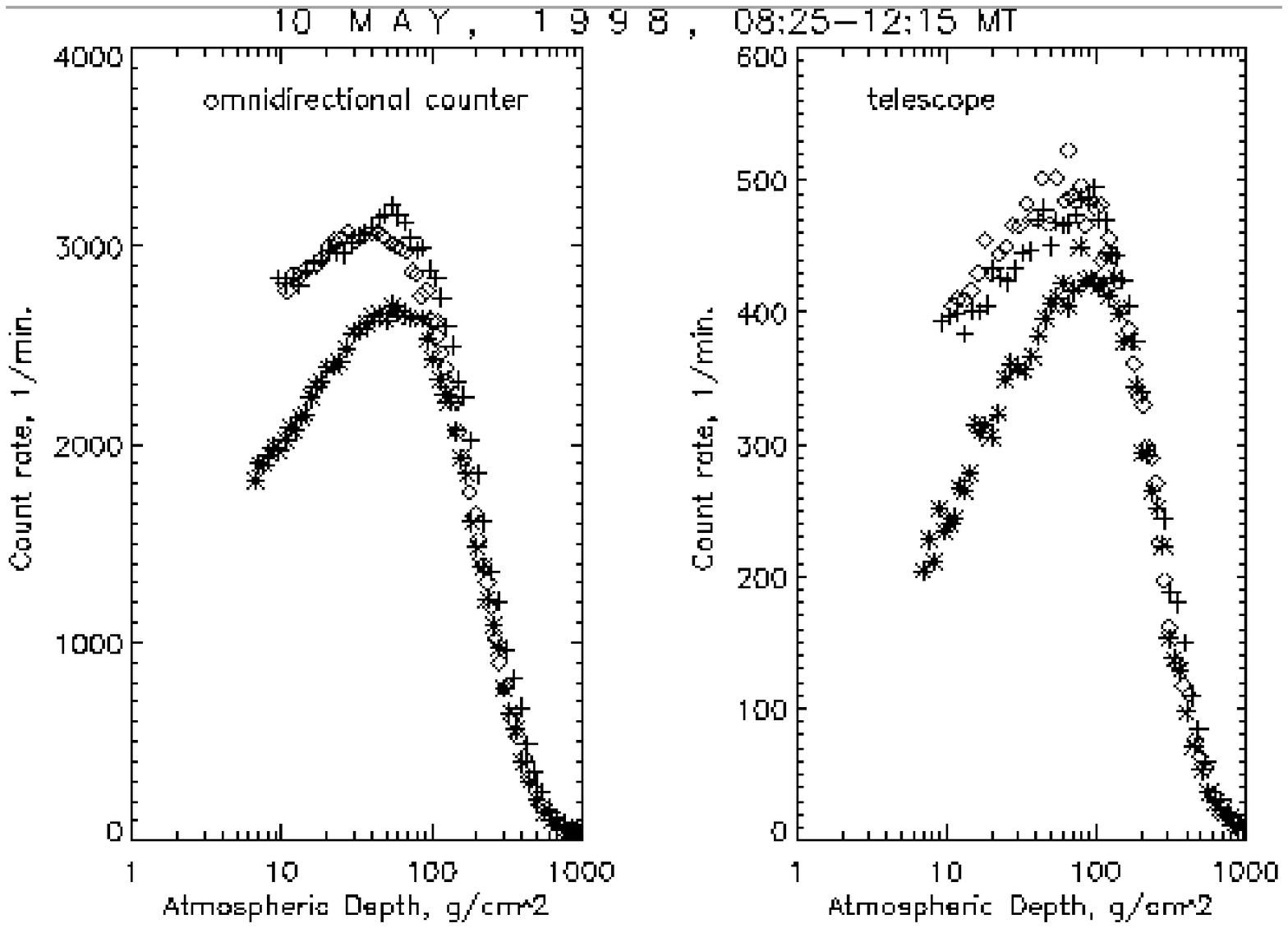,width=100mm}}
  \end{center}
   \vspace{-5mm}
  \caption{Balloon-borne measurements of cosmic ray intensities in
northern latitudes: Mirny (plus sign), Murmansk (diamond)  and Moscow
(asterisk). Data are shown for the  omnidirectional counter (left panel)
and the vertical telescope (right panel).  The data are obtained in May
1998, two years before a solar maximum (cosmic ray minimum).}
  \label{fig_cosmic_1}    
  \begin{center}
      \makebox{\epsfig{file=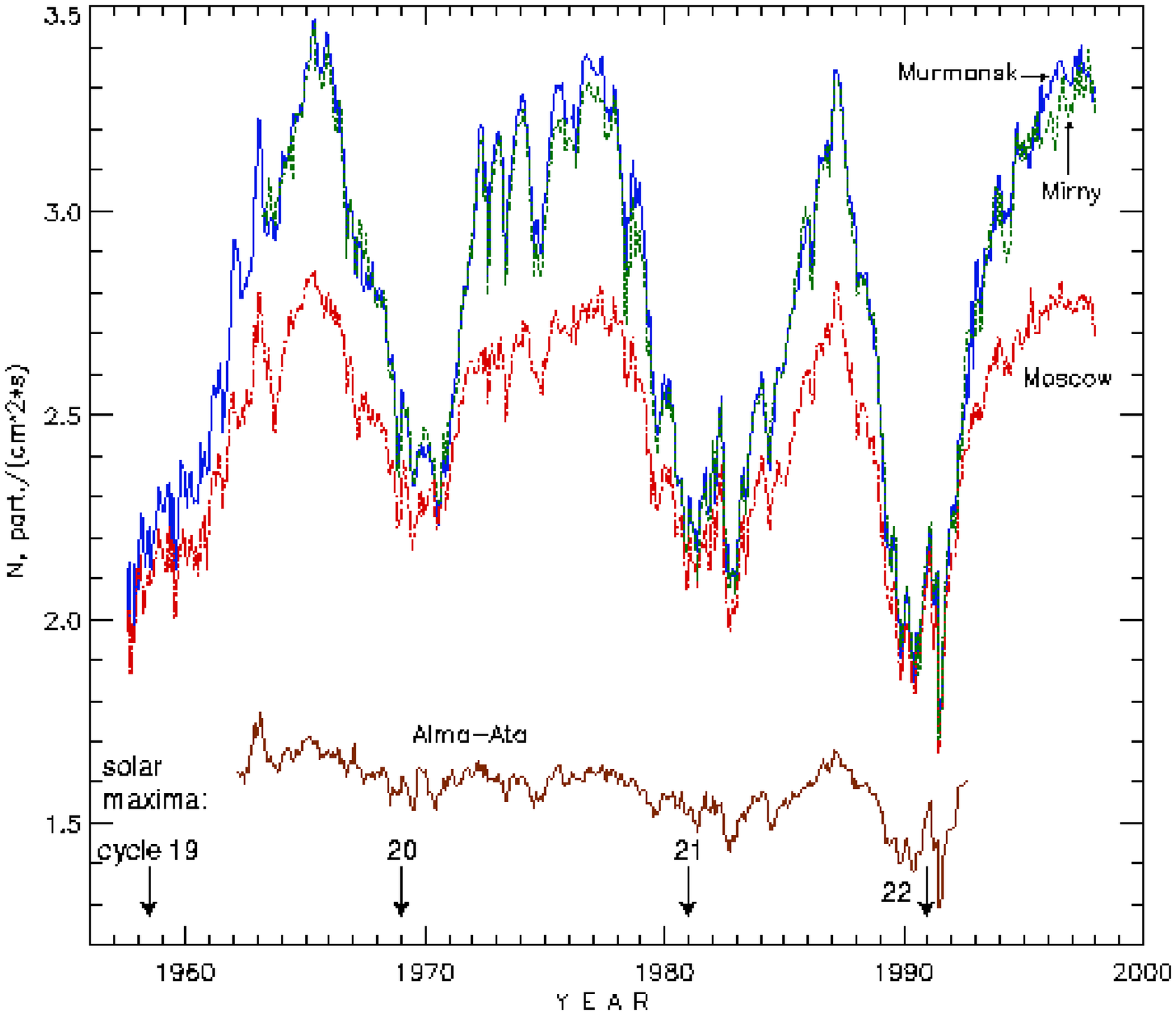,width=115mm}}
  \end{center}
   \vspace{-5mm}
  \caption{Cosmic ray intensity at the Pfotzer maximum in the atmosphere
as measured by the omnidirectional Geiger counter. The lines correspond
to different locations for the balloon flights: Mirny, 
 Murmansk, Moscow and  Alma-Ata. The data of Murmansk and Mirny
practically coincide with each other.  The approximate times of the
sunspot maxima for the last 4 solar cycles are indicated.}
  \label{fig_cosmic_2}    
\end{figure*}

\begin{table}[htbp]
  \begin{center}
  \caption{Summary of the monitoring stations used for long-term cosmic
ray measurements by the Lebedev Physical Institute.}
  \label{tab_stations}
  \vspace{5mm}
  \begin{tabular}{| l l c |}
  \hline
  \textbf{Station} & \textbf{Location} & \textbf{Geomagnetic cutoff} \\
       &  & \textbf{rigidity, R$_c$} \\
       &  & \textbf{[GeV/$c$]} \\
  \hline
    Mirny (Antarctica) & 66.34S, 92.55E  &  0.03  \\
    Murmansk & 68.57N, 33.03E  &  0.6  \\
    Moscow   & 55.56N, 37.11E  &  2.4  \\ 
    Alma-Ata & 43.25N, 76.92E  &  6.7 \\[0.5ex]
  \hline
  \end{tabular}
  \end{center}
\vspace{-8mm}
\end{table}

\begin{figure*}[htbp]
  \begin{center}
      \makebox{\epsfig{file=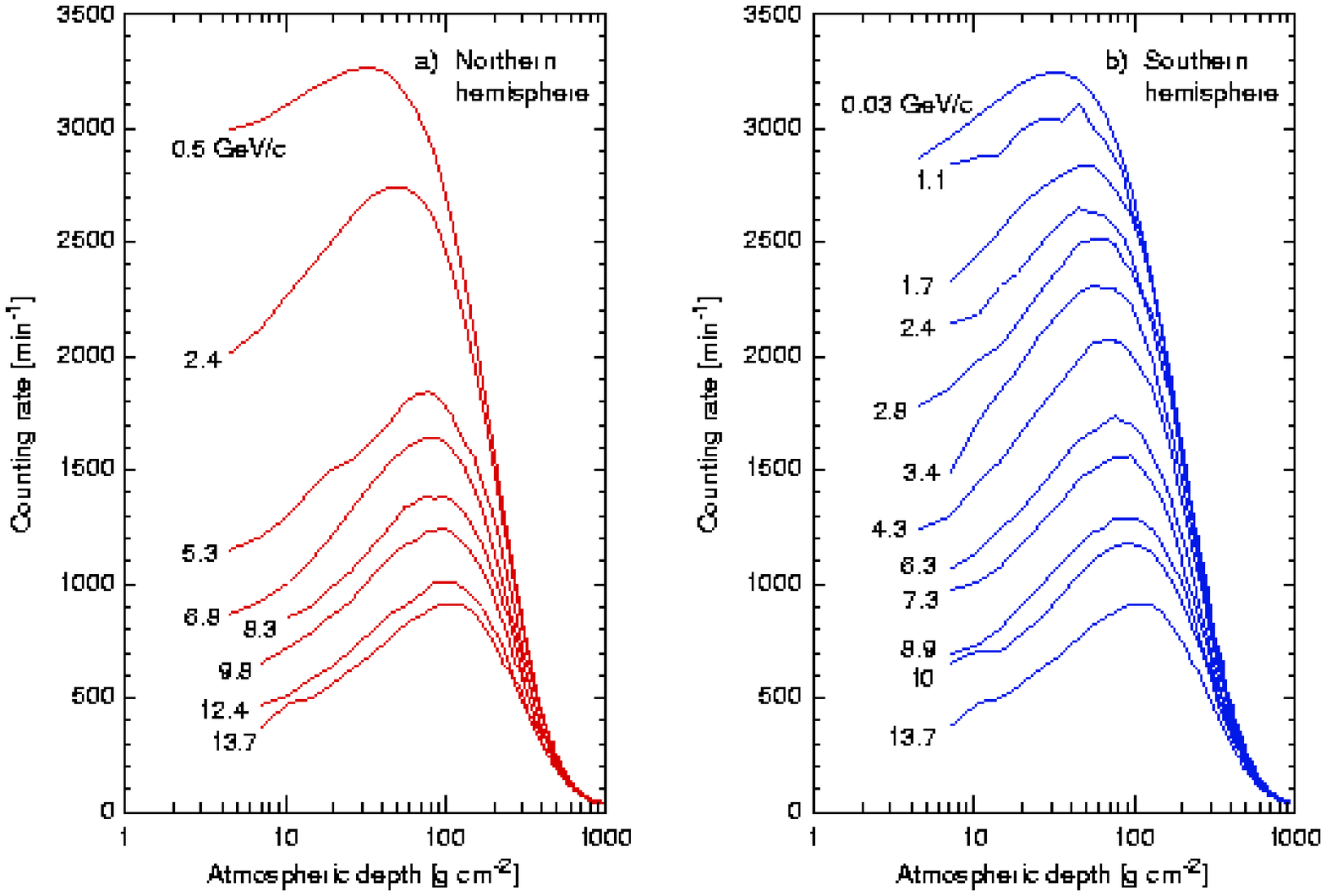,width=160mm}}
  \end{center}
  \caption{Cosmic ray intensities vs.\,altitude and geomagnetic
latitude.  Each curve corresponds to a different latitude with cutoff
rigidity, R$_c$ (GeV/$c$), as indicated. The data were recorded with the
omnidirectional counter during the 1987 (cosmic ray maximum) survey of
a) the northern and b) the southern  hemispheres
\cite{babarykin,golenkov}.}
  \label{fig_cosmic_3}    
\end{figure*}

The monitoring stations for the long-term cosmic ray study are
summarised in Table~\ref{tab_stations}. Balloon launches are made almost
every day at these sites, and have been continuously for about 40
years.  Despite the difference in their geomagnetic cutoff rigidities,
R$_c$, the measured intensities at Mirny and Murmansk are essentially 
equal to each other  because of the atmospheric material cutoff, which
is higher than R$_c$ at these latitudes.

Figure \ref{fig_cosmic_1} shows a typical measurement of CR transition
curves in the atmosphere obtained two years before a solar maximum
(cosmic ray minimum). The data obtained during quiet geomagnetic
conditions show the secondary flux in the atmosphere originating from
galactic CR's alone. The secondary CR fluxes grow with  increasing
altitude until they reach a maximum (which is known as the Pfotzer
maximum, N$_{max}$) at 16--25~km. The value of N$_{max}$ depends on the
geomagnetic latitude and phase of the solar activity cycle. The counting
rates are substantially increased in the upper atmosphere at times of
special events such as solar particle events, coronal mass ejections,
energetic electron precipitations, radioactive clouds, etc.

While propagating inside the heliosphere, the GCR intensity is modulated
by solar activity. The balloon CR observation provide experimental data
over 4 cycles of solar activity for the study of long-term modulation of
GCR's and its effects on atmospheric phenomena.  Figure
\ref{fig_cosmic_2} shows the monthly-averaged Pfotzer maxima measured 
with the omnidirectional counter at the sites  of permanent launchings.
The main feature of the GCR intensity is the 11-year solar cycle. It is
seen that the amplitude of the solar cycle modulation depends strongly
on R$_c$, being up to 70\% at the polar stations, 30\% at Moscow and
about 10\% at Alma-Ata.  These data also provide evidence for a
systematic decrease of cosmic ray intensity over the last 40 years
(Table \ref{tab_gcr_flux}) \cite{stozhkov00}.

To study the global distribution of CR in the atmosphere we carried out
{\it latitude} surveys of CR fluxes in the atmosphere. Such a surveys
were fulfilled during the sea expeditions in 1962--1965, 1968--1971,
1975--1976, 1979--1980 and in 1987. The measurements  covered the R$_c$
range from 0.03 to 15--17 GeV/$c$ \cite{babarykin,golenkov}. In Figure
\ref{fig_cosmic_3} we give an example of the CR transition  curves
obtained at different latitudes in the northern and southern hemispheres
during the 1987 solar minimum (cosmic ray maximum). Each curve averages
the data from the omnidirectional counter recorded over several
radiosonde flights. The ratio of N$_{max}$ at polar latitudes to its
value at the equator is about a factor 3.6.

\section{Cloud models}
\label{sec_cloud_models}

\paragraph{Marine stratus cloud model:}   The MISTRA
(\underline{mi}crophysical \underline{stra}tus) model
\cite{bott98} is a one-dimensional model of stratus clouds including a
detailed treatment of size-segregated aerosol and cloud microphysical
processes, which is essential in this study. Aerosols and cloud are
treated either in a 2-dimensional particle distribution or as internal
mixtures with 40 aerosol and 50 droplet size class. The hygroscopic
growth and evaporation of aerosols in sub-saturated air and their
activation to form droplets are explicitly calculated by solving the
droplet growth equation for all droplets at all humidities. Coalescence
of droplets necessary for the simulation of warm rain is included
\cite{trautmann99}. The model has a vertical resolution of 10 m from
ground level to 1~km (spanning the cloud level), a further 50
logarithmically spaced grid levels up to 2~km (to treat dynamics and
radiative transfer only) and a homogeneous layer up to 50~km (for
radiative transfer only). Entrainment of aerosols is accounted for using
a turbulent kinetic energy closure scheme. Radiative transfer is treated
using a $\delta$-two stream approximation (12 broad-band intervals from
4--100 $\mu$m and 6 intervals from 0.2--4 $\mu$m).

\paragraph{Cumulus cloud model:} Cumulus clouds will be simulated using
a two-dimensional slab-symmetric non-hydrostatic cloud model including
detailed cloud microphysics. Four hydrometeor classes are considered:
water drops, ice crystals, graupel particles and snowflakes (aggregates
of ice crystals). The warm microphysical processes included are 
nucleation of CCN, condensation and evaporation, collision-coalescence,
binary breakup, and sedimentation.  The ice microphysical processes
included are drop freezing, ice nucleation (deposition and
condensation-freezing, and contact nucleation), ice multiplication,
deposition and sublimation of ice, interactions of ice-ice and ice-drop
(aggregation, accretion and riming), melting of ice particles, and
sedimentation of ice particles.  All the microphysical processes are
formulated as kinetic equations and solved using the method of
Multi-Moments
\cite{tzivion87}--\cite{reisin96}.

Each type of particle is divided into 34 bins with mass doubling for
connected bins. The masses at the beginning of the first bin and the end
of last bin for both liquid and solid phases are 0.160$\times$10$^{-13}$
and 0.175$\times$10$^{-3}$ kg, which correspond to drop diameters of
3.13 and 8060 $\mu$m, respectively.

This model has been used in studies concerning the effect of natural and
artificial large and giant CCN on development of cloud particles and
precipitation in convective clouds (e.g. refs.~\cite{yin99a,yin99b}). It
is currently being extended to include the evolution of aerosols inside
various hydrometeors in order to study the venting of trace gases and
aerosols by convective clouds.

\end{document}